\documentclass[twocolumn,amsmath,showpacs,superscriptaddress,prb,aps]{revtex4-2}
\usepackage{graphicx}
\usepackage{bm}
\usepackage{xcolor}   
\usepackage{amssymb} 
\usepackage{amsmath}
\usepackage[utf8]{inputenc}
\usepackage{hyperref}
\usepackage{siunitx}
\hypersetup{
    colorlinks=true,
    linkcolor=blue,
    filecolor=magenta,
    urlcolor=cyan,
}

\usepackage[normalem]{ulem}

\begin{document}

\title{Quantum transport through a quantum dot side-coupled to a Majorana bound state pair in the presence of electron-phonon interaction}

\author{Levente M\'{a}th\'{e}}
\email[Corresponding author: ]{levente.mathe@itim-cj.ro}
\affiliation{National Institute for Research and Development of Isotopic
and Molecular Technologies, 67-103 Donat, 400293 Cluj-Napoca, Romania}
\affiliation{Faculty of Physics, Babeș-Bolyai University, 1 Kogălniceanu, 400084 Cluj-Napoca, Romania}

\author{Doru Sticlet}
\affiliation{National Institute for Research and Development of Isotopic
	and Molecular Technologies, 67-103 Donat, 400293 Cluj-Napoca, Romania}

\author{Liviu P. Z\^{a}rbo}
\affiliation{National Institute for Research and Development of Isotopic
	and Molecular Technologies, 67-103 Donat, 400293 Cluj-Napoca, Romania}

\date{\today}

\begin{abstract}
We theoretically study quantum transport through a quantum dot coupled to Majorana bound states confined at the ends of a topological superconducting nanowire. 
The topological superconductor forms a loop and is threaded by a tunable magnetic flux, which allows one to control the electron transport in the system.
In particular, we investigate phonon-assisted transport properties in the device when the central quantum dot interacts with a single long-wave optical phonon mode.
We find that when the two Majorana bound states are unhybridized, the zero-temperature linear conductance has a $2\pi$ periodicity as a function of magnetic flux phase, independent of the electron-phonon interaction, the quantum dot energy, or the finite values of dot-Majorana couplings. 
For a finite overlap between the Majorana bound states, the linear conductance periodicity generally changes to $4\pi$ either due to a finite electron-phonon coupling strength, or a dot energy level that is tuned away from the Fermi level.
Additionally, the differential conductance periodicity changes from $2\pi$ to $4\pi$ when the Majorana bound states hybridize and the electron-phonon coupling is finite.
Our results provide insight into transport signatures expected in topological quantum computational platforms that integrate quantum dots as a means for Majorana qubit readout.
The energy exchange with an environmental bath, here a single phonon mode, significantly alters the current signatures expected from Majorana modes.
\end{abstract}

\maketitle

\section{INTRODUCTION}
\label{sec:I}
Topological materials allow the realization of localized edge states with zero-energy excitations, called Majorana bound states (MBSs)~\cite{Kitaev2001}. The MBSs obey non-Abelian statistics and provide nonlocal electronic degrees of freedom to encode quantum information. Thus they could serve as a potential platform for topological quantum computation~\cite{Kitaev2001,Kitaev2006}.
Namely, the topological qubits realized by MBSs store quantum information nonlocally and provide robust protection against decoherence~\cite{Kitaev2001,Kitaev2006,Nayak2008,Alicea2012,Leijnse2012,Beenakker2013,Aguado2017}. 
References~\cite{Lutchyn2010, Oreg2010} have theoretically proposed realizing MBSs in a nanodevice containing a semiconducting nanowire with strong spin-orbit coupling, deposited in proximity to $s$-wave superconductors (SCs) in external magnetic fields. 
Signatures of MBSs were first demonstrated in conductance measurements in this heterostructure~\cite{Mourik2012}.
Such systems host MBSs at the ends of an InSb or InAs semiconducting nanowire and can be detected in transport measurements as a zero-bias anomaly in the differential conductance. Several theoretical and experimental studies have proposed alternatives to create MBSs, such as two-dimensional $p$-wave SCs~\cite{Read2000,Stone2004}, or $s$-wave SCs with topological insulators~\cite{Fu2008}, cold atom wires~\cite{Jiang2011}, half-metallic ferromagnets~\cite{Duckheim2011}, superfluids~\cite{Kopnin1991}, nanomagnets~\cite{Kjaergaard2012}, chains of magnetic impurities on $s$-wave SCs~\cite{NadjPerge2014}, vortex cores on SCs~\cite{Wang2018}, etc. 
A plethora of interesting phenomena, induced by MBSs, such as Andreev reflection~\cite{Nilsson2008,Law2009,Flensberg2010,Leumer2021,Duse2021} and the fractional Josephson effect~\cite{Fu2009,Jiang2011a}, have been demonstrated theoretically.
Unfortunately, the zero-bias peak can be due to other phenomena such as Kondo resonances~\cite{Lee2012}, Andreev bound states~\cite{Franz2013}, disorder~\cite{Liu2012}, confinement potential~\cite{Prada2012,Moore2018}, or weak antilocalization~\cite{Pikulin2012}.

The most common scheme proposed to probe MBSs in a topological superconducting nanowire (TSNW) requires embedding the nanowire in a mesoscopic circuit. A minimal device is made from MBSs coupled to regular fermionic degrees of freedom associated with a quantum dot (QD), which is then connected to normal leads. The presence of MBSs is seen in the conductance through the QD~\cite{Flensberg2011,Liu2011}. The study of transport through a noninteracting QD, symmetrically coupled to metallic leads, and connected to a TSNW, revealed that the presence of MBS dramatically influences the linear conductance~\cite{Liu2011}. 
When the dot couples to a regular fermionic zero mode, the conductance peak value is $0$. When the TSNW is in its topologically trivial phase, the conductance peak is $e^2/h$, while in the nontrivial phase, a signature of the MBS appears in the reduction of conductance to $e^2/2h$. It was also predicted that the linear conductance of a QD, connected to two MBSs confined at the ends of a TSNW, is a $2\pi$-periodic function of the threaded magnetic flux when the overlap energy between the MBSs is zero. 

In recent years, numerous QDs-MBSs devices have been proposed to detect the signature of MBSs by probing their transport properties, such as the conductance spectrum~\cite{Cao2012,Lee2013,Cheng2014,Vernek2014,Stefanski2015,Ramos2018}, current noise~\cite{Cao2012,Lu2012, Chen2014a,Lu2014,Liu2015,Zhao2016,Lu2016}, thermal conductance~\cite{Leijnse2014,Lopez2014,Khim2015,Ri2019,Chi2020a}, optical schemes~\cite{Chen2014b,Chen2018,Chi2020}, the Josephson effect~\cite{Lee2016}, and Fano resonance~\cite{Ueda2014, Dessotti2014,Jiang2015,Xia2015,Zeng2016a,Zeng2016b,Baranski2016,Schuray2017,Wang2018a,Wang2018b,Calle2020}. 
Zeng et al.~\cite{Zeng2016b} analyzed the transport inside a ring system formed by an asymmetrically biased QD coupled to two MBSs located at the ends of a TSNW. 
They found that the differential conductance shows a $2\pi$-period as a function of the threaded magnetic flux for a sufficiently long TSNW where MBS wave-function overlap is neglected.
In contrast, for a short TSNW with nonzero overlap energy between MBSs, the conductance has a $4\pi$ periodicity as a function of magnetic flux phase. In recent works, Calle et al.~\cite{Calle2020,Calle2013,Calle2017} have demonstrated that quantum transport in a conventional T-shaped double QD system connected to a topologically trivial superconducting lead can correspond to quantum transport in a QD-MBSs ring system when fine-tuning the systems parameters.
Other systems based on shot noise probing in monolayer graphene QDs~\cite{Huang2019} or in toroidal carbon nanotubes~\cite{Huang2020} have been proposed to detect the signature of MBSs.

More recently, researchers have demonstrated that entanglement can be achieved between two QDs coupled to a pair of MBSs confined at the ends of a TSNW owing to the nonlocal quantum nature of the MBSs~\cite{Fu2010a,Wang2013a,Wang2013b}. 
It has also been predicted that the nonlocality of the MBSs leads to two types of nonlocal processes, namely electron tunneling (ET) and crossed Andreev reflection (CAR)~\cite{Nilsson2008,Wang2013b,Zocher2013,Liu2013,Liu2014}. These nonlocal processes can be controlled by tuning the energy levels of the QDs through gate electrodes~\cite{Liu2014}. Another process, which takes place in such systems, is local Andreev reflection (LAR)~\cite{Nilsson2008,Zocher2013,Liu2013,Liu2014}.
The CAR process occurs when a pair of MBSs couple to two metallic leads in which an electron (hole) from one lead tunnels into the SC via one MBS and tunnels out as a hole (electron) via the other MBS at the other lead by splitting a Cooper pair over the leads.
In a LAR process, an electron (hole) tunnels from one lead into the SC via one MBS and is reflected as a hole (electron) in the same lead by injecting thus a Cooper pair into the SC. 
In an ET process, an electron from one lead is transmitted to the other lead without creating/annihilating Cooper pairs in the SC~\cite{Nilsson2008,Law2009,Lu2012,Zocher2013,Liu2013,Liu2014}.
It was found that, at sufficiently low excitation energies, the LAR is suppressed in favor of CAR~\cite{Nilsson2008}. 
Furthermore, only LAR processes survive when the overlap energy between the MBSs is zero~\cite{Liu2014}.
A Y-shaped junction device consisting of two metallic leads, each connected to a SC electrode through two QDs, played the role of a Cooper-pair splitter and allowed control over CAR and LAR processes when tuning the gate voltages~\cite{Hofstetter2009}.
The Andreev transport properties of double-QD Cooper-pair splitters~\cite{Eldridge2010,Hiltscher2011,Chevallier2011,Gong2016} or multi-QD-MBSs setups~\cite{Li2013,Gong2014a,Gong2014b,Jiang2014,Silva2016,Jiang2016,Ivanov2017,Cifuentes2019} have also been theoretically explored. 
In addition, the thermoelectric properties in Cooper-pair splitters with QDs have been explored both theoretically~\cite{Cao2015,Sanchez2018,Hussein2019} and experimentally with graphene-based QDs~\cite{Tan2021}. 
All these systems show more complicated transport behavior due to quantum interference and many system parameters~\cite{Gong2014a,Cifuentes2019}, and they provide a feasible platform to probe MBSs~\cite{Cifuentes2019}.

Up to now, various experimental realizations based on a normal lead (N)-SC junction~\cite{Mourik2012,Finck2013,Chen2017,Zhang2017,Guel2018,Lutchyn2018} or a QDs-TSNW junction~\cite{Deng2016,Sherman2016,Deng2018} have been proposed to detect MBSs and explore their transport properties.
However, due to the presence of decoherence processes in such experimental devices, significant deviations from theoretical predictions are observed in transport measurements. 
Decoherence occurs when the TSNW is coupled to an environmental bath~\cite{Aseev2018,Aseev2019,Zhang2019,Schulenborg2021}, which can be fermionic~\cite{Goldstein2011,Budich2012,Rainis2012} or bosonic~\cite{Goldstein2011,Pedrocchi2015,Knapp2018, Song2018,Aseev2019,Schmidt2013,Schmidt2013a,Lai2018,Aseev2018}. 
In this way, decoherence can be caused by quasiparticle poisoning \cite{Rainis2012}, electromagnetic field~\cite{Knapp2018,Antipov2018}, charge, and thermal fluctuations of the gate voltage~\cite{Schmidt2012,Lai2018,Knapp2018,Aseev2018} etc.

The vast majority of theoretical works in recent years have focused on the properties of phonon-assisted Andreev tunneling in N-SC~\cite{Setiawan2021}, N-QD-SC~\cite{Zhang2012,Baranski2015a,Baranski2015b,Stadler2016,Stadler2017,Cao2017,Dong2017}, or ferromagnet-QD-SC junctions~\cite{Zhang2009,Bai2011,Bocian2014,Bocian2015}. 
The phonon-assisted Andreev tunneling was likely observed in experiments for a carbon nanotube QD coupled to N and Nb SC leads~\cite{Gramich2015}. 
Moreover, only a few theoretical works have addressed the mechanisms of phonon-assisted transport at MBSs~\cite{Dai2019,Wang2021,Wang2021b,Wang2020}. 
Dai and Sun~\cite{Dai2019} showed that a N-SC junction can be viewed as an electron-lead/phonon-connected-MBS/hole-lead structure, which leads to converting an electron into a hole by absorbing/emitting phonons. 
The transport properties of a QD connected to one MBS in the presence of electron-phonon interaction (EPI) were studied before and resulted in diverging conclusions about the zero-temperature behavior of zero-bias conductance. 
The latter was purportedly shown either as dependent on the electron-phonon coupling strength~\cite{Wang2020}, or independent~\cite{Wang2021}.
It was found that the zero-temperature zero-bias conductance can be significantly modified by changing the strength of EPI or the values of dot-MBS coupling. At finite temperature, the EPI results in suppression of the magnitude of zero-bias conductance.

In the present work, we study the transport properties of a QD connected to two MBSs located at the ends of a TSNW loop, threaded by a magnetic field, in the presence of a single long-wave optical phonon mode when an opposite bias between the leads is applied. The central findings of our work are the following. 
We show that, without EPI, the usual line shape of the spectral function in the presence of MBSs can be reversed by tuning the gate voltage with respect to the Fermi level. 
We have demonstrated that the linear and differential conductances are due to both ET and LAR processes. The linear conductance manifests a $2\pi$ periodicity as a function of magnetic flux phase when the dot energy is at the Fermi level, independent of the MBSs overlap energy. 
The $2\pi$ periodicity changes to $4\pi$ by tuning the QD energy level or introducing EPI when the MBSs hybridize. In the presence of EPI, a series of phonon-assisted channels for transport are manifest in the spectral function as satellite peaks. 
The linear conductance is immune to the presence of EPI, the change in dot level, and finite values of the QD-MBS couplings for degenerate MBSs at zero temperature. 
The differential conductance periodicity switches from $2\pi$ to $4\pi$ for hybridized MBSs when varying the electron-phonon coupling strength or dot energy level.
The linear and differential conductance periodicity is invariant to the QD-lead couplings asymmetry in the absence of EPI. Therefore, this would represent a robust signature in future experiments.
\begin{figure}[t]
	\includegraphics[width =1\linewidth]{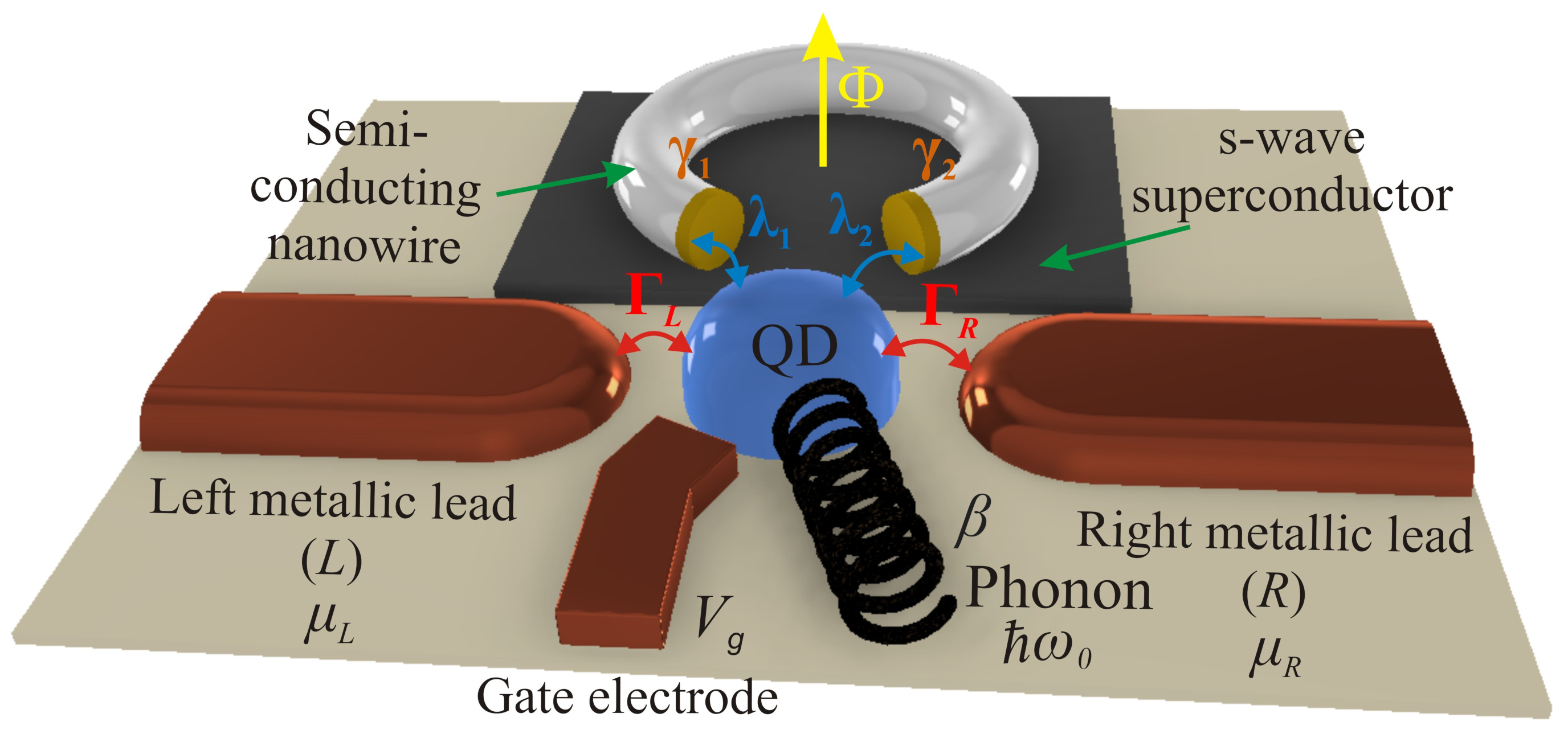}
	\centering
	\caption{Schematic representation of a quantum dot (QD) laterally coupled to two Majorana bound states (MBSs) located at the ends of a topological superconducting nanowire (TSNW) loop, threaded by a magnetic flux $\Phi$, with coupling strengths $\lambda_1$ and $\lambda_2$. The QD is connected to left ($L$) and right ($R$) metallic electrodes with the coupling strength $\Gamma_{L(R)}$. Here, $\mu_{L(R)}$ is the chemical potential in lead $L(R)$. The dot level is modulated by the gate voltage $V_g$ at the gate electrode. The localized electron in the QD interacts with a single optical phonon mode with frequency $\omega_0$ through the electron-phonon coupling $\beta$.}
	\label{fig:1}
\end{figure}

The article is structured as follows. 
In Sec.~\ref{sec:IIA}, we introduce the model (see the schematic setup in Fig.~\ref{fig:1}).
We then turn to a discussion of the relevant transport processes that take place in such a system, and we derive a current formula with the help of the spectral function of the QD in Sec.~\ref{sec:IIB}.
In Sec.~\ref{sec:IIC}, we use the canonical transformation in order to determine the spectral function via the Keldysh equation, and then we compute the relevant retarded Green's functions of the dot by applying the equation of motion technique (EOM). 
In Sec.~\ref{sec:III}, we present the numerical results in the absence and presence of EPI. Finally, we summarize the main results of our work in Sec. \ref{sec:IV}. 
Appendix~\ref{sec:A} contains the detailed calculation of conductance in the absence of EPI. We then calculate the retarded Green's functions for arbitrary QD-lead and QD-MBS couplings in Appendix~\ref{sec:B}, and we compute the phonon-assisted conductance in Appendix~\ref{sec:C}.
\section{Model and analytical results}
\label{sec:II}
\subsection{Theoretical model}
\label{sec:IIA}
We consider a QD laterally coupled to two MBSs located at the ends of a TSNW loop, threaded by an external magnetic flux $\Phi$, as shown in Fig.~\ref{fig:1}. 
It has been theoretically shown that a TSNW loop can host MBSs at its ends for magnetic fields applied perpendicularly to the surface enclosed by the loop~\cite{Lutchyn2010,Sau2010}. 
This arrangement of the magnetic field allows one to manipulate the QD system transport by modulating the magnetic flux.
The current through the QD is measured via the metallic electrodes. 
In order to analyze the influence of EPI on transport properties of the system, we consider localized electrons in the QD interacting with a single optical phonon mode.
The MBSs are coupled to the dot in the Coulomb blockade regime, such that the QD is modeled as a single fermion level~\cite{Flensberg2011,Lu2016}. 
The applied magnetic field should be large enough in order to drive the nanowire into the topological superconducting phase, which is achieved when $V_{\rm Z}>\sqrt{\Delta^2 + \mu^2}$, with the Zeeman energy $V_{\rm Z} = g\mu_B B/2$, the energy gap of the SC $\Delta$, and the chemical potential of the nanowire $\mu$. 
The Zeeman splitting is assumed to be the largest energy scale, larger than the system temperature $T$, the applied bias to the leads $|eV|$, the dot-lead coupling $\Gamma$, or the dot-MBS coupling strength $|\lambda_{j}|$.
Therefore, it is enough to consider a spinless single energy level in the QD~\cite{Liu2011,Cao2012}. In this case, the Hamiltonian of the system reads~\cite{Liu2011,Zeng2016a,Zeng2016b,Zhu2003,Chen2005,Ramos2018}
\begin{equation}
	\mathcal{H}=\mathcal{H}_{\text{leads}}+\mathcal{H}_{\text{MBS}}+\mathcal{H}_{\text{ph}}+\mathcal{H}_{\text{D}}+\mathcal{H}_{\text{T}}.
	\label{eq:1}
\end{equation}

The first term in Eq.~\eqref{eq:1}, $\mathcal{H}_{\text{leads}}$, represents the Hamiltonian of the metallic electrodes, which describes the free electrons in the leads, and it reads
\begin{equation}
	\mathcal{H}_{\text{Leads}}=\sum_{\alpha,k}\varepsilon_{\alpha k}\,c_{\alpha k}^{\dagger}c_{\alpha k},
	\label{eq:2}
\end{equation}
where $c_{\alpha k}^{\dagger}$ and $c_{\alpha k}$ denote the creation and annihilation operators for noninteracting electrons with momentum \textit{k} in the lead $\alpha$ [$\alpha$ $\equiv$ left (\textit{L}), right (\textit{R})].
The $\varepsilon_{\alpha k}=\varepsilon_{k}-\mu_{\alpha}$ are the single-particle energies where $\mu_\alpha$ is the chemical potential in the lead $\alpha$ and it is assumed to be temperature-independent. 
We consider equal temperatures in the leads as $T_\alpha=T$.

The second term in Eq.~\eqref{eq:1}, $\mathcal{H}_{\text{MBS}}$, is the MBSs Hamiltonian and models the coupling between the two MBSs, $\gamma_1$ and $\gamma_2$, located at the ends of the TSNW loop,
\begin{equation}
	\mathcal{H}_{\text{MBS}}=i \varepsilon_M \gamma_1 \gamma_2,
	\label{eq:3}
\end{equation}
where $\varepsilon_M$ is the overlap energy that exponentially depends on the length of TSNW loop ($L$) as $\varepsilon_M \propto e^{-L/\xi}$, with $\xi$ being the superconducting coherence length.

The third term in Eq.~\eqref{eq:1}, $\mathcal{H}_{\text{ph}}$, describes the longitudinal optical phonon mode:
\begin{equation}
	\mathcal{H}_{\text{ph}}=\hbar \omega_0 a^\dagger a,
	\label{eq:4}
\end{equation}
where $\omega_0$ is the frequency of the single-level phonon mode, with the $a^{\dagger}(a)$ phonon creation (annihilation) operator. The phonon distribution is given by the Bose-Einstein function $N_{\text{ph}}=1/(e^{\hbar \omega_0/k_B T}-1)$. Throughout the paper, we will set $\hbar$ and the Boltzmann constant $k_B$ to unity.

The fourth term in Eq.~\eqref{eq:1}, $\mathcal{H}_\text{D}$, models the QD and it is expressed as
\begin{equation}
	\mathcal{H}_\text{D}=\varepsilon_{d}d^{\dagger}d +\beta (a+a^\dagger)d^\dagger d,
	\label{eq:5}
\end{equation}
where $\varepsilon_{d}$ is the energy of the dot level and $d^{\dagger}(d)$ represents the fermionic creation (annihilation) operator for the localized electron in the QD. 
The dot level $\varepsilon_d$ is tuned by the gate voltage $V_g$ through the gate electrode. The second term in Eq.~\eqref{eq:5} models the interaction between the localized electron in dot and the phonon mode, with $\beta$ being the electron-phonon coupling strength.

The last term in Eq.~\eqref{eq:1}, $\mathcal{H}_\text{T}$, is the tunneling Hamiltonian and represents the coupling of the QD to the MBSs and to the metallic electrodes:
\begin{eqnarray}
	\mathcal{H}_\text{T}&=&\big[(\lambda_1 d - \lambda_1^* d^\dagger)\gamma_1+ i (\lambda_2 d + \lambda_2^* d^\dagger)\gamma_2\big]\notag\\
	&&+\sum_{\alpha,k}\big (V_{\alpha k}c_{\alpha k}^{\dagger}d +V^*_{\alpha k}d^{\dagger}c_{\alpha k}\big ).
	\label{eq:6}
\end{eqnarray}
The first term in Eq.~\eqref{eq:6} describes the coupling between the dot and the two MBSs ($\gamma_1$ and $\gamma_2$) of the TSNW loop with the coupling strengths $\lambda_1$ and $\lambda_2$. 
Here, we assume that $\lambda_1$ and $\lambda_2$ are complex, setting as $\lambda_1=|\lambda_1|e^{i\phi/4}$ and $\lambda_2=|\lambda_2|e^{-i\phi/4}$~\cite{Zeng2016b}. 
The magnetic flux phase difference between the two couplings $\phi$ is determined by the threading magnetic flux $\Phi$ via $\phi=\pi \Phi/\Phi_0=2\arg(\lambda_1/\lambda_2)$, where $\Phi_0=h/(2e)$ is the magnetic flux quantum. The second term in Eq.~\eqref{eq:6} models the coupling between the free electrons in leads and the localized electron in QD, where the $V_{\alpha k}$ tunneling amplitude determines the coupling strength between the QD and the lead $\alpha$.

The MBSs are expressed with the help of regular fermionic operators as $\gamma_1=(f^\dagger +f)/ \sqrt{2}$ and $\gamma_2=i(f^\dagger -f)/ \sqrt{2}$ such that $\mathcal{H}_\text{MBS}$ and $\mathcal{H}_\text{T}$ read
\begin{equation}
	\mathcal{H}_{\text{MBS}}=\varepsilon_M \Big(f^\dagger f -\frac{1}{2}\Big)
	\label{eq:7}
\end{equation}
and
\begin{eqnarray}
	\mathcal{H}_\text{T}&&= \frac{1}{\sqrt{2}}[ (\lambda_1 + \lambda_2)  d f + (\lambda_1 - \lambda_2) d f^{\dagger} - (\lambda_1^* - \lambda_2^*)  d^{\dagger} f \notag\\
	&& - (\lambda_1^* + \lambda_2^*)  d^\dagger f^{\dagger}]+ \sum_{\alpha,k}\big (V_{\alpha k}c_{\alpha k}^{\dagger}d +V^*_{\alpha k}d^{\dagger}c_{\alpha k}\big ).
	\label{eq:8}
\end{eqnarray}
\subsection{Current formulas}
\label{sec:IIB}
The current from the $\alpha$ metallic electrode is given by \cite{Michalek2013,Michalek2015,Bocian2016,Cao2017} 
\begin{equation}
	I_\alpha = \frac{ie}{h} \int d\varepsilon \big \{ \mathbf{\Gamma}_{\alpha} \big [\mathbf{f}_{\alpha}(\varepsilon) [\mathbf{G}_{d}^{r}(\varepsilon) - \mathbf{G}_{d}^{a}(\varepsilon)] + \mathbf{G}_{d}^{<}(\varepsilon) \big ] \big\}_{11},
	\label{eq:9}
\end{equation}
where $\mathbf{G}_{d}^{r(a)}(\varepsilon)$ is the retarded (advanced) Green's function matrix of the dot defined in the Nambu representation as \cite{Hwang2015}
\begin{equation}
	\mathbf{G}_d^{r(a)}(\varepsilon)=
	\begin{pmatrix}
		G_{d11}^{r(a)}(\varepsilon)&G_{d12}^{r(a)}(\varepsilon)\\
		G_{d21}^{r(a)}(\varepsilon)&G_{d22}^{r(a)}(\varepsilon)\\
	\end{pmatrix},
	\label{eq:10}
\end{equation}
where the non-diagonal matrix elements, $G_{d12}^{r(a)}(\varepsilon)$ and $G_{d21}^{r(a)}(\varepsilon)$, are the anomalous Green's functions, with the subscript notation $1 \,(2)$ referring to the electron (hole) sector. The QD-lead coupling matrix is defined as
\begin{equation}
	\mathbf{\Gamma}_{\alpha}=
	\begin{pmatrix}
		\Gamma_{\alpha}^{e}&0\\
		0&\Gamma_{\alpha}^{h}\\
	\end{pmatrix},
	\label{eq:11}
\end{equation}
with $\Gamma_{\alpha}^{e(h)} = 2\pi \sum_k |V_{\alpha k }|^2 \delta (\varepsilon \mp \varepsilon _{\alpha k})$ being the coupling strength between the QD and the metallic lead $\alpha$ for electrons (holes).
Therefore, $\mathbf{f}_{\alpha}(\varepsilon)$ represents the Fermi-Dirac distribution matrix in lead $\alpha$, 
\begin{equation}
	\mathbf{f}_{\alpha}(\varepsilon)=
	\begin{pmatrix}
		f_{\alpha}^{e}(\varepsilon)&0\\
		0&f_{\alpha}^{h}(\varepsilon)\\
	\end{pmatrix},
	\label{eq:12}
\end{equation}
where $f_{\alpha}^{e}(\varepsilon)=1/[e^{(\varepsilon-\mu_\alpha)/T}+1]$ and $f_{\alpha}^{h} (\varepsilon)=1-f_{\alpha}^{e}(-\varepsilon)$ are the Fermi-Dirac distribution functions for electrons and holes \cite{Michalek2013,Michalek2015,Bocian2015,Stadler2016}. Here, $\mathbf{G}_{d}^{<}(\varepsilon)$ is the lesser Green's function matrix. 
The current formula from Eq.~\eqref{eq:9} reduces then to
\begin{equation}
	I_\alpha = \frac{ie}{h} \int d\varepsilon \, \Gamma_\alpha^e \big [f_{\alpha}^e(\varepsilon) G_{d11}^> (\varepsilon) -  [f_\alpha^e(\varepsilon) - 1] G_{d11}^<(\varepsilon) \big],
	\label{eq:13}
\end{equation}
where $G_{d11}^>(\varepsilon)$ is the $11$ component of the greater Green's function matrix $\mathbf{G}_{d}^{>}(\varepsilon)$. We apply the Keldysh equation $\mathbf{G}_d^{<(>)}=\mathbf{G}_d^r \mathbf{\Sigma}^{<(>)}\mathbf{G}_d^a$ and consider that the EPI contribution to the lesser (greater) self-energy $\mathbf{\Sigma}^{<(>)}$~\cite{Huang2019} is negligible in the relatively weak electron-phonon coupling limit~\cite{Zhu2003}. Since we are interested in an energy window around the Majorana zero-energy narrower than the superconducting gap, we consider only the subgap regime $|eV|<\Delta$, where the current from lead $\alpha$ given by Eq.~\eqref{eq:13} is decomposed as follows:
\begin{equation}
	I_\alpha = I_{\alpha}^{ET} + I_{\alpha}^{LAR}  +I_{\alpha}^{CAR},
	\label{eq:14}
\end{equation}
\begin{equation}
	I_{\alpha}^{ET}= \frac{e}{h}\int d\varepsilon \mathcal{T}_{\alpha \alpha '}^{ee}(\varepsilon)[f_{\alpha}^{e}(\varepsilon)-f_{\alpha '}^{e}(\varepsilon)],
	\label{eq:15}
\end{equation}
\begin{equation}
	I_{\alpha}^{LAR} = \frac{e}{h}\int d\varepsilon \mathcal{T}_{\alpha \alpha}^{eh}(\varepsilon)[f_{\alpha}^{e} (\varepsilon)-f_{\alpha}^{h} (\varepsilon)],
	\label{eq:16}
\end{equation}
and
\begin{equation}
	I_{\alpha}^{CAR} = \frac{e}{h}\int d\varepsilon \mathcal{T}_{\alpha \alpha '}^{eh}(\varepsilon)[f_{\alpha}^{e} (\varepsilon)-f_{\alpha '}^{h} (\varepsilon)],
	\label{eq:17}
\end{equation}
where $I_{\alpha}^{ET}$ represents the current generated in the ET process, while $I_{\alpha}^{LAR}$ and $I_{\alpha}^{CAR}$ are currents due to the LAR and CAR processes.
Here, $\mathcal{T}_{\alpha \alpha '}^{ee}(\varepsilon)=\Gamma_{\alpha}^{e} \Gamma_{\alpha '}^{e}|G_{d11}^r(\varepsilon)|^2$, $\mathcal{T}_{\alpha \alpha}^{eh}(\varepsilon)=\Gamma_{\alpha}^{e}\Gamma_{\alpha}^{h}|G_{d12}^r(\varepsilon)|^2$ and $\mathcal{T}_{\alpha \alpha '}^{eh}(\varepsilon)=\Gamma_{\alpha}^{e}\Gamma_{\alpha '}^{h}|G_{d12}^r(\varepsilon)|^2$ are the corresponding transmission probabilities, with $G_{d11}^r(\varepsilon)$ and $G_{d12}^r(\varepsilon)$ being the electron-electron and electron-hole parts of the retarded Green's function $\mathbf{G}_d^r(\varepsilon)$, respectively. The current $I_S$ flowing from the SC is determined from the Kirchhoff's law as $I_L + I_R + I_S = 0$~\cite{Michalek2013}. 

In this article, we work in the wide-band limit by assuming electron-hole symmetry in the system, $\Gamma_{\alpha}^{e} = \Gamma_{\alpha}^{h} = \Gamma_{\alpha}$~\cite{Jauho1994,Ramos2018}, and we restrict the calculations for a symmetrically coupled QD with $\Gamma_{\alpha} = \Gamma$. 
In addition, we investigate the transport properties of the QD-MBSs system in absence and presence of EPI. 
We also consider that the SC is grounded, i.e., $\mu_S = 0$, and there is an opposite bias applied between the metallic electrodes, $\mu_L=-\mu_R=eV/2$. 
In such a case, $f_{\alpha}^{e} (\varepsilon) = f_{\alpha '} ^{h}(\varepsilon)$, which allows only for ET and LAR processes to occur (and inhibits CAR processes)~\cite{Wang2018a}. Using the Kirchhoff's law, the current from the SC is expressed as $I_S= -(I_L+I_R)$. Applying the current formula for metallic leads, given by Eqs.~\eqref{eq:14}-\eqref{eq:17}, with the above restrictions, one obtains  $I_L = - I_R$, which implies $I_S = 0$.
The calculations of the corresponding differential and linear conductances are detailed for finite and zero temperatures, without EPI, in Appendix~\ref{sec:A}.

We present throughout the paper the EPI-related relevant calculations. 
Since the total current is $I_L + I_R = 0$, and using Eq.~\eqref{eq:13}, we symmetrize the current as $I = (I_L - I_R)/2$,
\begin{equation}
	I = \frac{e}{h} \int d\varepsilon \mathcal{A}_d(\varepsilon) \big[f_{L}^{e}(\varepsilon) - f_{R}^{e}(\varepsilon)\big],
	\label{eq:18}
\end{equation}
where $\mathcal{A}_d(\varepsilon)$ is the spectral function of the QD,
\begin{equation}
	\mathcal{A}_d(\varepsilon)= i \frac{\Gamma}{2}[\mathbf{G}_d^>(\varepsilon) - \mathbf{G}_d^<(\varepsilon)]_{11}= i \frac{\Gamma}{2} [\mathbf{G}_d^r(\varepsilon) - \mathbf{G}_d^a(\varepsilon)]_{11}.
	\label{eq:19}
\end{equation}
Consequently, the current is evaluated by performing a decoupling treatment for the electron-phonon system on the spectral function.

\subsection{Canonical transformation: The presence of EPI}
\label{sec:IIC}
In the following, we determine the spectral function for the dot, $\mathcal{A}_d(\varepsilon)$, given by Eq.~\eqref{eq:19}. To find the lesser and greater Green's functions in the presence of EPI, we eliminate the electron-phonon coupling term in the Hamiltonian, given by Eq.~\eqref{eq:1}, by applying a canonical transformation. 
The transformed Hamiltonian $\mathcal{\bar H} = e^S \mathcal{H} e^{-S}$ is determined by using the operator $S=(\beta/\omega_0)d^{\dagger} d (a^{\dagger} - a)$~\cite{Zhu2003,Chen2005,Swirkowicz2008}. 
Thus, the new Hamiltonian is decoupled into an electron and phonon part as $\mathcal{\bar H} = \mathcal{\bar H}_{\text{el}} + \mathcal{H}_{\text{ph}}$. While the phonon part remains unchanged, the electron part $\mathcal{\bar H}_{\text{el}}$ reads
\begin{equation}
	\mathcal{\bar H}_{\text{el}} = \mathcal{H}_{\text{leads}} + \mathcal{H}_{\text{MBS}} +\mathcal{\tilde H}_{\text{D}} + \mathcal{\bar H}_{\text{T}},
	\label{eq:20}
\end{equation}
where
\begin{eqnarray}
	\mathcal{\bar H}_\text{T}&&= \frac{1}{\sqrt{2}}[ (\tilde{\lambda}_1 + \tilde{\lambda}_2)  d f + (\tilde{\lambda}_1 - \tilde{\lambda}_2) d f^{\dagger} - (\tilde{\lambda}_1^* - \tilde{\lambda}_2^*)  d^{\dagger} f \notag\\
	&& - (\tilde{\lambda}_1^* + \tilde{\lambda}_2^*)  d^\dagger f^{\dagger}]+ \sum_{\alpha,k}\big (\tilde{V}_{\alpha k}c_{\alpha k}^{\dagger}d +\tilde{V}^*_{\alpha k}d^{\dagger}c_{\alpha k}\big )
	\label{eq:21}
\end{eqnarray}
and $\mathcal{\tilde H}_\text{D}=\tilde {\varepsilon}_{d}d^{\dagger}d$, with the renormalized QD energy level $\tilde {\varepsilon}_d = \varepsilon_d - g \omega_0$ with $g=(\beta/\omega_0)^2$. 
Now, $\mathcal{\bar H}_{\text{T}}$ includes the phonon operators via the renormalized tunneling amplitude between the QD and lead $\alpha$ as $\tilde{V}_{\alpha k} = V_{\alpha k} X$ and the renormalized coupling between the QD and the $j$th MBS as $\tilde{\lambda}_j = \lambda_j X$ with $X = \exp(-\frac{\beta}{\omega_0}(a^{\dagger}-a))$.
The operator $X$ is then replaced by its expectation value in thermal equilibrium, $X \approx \langle X \rangle = \exp(-g(N_{\text{ph}}+\frac{1}{2}))$ in a broadly used approximation~\cite{Chen2005,Swirkowicz2008}. 
Then the Hamiltonian $\mathcal{\bar H}_{\text{el}}$, given by Eq.~\eqref{eq:20}, will be decoupled from the phonon operator. 
One obtains $\tilde V_{\alpha k} \approx V_{\alpha k} \langle X \rangle$ and $\tilde \lambda_j \approx \lambda_j \langle X \rangle$. 
This approximation is valid when $V_{\alpha k}, \, \lambda_{j} \ll \text{min}(\beta,\Delta)$ or $\beta \ll \text{min}(V_{\alpha k},\lambda_{j},\Delta)$, which means that the $V_{\alpha k}$ and $\lambda_{j}$ or $\beta$ are the smallest energy scale. 
In addition, it is exact when $V_{\alpha k} = \lambda_{j} = 0$ or $\beta = 0$ \cite{Chen2005,Swirkowicz2008,Zhang2012,Dai2019}. 
Consequently, we separate the greater and lesser Green's functions into two parts:
\begin{eqnarray}
	\mathbf{G}_d^{>}(t)&=&-i \langle d(t)d^{\dagger} \rangle\notag\\
	&=& - i \langle e^{i\mathcal{\bar H}_{\text{el}}t}de^{-i\mathcal{\bar H}_{\text{el}}t}  d^{\dagger} \rangle_{\text{el}}
	\langle e^{i\mathcal{H}_{\text{ph}}t}Xe^{-i\mathcal{H}_{\text{ph}}t} X^{\dagger} \rangle_{\text{ph}}\notag\\
	&=&\tilde{\mathbf{G}}_d^{>}(t) \langle X(t) X^{\dagger} \rangle_{\text{ph}},
	\label{eq:22}\\
	\mathbf{G}_d^{<}(t)&=&i \langle d^{\dagger}d(t) \rangle =\tilde{\mathbf{G}}_d^{<}(t) \langle X^{\dagger}X(t) \rangle_{\text{ph}},
	\label{eq:23}
\end{eqnarray}
where $\tilde{\mathbf{G}}_d^{<(>)}(t)$ is the lesser (greater) Green's function for the electron governed by $\mathcal{\bar H}_{\text{el}}$. The phonon parts are $\langle X(t) X^{\dagger} \rangle_{\text{ph}}=e^{-\varphi (t)}$ and $\langle X^{\dagger}X(t) \rangle_{\text{ph}}=e^{-\varphi (-t)}$ with $\varphi (t) = g[N_{\text{ph}}(1-e^{i \omega_0 t}) + (N_{\text{ph}} + 1)(1 - e^{-i \omega_0 t})]$~\cite{Chen2005,Swirkowicz2008}. 
Note here that the approximation $\langle X(t) X^{\dagger} \rangle_{\text{ph}} = \langle X^{\dagger}X(t) \rangle_{\text{ph}}$ used in previous works~\cite{Zhu2003,Chen2005,Yang2009,Mathe2018}, which directly decouples the retarded Green’s function $\mathbf{G}^r_d(t)$, is valid only at high temperatures, where the phonon number $N_{\text{ph}}$ is much larger than one. At lower temperatures, where $N_{\text{ph}}\sim\mathcal O(1)$, the approximation $\langle X(t) X^{\dagger} \rangle_{\text{ph}} = \langle X^{\dagger}X(t) \rangle_{\text{ph}}$ is no longer applicable~\cite{Chen2005}. 

The phonon term is determined via the identity $e^{-\varphi(\mp t)}=\sum_{p=-\infty}^{\infty}\mathcal{L}_p e^{\pm i p\omega_0 t}$, where $\mathcal{L}_p$ are coefficients, with an integer index $p$, that depend on the electron-phonon coupling strength $\beta$ and temperature $T$. At finite temperatures, the Franck-Condon factor $\mathcal{L}_p$ has the form \cite{Chen2005,Swirkowicz2008}
\begin{equation}
	\mathcal{L}_p=e^{-g(2N_{\text{ph}}+1)}e^{p\omega_0/(2T)}I_p\big(2g\sqrt{N_\text{ph}(N_{\text{ph}}+1)}\big),
	\label{eq:24}
\end{equation}
where $I_p(z)$ is the \textit{p}th-order modified Bessel function of the first kind. At zero temperature, $\mathcal{L}_p$ reduces to
\begin{equation}
	\mathcal{L}_p = \left\{ \begin{array}{rl}
		e^{-g}g^p/p!, &\mbox{$p \ge$0,} \\
		0, &\mbox{$p<0$.}
	\end{array} \right.
	\label{eq:25}
\end{equation}
Using the above relations, the Fourier transforms of the greater and the lesser Green's functions given by Eqs.~\eqref{eq:22} and~\eqref{eq:23} read
\begin{equation}
	\begin{aligned}
		\mathbf{G}_d^>(\varepsilon)&=\sum_{p=-\infty}^{\infty}\mathcal{L}_p \mathbf{ \tilde G}_d^>(\varepsilon-p\omega_0)\\
		\mathbf{G}_d^<(\varepsilon)&=\sum_{p=-\infty}^{\infty}\mathcal{L}_p \mathbf{ \tilde G}_d^<(\varepsilon+p\omega_0),
	\end{aligned}\label{eq:26}
\end{equation}
which results in the spectral function for the dot:
\begin{equation}
	\mathcal{A}_d(\varepsilon)=i \frac{\Gamma}{2} \sum_{p=-\infty}^{\infty} \mathcal{L}_p \big[ \mathbf{ \tilde G}_d^>(\varepsilon-p\omega_0) - \mathbf{ \tilde G}_d^<(\varepsilon+p\omega_0) \big]_{11}.
	\label{eq:27}
\end{equation}

The dressed lesser (greater)  Green's function $\mathbf{ \tilde G}_d^{<(>)}$ is determined from the corresponding Keldysh equation. The general relations for the $\mathbf{G}_d^{<(>)}$ 
are also fulfilled by $\mathbf{ \tilde G}_d^{<(>)}$~\cite{Swirkowicz2008}. 
Thus the Keldysh equation is $\mathbf{ \tilde G}_d^{<(>)}=\mathbf{ \tilde G}_d^r \mathbf{\tilde\Sigma}^{<(>)}\mathbf{ \tilde G}_d^a$, where the lesser (greater) self-energy $\mathbf{\tilde\Sigma^{<(>)}}$ reads
\begin{equation}
	\mathbf{\tilde \Sigma} ^{<(>)}=\\
	\begin{pmatrix}
		\tilde \Sigma_{11}^{<(>)}&0\\
		0&\tilde \Sigma_{22}^{<(>)}\\
	\end{pmatrix}.
	\label{eq:28}
\end{equation}
Here, $\tilde \Sigma ^{<(>)}_{11}$ and $\tilde \Sigma ^{<(>)}_{22}$ are the electron and hole self-energies, with $\tilde \Sigma_{11}^< = i\sum_\alpha \tilde \Gamma_{\alpha}^{e} f_{\alpha}^{e}$, $\tilde \Sigma_{22}^< = i\sum_\alpha \tilde \Gamma_{\alpha}^{h} f_{\alpha}^{h}$, $\tilde \Sigma_{11}^> = i\sum_\alpha \tilde \Gamma_{\alpha}^{e} (f_{\alpha}^{e}-1)$, and $\tilde \Sigma_{22}^> = i\sum_\alpha \tilde \Gamma_{\alpha}^{h} (f_{\alpha}^{h}-1)$. 
Using the above relations, the dressed greater and lesser Green's functions take the form
\begin{eqnarray}
	\begin{aligned}
		\tilde G_{d11}^>&= \tilde \Sigma_{11}^> |\tilde G_{d11}^r|^2 + \tilde \Sigma_{22}^> |\tilde G_{d12}^r|^2, \\
		\tilde G_{d11}^<&= \tilde \Sigma_{11}^< |\tilde G_{d11}^r|^2 + \tilde \Sigma_{22}^< |\tilde G_{d12}^r|^2.
	\end{aligned}
	\label{eq:29}
\end{eqnarray}
The retarded $\tilde G_{d11}^r(\varepsilon)$ and $\tilde G_{d12}^r(\varepsilon)$ Green's functions are calculated using the EOM technique, presented in Appendix \ref{sec:B} for arbitrary values of $\tilde \lambda_1$ and $\tilde \lambda_2$.
Here, we give the relevant Green's functions for the case when $\tilde \lambda_1=|\tilde \lambda_1|e^{i\phi/4}$ and $\tilde \lambda_2=|\tilde \lambda_2|e^{-i\phi/4}$:
\begin{subequations}
	\begin{eqnarray}
		\tilde{G}_{d11}^{r}(\varepsilon)&=&\Big\{ (\varepsilon + \tilde \varepsilon_d + i \tilde \Gamma) - \Big[ |\tilde\lambda_1|^2 + |\tilde\lambda_2|^2\label{eq:30a}\\
		&&+\frac{2\varepsilon_M}{\varepsilon}|\tilde\lambda_1\tilde\lambda_2|\cos\frac{\phi}{2} \Big]\mathcal{K}(\varepsilon) \Big\} \mathcal{\tilde F}(\varepsilon)^{-1},\notag\\
		\tilde{G}_{d12}^{r}(\varepsilon)&=&(|\tilde\lambda_2|^2 e^{i\phi/2}-|\tilde\lambda_1|^2 e^{-i\phi/2})\mathcal{K}(\varepsilon) \mathcal{\tilde F}(\varepsilon)^{-1},
		\label{eq:30b}\\
		\mathcal{\tilde F}(\varepsilon)&=&(\varepsilon - \tilde \varepsilon_d + i \tilde \Gamma)(\varepsilon + \tilde \varepsilon_d + i \tilde \Gamma) - 2(\varepsilon + i \tilde \Gamma)\notag\\
		&&\times (|\tilde\lambda_1|^2 + |\tilde\lambda_2|^2)\mathcal{K}(\varepsilon) 
		+ \frac{4 |\tilde \lambda_1\tilde\lambda_2|^2}{\varepsilon}\mathcal{K}(\varepsilon) \cos^2\frac{\phi}{2}\notag\\
		&&+ \frac{4\varepsilon_M}{\varepsilon}\tilde\varepsilon_d |\tilde\lambda_1\tilde\lambda_2| \mathcal{K}(\varepsilon) \cos\frac{\phi}{2},
		\label{eq:30c}
	\end{eqnarray}
\end{subequations}
where $\mathcal{K}(\varepsilon)=\varepsilon/(\varepsilon^2 - \varepsilon_{M}^2)$ and $\tilde \Gamma=\frac{1}{2}\sum_\alpha \tilde \Gamma_{\alpha}$ with the restriction $\tilde \Gamma_{\alpha}^{e} = \tilde \Gamma_{\alpha}^{h}=\tilde \Gamma_{\alpha}$. Note that the Green's function $\tilde{G}_{d11}^{r}(\varepsilon)$ can be written in the following form:
\begin{widetext}
\begin{subequations}
	\begin{eqnarray}
		G_{d11}^{r}(\varepsilon)&=&\Big[ \varepsilon - \varepsilon_d + i\frac{\Gamma}{2} - A(\varepsilon) - B(\varepsilon) \Big]^{-1},\label{eq:31a}\\
		A(\varepsilon)&=& \mathcal{K}(\varepsilon)\Big[ |\lambda_1|^2 + |\lambda_2|^2 -\frac{2 \varepsilon_M}{\varepsilon} |\lambda_1||\lambda_2|\cos\frac{\phi}{2}\Big],\label{eq:31b}\\
		B(\varepsilon)&=&\frac{\mathcal{K}(\varepsilon)^2 \big( |\lambda_1|^4 + |\lambda_2|^4 - 2 |\lambda_1|^2 |\lambda_2|^2 \cos\phi \big)}{\varepsilon + \varepsilon_d + i\frac{\Gamma}{2} - \mathcal{K}(\varepsilon) \big[ |\lambda_1|^2 + |\lambda_2|^2 +\frac{2 \varepsilon_M}{\varepsilon} |\lambda_1||\lambda_2|\cos\frac{\phi}{2}\big]},\label{eq:31c}
	\end{eqnarray}
\end{subequations}
\end{widetext}
where the EPI has not been taken into account and the changing of variable $\Gamma \to \Gamma/2$ has been performed.
The form of $G_{d11}^{r}(\varepsilon)$ given by Eq.~\eqref{eq:31a} agrees with the results of Zeng et al.~\cite{Zeng2016b}  with the exception of the sign of the overlap energy $\varepsilon_M$, which is due to a different choice of the $\gamma_1$ sign. 
When $\varepsilon_M = 0$, $\lambda_1=|\lambda|$, $\lambda_2=0$, and $\phi=0$, the system reduces to a system composed of one QD with EPI connected to one MBS, and the retarded Green's functions change to
\begin{subequations}
	\begin{equation}
		\tilde{G}_{d11}^{r}(\varepsilon)
		=\frac{\varepsilon (\varepsilon + \tilde \varepsilon_d + i \tilde \Gamma) - |\tilde\lambda|^2}{\varepsilon (\varepsilon - \tilde \varepsilon_d + i \tilde \Gamma)(\varepsilon + \tilde \varepsilon_d + i \tilde \Gamma)-2|\tilde\lambda|^2 (\varepsilon + i \tilde \Gamma)},
		\label{eq:32a}
	\end{equation}
	\begin{equation}
		\tilde{G}_{d12}^{r}(\varepsilon)
		=\frac{- |\tilde\lambda|^2}{\varepsilon (\varepsilon - \tilde \varepsilon_d + i \tilde \Gamma)(\varepsilon + \tilde \varepsilon_d + i \tilde \Gamma)-2|\tilde\lambda|^2 (\varepsilon + i \tilde \Gamma)}.
		\label{eq:32b}
	\end{equation}
\end{subequations}
which agree with the calculations of Wang et al.~\cite{Wang2020}. 
 
The current from Eq.~\eqref{eq:18} is determined in Appendix~\ref{sec:C} by substituting the corresponding Green's functions given by Eqs.~\eqref{eq:30a}-\eqref{eq:30c} into the formula for the dressed greater and lesser Green's functions~\eqref{eq:29}, with the help of the spectral function relation~\eqref{eq:27}.

\section{Numerical results and discussion}
\label{sec:III}
\begin{figure*}[ht]
\includegraphics[width =1\linewidth]{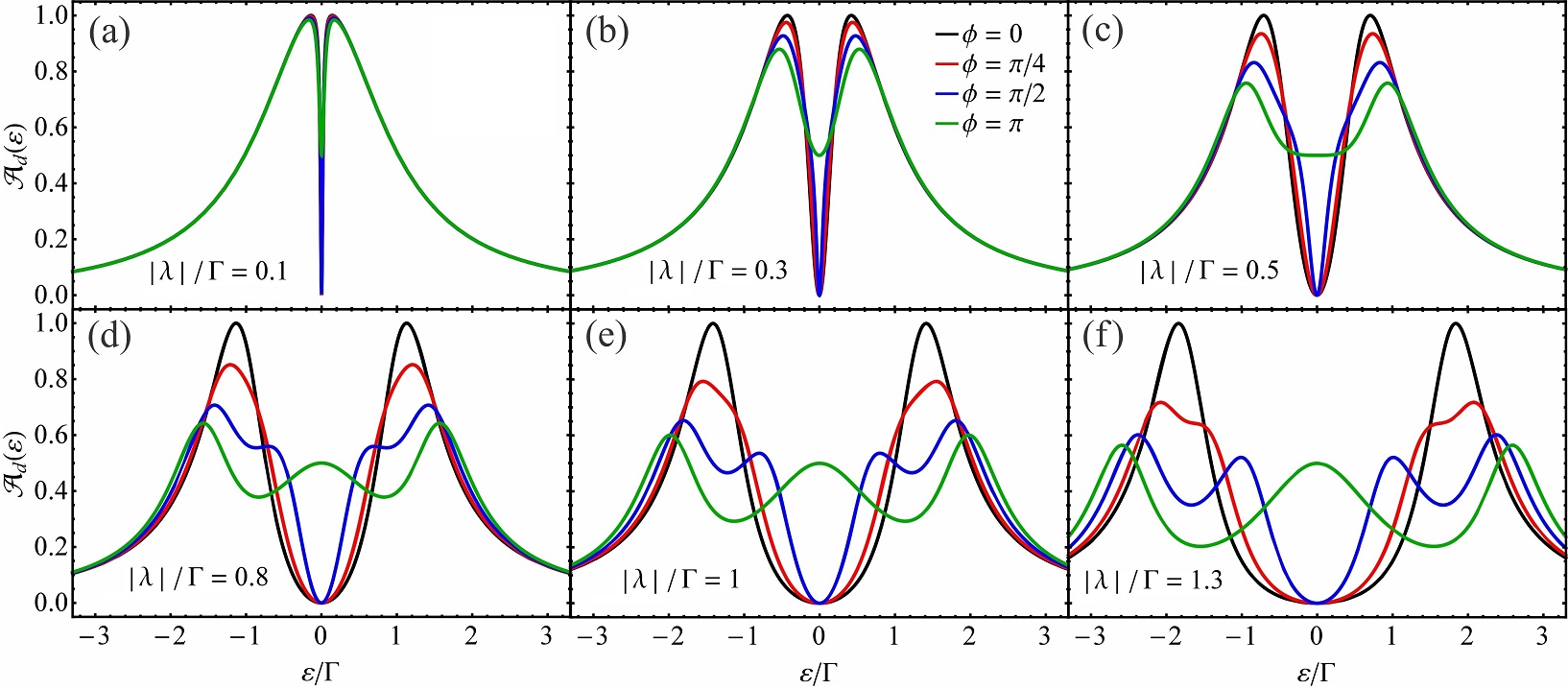}
\centering
\caption{The spectral function of the QD $\mathcal{A}_d(\varepsilon)$ for different values of the magnetic flux phase $\phi$, with unhybridized MBSs $\varepsilon_M/\Gamma = 0$, at the QD energy level $\varepsilon_d/\Gamma = 0$, when the symmetrical QD-MBS couplings $|\lambda_j| = |\lambda|$ are (a) $|\lambda|/\Gamma = 0.1$, (b) $|\lambda|/\Gamma = 0.3$, (c) $|\lambda|/\Gamma = 0.5$, (d) $|\lambda|/\Gamma = 0.8$, (e) $|\lambda|/\Gamma = 1$, and (f) $|\lambda|/\Gamma = 1.3$.}
\label{fig:2}
\end{figure*}
In this section, we present the numerical results for the transport properties of our QD-MBSs system in the absence and presence of EPI. As we mentioned above, all the system parameters should be much smaller than the superconducting gap $\Delta$ which in a typical experimental setup is on the order of \SI{250}{\micro eV} for a TSNW realized from InSb with a strong Rashba-type spin-orbit interaction~\cite{ Mourik2012}.  
The values for the QD-lead coupling $\Gamma$ and the QD-MBS couplings $|\lambda_j|$ in experiments are on the order of a few \si{\micro eV}~\cite{Cao2012}.
We consider longitudinal optical phonons with energy on the order of \SI{100}{\micro eV}~\cite{Sapmaz2006,Leturcq2009,Zhang2012}, smaller than the induced superconducting gap $\Delta$.
In our case, we work in the limit where electron-phonon coupling strength $\beta$ is stronger than the QD-lead coupling $\Gamma$ and the QD-MBS couplings $|\lambda_j|$ (see Sec.~\ref{sec:IIC}).
In numerical calculation, all energies are measured in units of $\Gamma$.
In the EPI-related numerical computations, the phonon energy is fixed as $\omega_0/\Gamma = 5$.

In such a system, where a QD connects to MBSs and the leads, different paths are allowed for charge carriers that travel from one lead to the other one through the dot. Paths involving Andreev reflection at the TSNW interface can constructively and destructively interfere with paths that pass only through the dot, and this produces a Fano line shape in the transport spectrum~\cite{Zeng2016b,Schuray2017,Calle2020}. Such a Fano structure shows up when the dot couples to the side-attached MBSs and thus the electrons can resonantly scatter on those. These fractional Fano-type resonances can give clear evidence of the presence of MBSs in a QD side-coupled to edges of a TSNW~\cite{Baranski2016}. 

\subsection{Numerical results in the absence of EPI}
\label{sec:IIIA}
As already mentioned above, we first examine the spectral function, and the contributions of the ET and LAR processes to the total linear and differential conductance when an opposite bias between the leads ($\mu_L = -\mu_R = eV/2$) is applied in the absence of EPI.
The spectral function given in Eq.~\eqref{eq:19} is also expressed in the form $\mathcal{A}_d(\varepsilon) = -\Gamma \text{Im}G_{d11}^r(\varepsilon)$, which is equivalent to $\mathcal{A}_d(\varepsilon) = \mathcal{T}_{LR}^{ee}(\varepsilon) + \mathcal{T}_{LL}^{eh}(\varepsilon)$. 
Note also that in the approximation scheme employed here within the EOM method (see App.~\ref{sec:B}), the retarded Green's functions $G_{d11}^r(\varepsilon)$ and $G_{d12}^r(\varepsilon)$ do not depend on temperature and bias voltage in the absence of EPI, which results in a temperature- and voltage-independent spectral function $\mathcal{A}_d(\varepsilon)$. 
\begin{figure*}[ht]
	\includegraphics[width =1\linewidth]{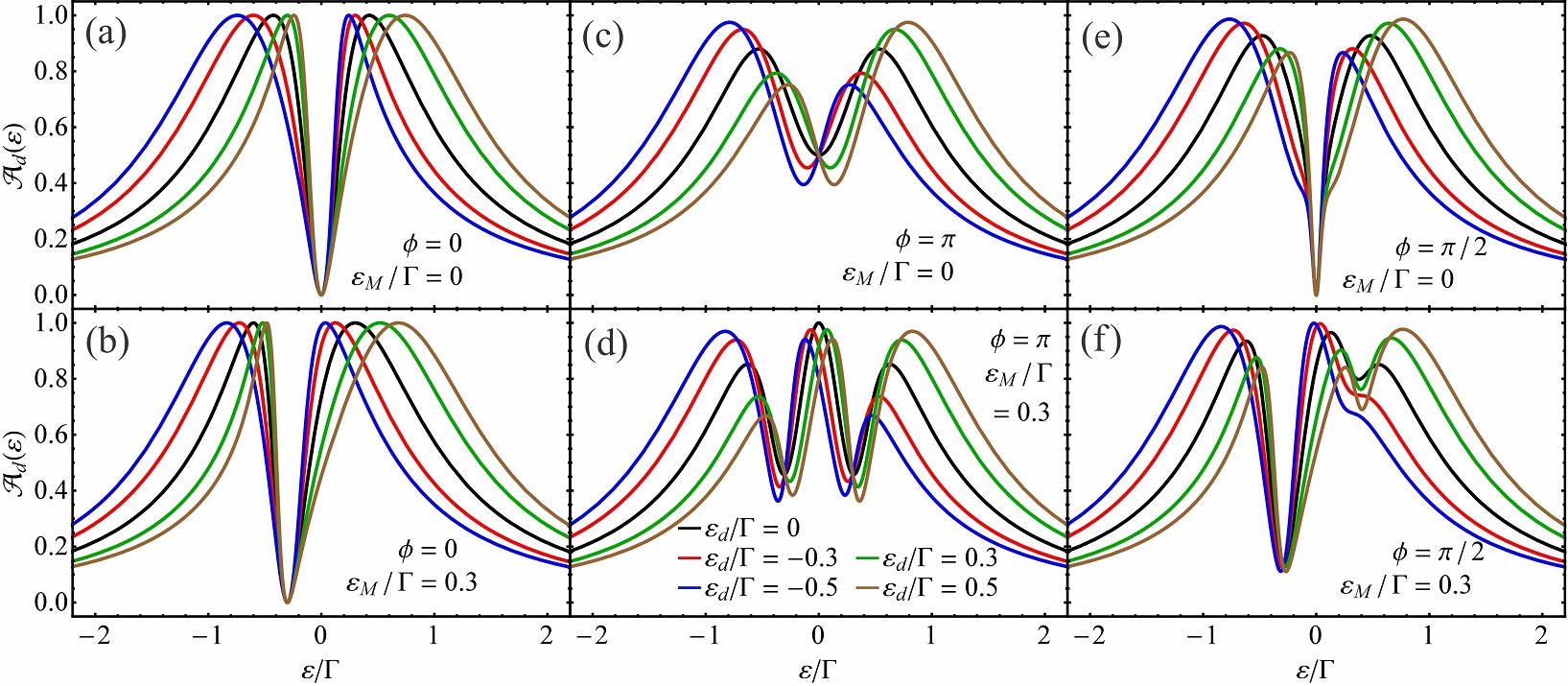}
	\centering
	\caption{The spectral function of the QD $\mathcal{A}_d(\varepsilon)$ for $\varepsilon_d$ and $\varepsilon_M$ with the symmetrical QD-MBS couplings $|\lambda| /\Gamma = 0.3$ for different magnetic flux phases $\phi$. The rest of parameters are (a) $\phi = 0$, $\varepsilon_M/\Gamma = 0$; (b) $\phi = 0$, $\varepsilon_M/\Gamma = 0.3$; (c) $\phi = \pi$, $\varepsilon_M/\Gamma = 0$; (d) $\phi = \pi$, $\varepsilon_M/\Gamma = 0.3$; (e) $\phi = \pi/2$, $\varepsilon_M/\Gamma = 0$; and (f) $\phi = \pi/2$, $\varepsilon_M/\Gamma = 0.3$.}
	\label{fig:3}
\end{figure*}

Figure~\ref{fig:2} shows the spectral function $\mathcal{A}_d(\varepsilon)$ of the QD as a function of energy for the case of a sufficiently long TSNW with vanishing overlap energy between MBSs, i.e., $\varepsilon_M = 0$, for different values of the symmetrical dot-MBS couplings $|\lambda_j|=|\lambda|$, at four different magnetic flux phase $\phi$ values and dot level $\varepsilon_d = 0$. 
The $\mathcal{A}_d(\varepsilon)$ peak position is approximated analytically from the poles of the retarded Green's function $G_{d11}^r(\varepsilon)$ [see Appendix \ref{sec:B}, Eq.~\eqref{eq:B26}].
We see that for small values of the symmetrical QD-MBS couplings $|\lambda|$, the spectral function presents two maxima at the positions $\varepsilon \approx \pm \sqrt{2}|\lambda| \sqrt{1 + |\sin(\phi/2)|}$. 
With increasing QD-MBS coupling $|\lambda|$, the maxima of $\mathcal{A}_d(\varepsilon)$ shift to higher values of $|\varepsilon|$. 
In particular, the spectral function at $\varepsilon=0$ vanishes for almost any magnetic phase value, except for $\phi = (2n + 1)\pi$ with $n = 0, \pm 1, \pm 2, ...$ when $\mathcal{A}_d(0) = 1/2$.
In addition, when the dot-MBS couplings reach the value $|\lambda|/\Gamma \approx 1/2$, the spectral function presents a narrow flat region near zero energy, i.e., $|\varepsilon|/\Gamma \lesssim 1/3$, for magnetic flux phase $\phi = (2n + 1)\pi$, in agreement with the literature~\cite{Ramos2018}. 
After the symmetrical dot-MBS coupling exceeds the value $|\lambda|/\Gamma \approx 1/2$, the spectrum shows a robust three-peak structure for $\phi = (2n + 1)\pi$.
The central peak at $\varepsilon=0$ is clear evidence for the presence of MBSs in the system.
Consequently, for any finite values of $|\lambda|$, at $\phi = (2n + 1)\pi$, the zero-energy spectral function $\mathcal{A}_d(0)$ is always $1/2$ for unhybridized MBSs $\varepsilon_M = 0$, thus yielding a linear conductance $e^2/2h$. The $e^2/2h$ value of linear conductance is a clear signature of the presence of MBSs~\cite{Liu2011}.
In the cases, $\phi = \pi/4$ and $\phi = \pi/2$, when the QD-MBS coupling exceeds the value $|\lambda|/\Gamma > 1/2$, the two-peak structure begins to evolve into a four-peak structure.
The positions of the side peaks are given by the relation $\varepsilon \approx \pm \sqrt{2}|\lambda| \sqrt{1 + |\sin(\phi/2)|}$ while the new intermediate peaks are located at $\varepsilon \approx \pm \sqrt{2}|\lambda| \sqrt{1 - |\sin(\phi/2)|}$. 
As a consequence, the increasing $|\lambda|$ significantly changes the peak structure of the spectral function, but remarkably it shows little influence on the spectrum near zero energy for $\phi \neq (2n + 1)\pi$.
Note that the spectral function $\mathcal{A}_d(\varepsilon)$ is a $2\pi$-periodic function of the magnetic flux phase $\phi$.
\begin{figure*}[ht]
	\includegraphics[width =1\linewidth]{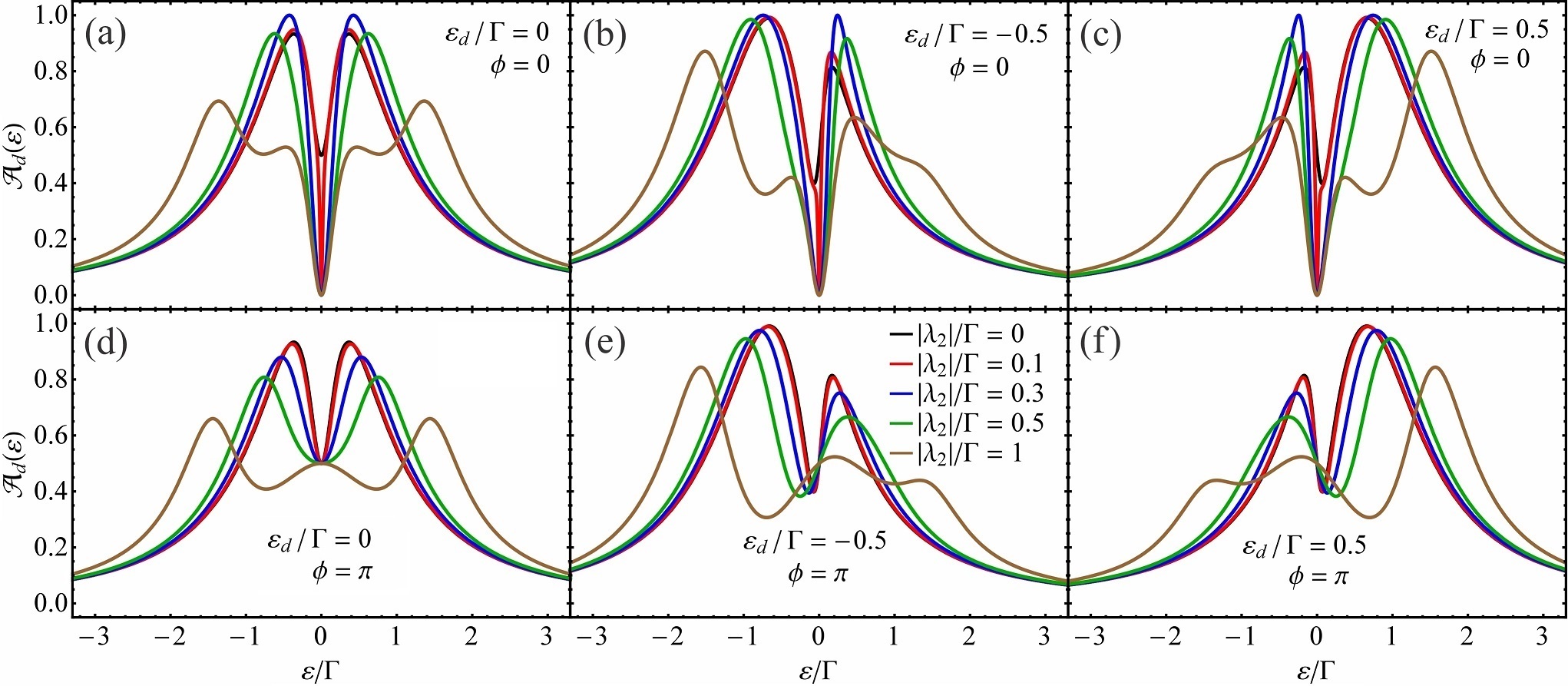}
	\centering
	\caption{The spectral function of the QD $\mathcal{A}_d(\varepsilon)$ as the differently tuned QD energy level $\varepsilon_d$ with unhybridized MBSs $\varepsilon_M/\Gamma = 0$, where the QD-MBS coupling $|\lambda_1| /\Gamma = 0.3$ is fixed at different values of magnetic flux phase $\phi$ and QD-MBS coupling $|\lambda_2|$. The rest of parameters are (a) $\phi = 0$, $\varepsilon_d/\Gamma = 0$; (b) $\phi = 0$, $\varepsilon_d/\Gamma = -0.5$; (c) $\phi = 0$, $\varepsilon_d/\Gamma = 0.5$; (d) $\phi = \pi$, $\varepsilon_d/\Gamma = 0$; (e) $\phi = \pi$, $\varepsilon_d/\Gamma = -0.5$; and (f) $\phi = \pi$, $\varepsilon_d/\Gamma = 0.5$.}
	\label{fig:4}
\end{figure*}

To investigate the influence of dot level $\varepsilon_d$ and overlap energy $\varepsilon_M$ on the spectral function of the QD, we plot in Fig.~\ref{fig:3} $\mathcal{A}_d(\varepsilon)$ as a function of energy at three different magnetic flux phase $\phi$ values and dot level $\varepsilon_d \neq 0$ energies with finite MBS-MBS overlap $\varepsilon_M \neq 0$ and weak symmetrical QD-MBS couplings $|\lambda_j|=|\lambda|$.
In experiments, the energy level of the QD is tuned by metallic gate electrodes. 
When the dot level shifts away from the Fermi level $\varepsilon_d \neq \varepsilon_F$, where $\varepsilon_F = 0$, the line shape in the spectral function profile is clearly modified. For magnetic flux phase $\phi = 0$ [see Figs.~\ref{fig:3}(a) and~\ref{fig:3}(b)], the antiresonance point of the spectrum always decreases to zero and shifts to the point of $\varepsilon \approx -\varepsilon_M$ for any values of the dot level $\varepsilon_d$~\cite{Wang2018b}. 
In this case, two maxima of $\mathcal{A}_d(\varepsilon)$ emerge in the spectrum [see also Eq.~\eqref{eq:B26}]. 
The broadened peak at $\varepsilon\approx \varepsilon_d$ is due to resonant transmission through a QD at the tuned dot level ($\varepsilon_d \neq 0$) even in the absence of MBSs.
The sharp peaks are mainly attributed to the regular fermionic states originating from the MBSs. 
In addition, for $\varepsilon_d = - \varepsilon_M$, the two resonant peaks of spectral function become identical, as was also found in Ref.~\cite{Wang2018b}.
When the magnetic flux phase is $\phi = \pi$ [see Figs.~\ref{fig:3}(c) and~\ref{fig:3}(d)], the two-peak structure in the spectrum (at $\varepsilon_M = 0$) evolves when increasing $\varepsilon_{M}$ into a three-peak structure showing the presence of the MBSs in the system~\cite{Liu2011,Chi2020}.
Moreover, the zero-energy spectral function takes the value of $1/2$ for unhybridized MBSs ($\varepsilon_M = 0$) regardless of QD level tuning, while the antiresonance point is $\varepsilon_d$-dependent. It shifts to the negative or positive energy region for $\varepsilon_d < 0$ or $\varepsilon_d > 0$, respectively. 
For hybridized MBSs ($\varepsilon_M \neq 0$), increasing the value of $|\varepsilon_d|$, the intermediate peak magnitude decreases from $1$ and its location shifts away from the zero-energy point $\varepsilon = 0$. Thus, the main peak is shifted from the lower- to the higher-energy region by tuning the dot level from the filled ($\varepsilon_d < 0$) to the empty region ($\varepsilon_d > 0$).
Beside the main peak, two side peaks develop in the spectrum with amplitudes significantly affected by the dot level. 
The broadened peaks due to the resonant tunneling of electrons at the QD energy level $\varepsilon_d$ present an enhanced magnitude for large values of $|\varepsilon_d|$. 
The sharp peaks are destructively affected by the tuned QD level resulting in smaller amplitude peaks.
Note that at $\varepsilon_d = 0$ for weak QD-MBS couplings considered here the spectral function presents two minima at energies $\varepsilon \approx \pm \varepsilon_M$. As $|\lambda|$ increases, for any nonzero values of the MBS-MBS coupling $\varepsilon_M \neq 0$, the zero-energy spectral function always gives $\mathcal{A}_d(0) = 1$ instead of $1/2$ (not shown here). The increase of $|\lambda|$ results in the minima of $\mathcal{A}_d(\varepsilon)$ moving away from $\varepsilon \approx \pm \varepsilon_M$. The full width at half-maximum of the zero-energy peak is proportional to $\varepsilon_M$. 
For magnetic flux phase $\phi = \pi/2$ [Figs.~\ref{fig:3}(e) and~\ref{fig:3}(f)], the antiresonance point of $\mathcal{A}_d(\varepsilon)$ is zero at $\varepsilon = 0$ for unhybridized MBSs and is independent of $\varepsilon_d$. 
At finite values of $\varepsilon_M$, the antiresonance point shifts its location to $\varepsilon \approx - \varepsilon_M$, and becomes slightly dependent on $\varepsilon_d$.
As was discussed above, for $\varepsilon_M = 0$, the side peaks locations are analytically determined in Eq.~\eqref{eq:B26}. 
In the case of $\varepsilon_M \neq 0$, the spectrum becomes more complicated, showing a three-peak structure.
Note that for any $\phi$ and $\varepsilon_M=0$, the Fano line shape at finite $\varepsilon_d$ and the one at $-\varepsilon_d$ are related by a mirror reflection symmetry around the $\varepsilon=0$ axis~\cite{Wang2018b}.
Consequently, the applied magnetic field could be used in flux-controlled operation in TSNW-based quantum computation~\cite{Flensberg2011,Ramos2018}.
\begin{figure*}[ht]
\includegraphics[width =1\linewidth]{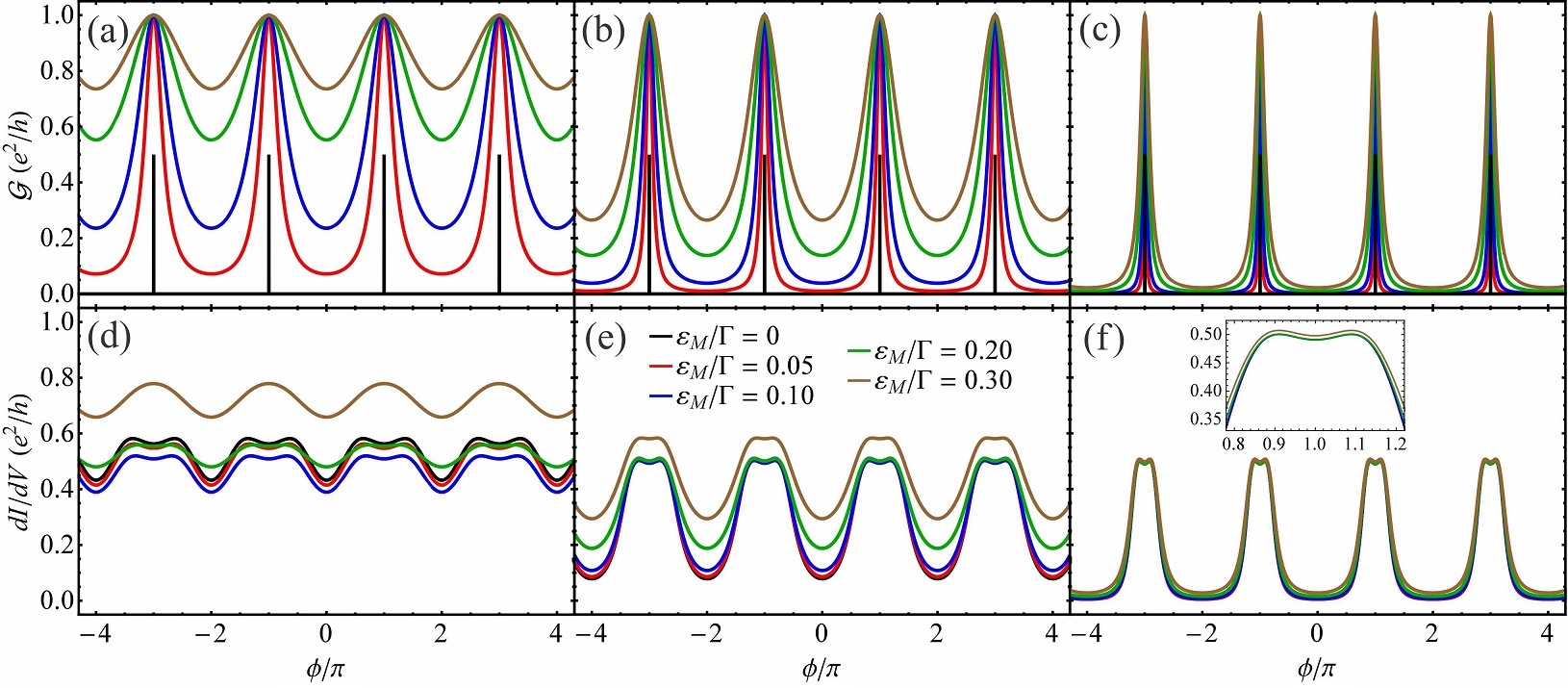}
\centering
\caption{Spectra of the linear $\mathcal{G}$ (a)-(c) and differential conductance $dI/dV$ with the bias voltage $eV/\Gamma = 0.28$ (d)-(f) as a function of magnetic flux phase $\phi$ for different values of the overlap energy $\varepsilon_M$ at three different symmetrical QD-MBS couplings $|\lambda_j| = |\lambda|$ at dot level $\varepsilon_d/\Gamma = 0$, and zero temperature. The rest of parameters are (a) $|\lambda| /\Gamma = 0.3$, (b) $|\lambda| /\Gamma = 0.5$, (c) $|\lambda| /\Gamma = 1$, (d) $|\lambda| /\Gamma = 0.3$, (e) $|\lambda| /\Gamma = 0.5$, and (f) $|\lambda| /\Gamma = 1$.
Inset: Zoom at a differential conductance peak.}
\label{fig:5}
\end{figure*}

We now study the influence of different QD-MBS coupling strengths $|\lambda_j|$ on the peak structure of the spectral function $\mathcal{A}_d(\varepsilon)$ at two different magnetic flux phase $\phi$ values when the dot level is tuned away from the Fermi level $\varepsilon_d \neq \varepsilon_F$ with unhybridized MBSs ($\varepsilon_M = 0$).
The numerical results are shown in Fig.~\ref{fig:4}.
In the following, we analyze the features of the spectral function by fixing one of the couplings $|\lambda_1|/\Gamma = 0.3$ and varying $|\lambda_2|$.
We observe that when the QD is at the Fermi level ($\varepsilon_d  = 0$) and the magnetic flux is turned off ($\phi = 0$), two or four identical resonant peaks develop in the spectrum as a function of the values of $|\lambda_2|$ [see Fig.~\ref{fig:4}(a)]. 
For $|\lambda_2| = 0$, the zero-energy spectral function takes the expected $1/2$ value, i.e., $\mathcal{A}_d(0) = 1/2$, which decreases to zero when $|\lambda_2|$ is finite.
With increasing $|\lambda_2|$, the resonant peaks are shifted from the antiresonance point $\varepsilon = 0$ and their amplitude has a maximum at $|\lambda_1|=|\lambda_2|$. 
When $|\lambda_2| > |\lambda_1|$, the antiresonance valley further widens, which has been established in a previous work~\cite{Wang2018b}. 
With increasing $|\lambda_2|$, the two-peak structure holds until $|\lambda_2|$ approaches $\approx 0.8\Gamma$. 
With further increase of $|\lambda_2|$, the spectrum of $\mathcal{A}_d(\varepsilon)$ presents a four-peak structure. 
In addition, the minimum of $\mathcal{A}_d(\varepsilon)$ for $|\lambda_2| = 0$ shifts from $\varepsilon = 0$ to the negative energy region for $\varepsilon_d < 0$ and to the positive energy region for $\varepsilon_d > 0$, respectively [see Figs.~\ref{fig:4}(b) and~\ref{fig:4}(c)]. 
The position of the antiresonance point $\varepsilon = 0$ remains unchanged when $|\lambda_2| \neq 0$ and $\varepsilon_d \neq 0$. 
In agreement with the results of Fig.~\ref{fig:3} where symmetrical QD-MBS couplings $|\lambda_j| = |\lambda|$ are considered, the amplitude of the broadened peaks attributed to the resonant tunneling at the dot level $\varepsilon_d$ and the narrow peaks caused by the MBSs, increase to unity when $|\lambda_2| = |\lambda_1|$. 
When $|\lambda_2|$ exceeds $|\lambda_1|$, the peaks' amplitude decreases.
Additionally, the two-peak structure of $\mathcal A_d(\varepsilon)$ at small $|\lambda_2|$ evolves into a four-peak structure visible at $|\lambda_2|/\Gamma\approx 1$.
Therefore, when a magnetic flux threads the MBS ring system with the magnetic flux phase $\phi = \pi$ [see Figs.~\ref{fig:4}(d)-\ref{fig:4}(f)], the zero-energy spectral function at $\varepsilon_M=0$ always takes the value $1/2$, $\mathcal{A}_d(0) = 1/2$, for any values of $|\lambda_2|$ and $\varepsilon_d$, respectively.
In this case, the MBS signature is independent of the finite values of couplings $|\lambda_j|$, and this can be interpreted as robustness to disorder~\cite{Li2013}.
Similar to the case $\phi = 0$, the spectrum exhibits two peaks for small $|\lambda_2|$, and a clear three-peak structure is visible in the spectrum at $\varepsilon_d = 0$ for $|\lambda_2|/\Gamma \gtrsim 0.7$, while it is visible at $\varepsilon_d \neq 0$ for $|\lambda_2|/\Gamma \gtrsim 0.9$.
Therefore, when the dot level is tuned away from the Fermi level ($\varepsilon_d \neq 0$), the antiresonance point behaves as in the case of $\phi = 0$ for $|\lambda_2| = 0$. Moreover, in contrast to the results from Figs.~\ref{fig:4}(b) and~\ref{fig:4}(c), here when $\varepsilon_d \neq 0$, the position of the antiresonance point at small values of $|\lambda_2|$ changes as in the case of $|\lambda_2| = 0$.

As a consequence, we conclude from Figs.~\ref{fig:3} and~\ref{fig:4} that, in the case of unhybridized MBSs, $\varepsilon_M = 0$, the zero-energy spectral function is $\mathcal{A}_d(0) = 1/2$ if either of the QD-MBS couplings is finite and the other one is zero for any $\phi$, and $\mathcal{A}_d(0) = 1/2$ or $\mathcal{A}_d(0) = 0$ if both of the QD-MBS couplings have finite values, $|\lambda_j| \neq 0$, for $\phi = (2n + 1)\pi$ or $\phi \neq (2n + 1)\pi$, respectively, regardless of the change in $\varepsilon_d$.
We will see next that also the linear conductance is independent of $\varepsilon_d$ at $\varepsilon_M=0$ and zero temperature, which indicates again the robustness of MBSs signatures~\cite{Liu2011,Lee2013,Lopez2014}.
\begin{figure*}[ht]
\includegraphics[width =1\linewidth]{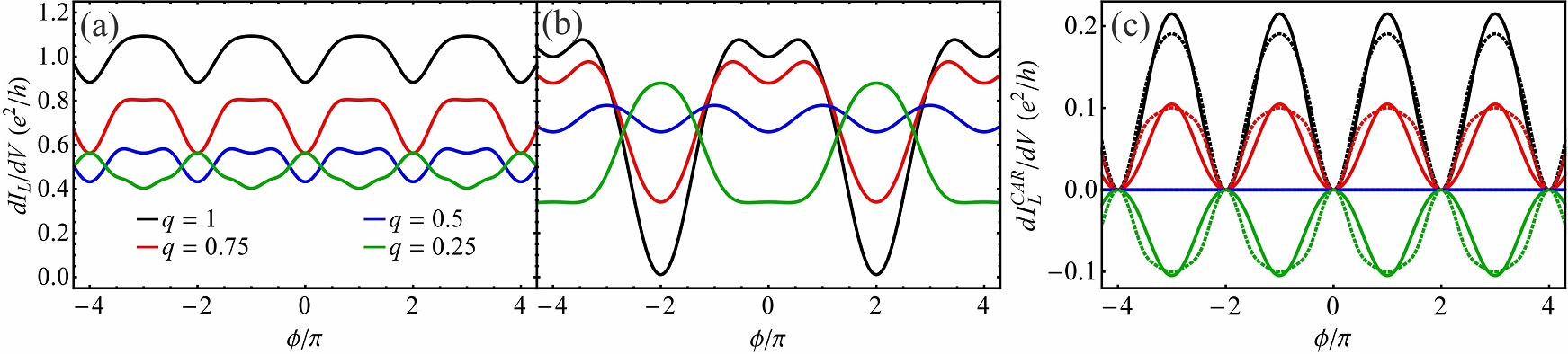}
\centering
\caption{Spectra of the differential conductance from the left ($L$) lead $dI_L/dV$ as a function of magnetic flux phase $\phi$ for (a) unhybridized ($\varepsilon_M/\Gamma = 0$) and (b) hybridized MBSs with $\varepsilon_M/\Gamma = 0.3$. (c) The CAR component of the differential conductance for lead $L$ $dI_L^{CAR}/dV$ where the solid lines correspond to $\varepsilon_M/\Gamma = 0.3$ while the dashed lines correspond to the case of $\varepsilon_M/\Gamma = 0$, respectively.
The system is biased for different values of $\mu_L=qeV$ with $\mu_L-\mu_R=eV$ and $eV/\Gamma = 0.28$.
The rest of parameters are the symmetrical QD-MBS couplings $|\lambda|/\Gamma = 0.3$, dot level $\varepsilon_d/\Gamma = 0$, and zero temperature.}
\label{fig:6}
\end{figure*}

Figures~\ref{fig:5}(a)-~\ref{fig:5}(c) illustrate the zero-temperature total linear conductance $\mathcal{G}$, and~\ref{fig:5}(d)-~\ref{fig:5}(f) illustrate the zero-temperature total differential conductance $dI/dV$ with fixed $eV$ as a function of magnetic flux phase $\phi$ at zero and finite values of the overlap energy $\varepsilon_M$ between MBSs and for three different strengths of the symmetrical QD-MBS couplings $|\lambda_j|=|\lambda|$ and dot level $\varepsilon_d = 0$. 
The linear conductance $\mathcal{G}$ shows a $2\pi$ periodicity for $\varepsilon_M =0$~\cite{Liu2011} and $\varepsilon_M \neq 0$ with maxima located at $\phi = (2n + 1)\pi$, with $n$ an integer. 
When $\phi \neq (2n + 1)\pi$, the linear conductance is $\mathcal{G} = 0$ for $\varepsilon_M = 0$, regardless of the finite values of symmetrical QD-MBS couplings $|\lambda|$.
The minima of $\mathcal{G}$ increase with $\varepsilon_M$. 
As $|\lambda|$ increases, the minima of $\mathcal{G}$ decrease and tend to zero even for strong MBS-MBS couplings $\varepsilon_M$, while the conductance peaks become sharper. Note that $\varepsilon_M$ has an observable influence on the peaks width when the QD weakly couples to the MBSs. The peak height is $e^2/h$ for $\varepsilon_M \neq 0$ and $e^2/2h$ for $\varepsilon_M = 0$ and is not affected by the symmetrical QD-MBS coupling $|\lambda|$.
The differential conductance $dI/dV$ also presents a $2\pi$ periodicity for $\varepsilon_M =0$ and $\varepsilon_M \neq 0$, with global minima located at $\phi = 2n\pi$, with $n$ an integer, as in the case of $\mathcal{G}$.
For small $\varepsilon_{M}$, an additional set of local minima exist at $\phi=(2n+1)\pi$.
However, at higher values of $\varepsilon_{M}$ the local minima evolve into maxima and the entire $dI/dV$ curve shifts upward.
The critical value of $\varepsilon_{M}$ where the local minima vanish increases with the couplings $|\lambda|$.
With increasing the QD-MBS couplings $|\lambda|$, the $dI/dV$ curves tend to the curve corresponding to $\varepsilon_M = 0$, and thus the overlap energy only weakly influences the spectrum at larger $|\lambda|$. In addition, the width of plateau-like maxima is suppressed and the height is approximately $e^2/2h$. 
Both $2\pi$ and $4\pi$ periodicity of the differential conductance as a function of magnetic flux phase $\phi$ have been reported in the literature for TSNWs, with $\varepsilon_{M}=0$ and $\varepsilon_{M}\neq0$, connected to a QD~\cite{Zeng2016b}. To clarify the influence of biasing on the magnetic flux phase periodicity of differential conductance, we consider that the system is biased as follows: $\mu_L=q eV$ and $\mu_R = (q-1)eV$ such that $\mu_L - \mu_R = eV$ with a choice of $0 \le q \le 1$.
Next, we present our findings for the zero-temperature differential conductance $dI_L/dV$ determined for the current $I_L$ flowing from the left ($L$) lead. The analytical results for differential conductance are presented in Appendix~\ref{sec:A}. We plot the total differential conductance for lead $L$ $dI_L/dV$ as a function of the magnetic flux phase $\phi$ for $\varepsilon_{M}=0$ and $\varepsilon_{M}\neq0$ in Figs.~\ref{fig:6}(a) and~\ref{fig:6}(b), respectively. Figure~\ref{fig:6}(c) shows the component of the total differential conductance from the CAR processes, i.e., $dI_L^{CAR}/dV$. In order to avoid all possible consequences caused by the symmetry of the system, the bias voltage is set to $eV/\Gamma = 0.28$. 
The periodicity of differential conductance as a function of $\phi$ is $2\pi$ for $\varepsilon_M=0$ and remains invariant when changing the biasing of the system. 
However, the differential conductance period can be $2\pi$ and $4\pi$ depending on the choice of biasing for $\varepsilon_M \neq 0$. 
Namely, when the system is biased as $\mu_L = eV$ and $\mu_R = 0$ all the ET, LAR and CAR processes are involved, resulting in an enhancement in the amplitude of the differential conductance.  
The contribution of CAR processes is generally finite except at $\mu_L = -\mu_R = eV/2$, as discussed in Sec.~\ref{sec:IIB}.
Note, therefore, that the differential conductance periodicity remains unchanged under asymmetric dot-lead couplings ($\Gamma_L \neq \Gamma_R$).

\begin{figure}[ht]
	\includegraphics[width =1\linewidth]{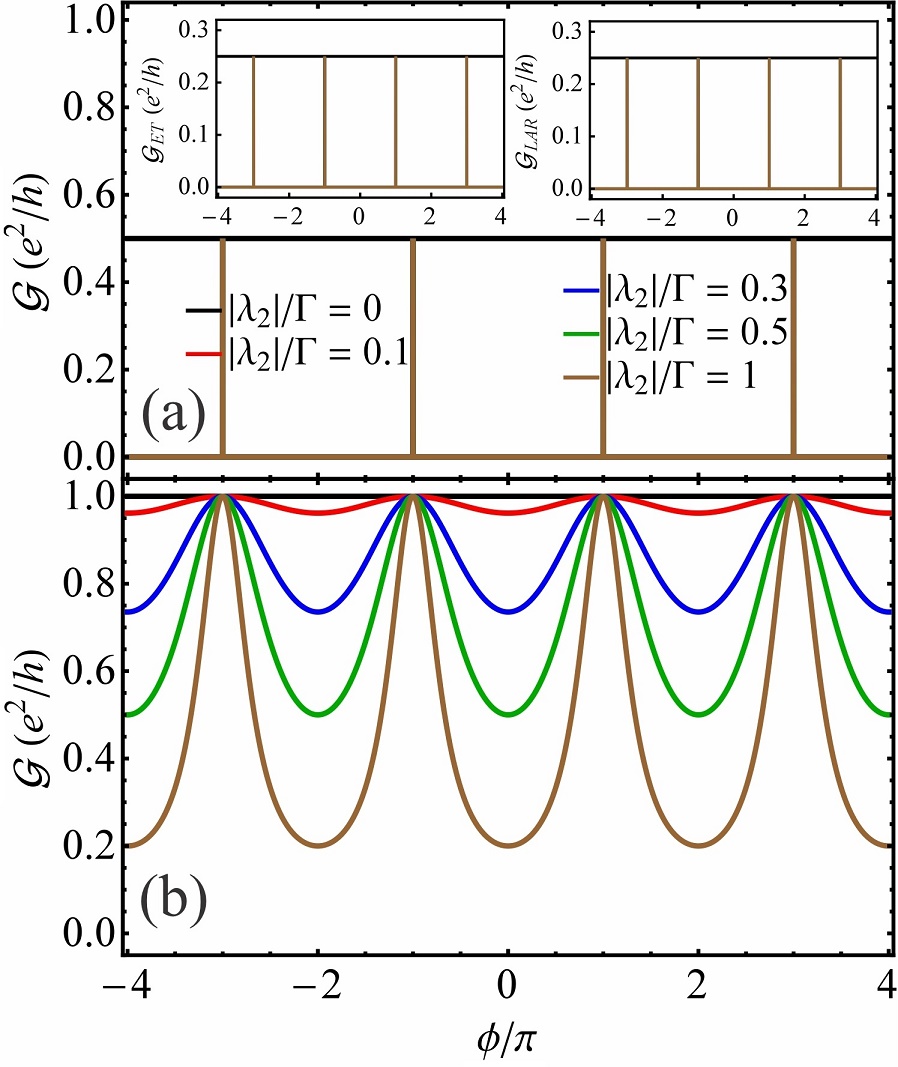}
	\centering
	\caption{(a) The total linear conductance $\mathcal{G}$ as a function of magnetic flux phase $\phi$ for $\varepsilon_M/\Gamma = 0$ at zero temperature. 
		The left and right insets show the results for $\mathcal{G}_{ET}$ and $\mathcal{G}_{LAR}$, respectively, as a function of $\phi$ for different values of $|\lambda_2|$. (b) The total linear conductance $\mathcal{G}$ as a function of $\phi$ for hybridized MBSs, $\varepsilon_M/\Gamma = 0.3$, at zero temperature. 
		The QD-MBS1 coupling is fixed, $|\lambda_1|/\Gamma = 0.3$, and the QD-MBS2 coupling $|\lambda_2|$ varies. 
		The QD energy level is $\varepsilon_d/\Gamma = 0$.}
	\label{fig:7}
\end{figure}

As we have seen above, when the system is biased as $\mu_L = - \mu_R = eV/2$, ET and LAR processes contribute to the differential and linear conductance. 
For this reason, we examine the ET and LAR components of the zero-temperature total linear conductance for different QD-MBS2 coupling strengths $|\lambda_2|$ with fixed QD-MBS1 coupling $|\lambda_1|$. 
We plot in Fig.~\ref{fig:7}(a) the total linear conductance $\mathcal{G}$, together with ET ($\mathcal{G}_{ET}$) and LAR ($\mathcal{G}_{LAR}$) linear conductances, as a function of magnetic flux phase $\phi$ at $\varepsilon_M = 0$ and zero temperature with $\varepsilon_d = 0$. 
The numerical results show that the linear conductances $\mathcal{G}_{ET}$ and $\mathcal{G}_{LAR}$ exhibit similar properties to the total linear conductance presented in Fig.~\ref{fig:5}.
Namely, the ET and LAR linear conductances oscillate with a $2\pi$ period, having maxima at $\phi = (2n + 1)\pi$, with magnitudes $\mathcal{G}_{ET} = \mathcal{G}_{LAR} = e^2/4h$, regardless of the finite magnitude of coupling $|\lambda_2|$. 
The total linear conductance reaches the value $e^2/2h$. 
When the MBS2 is not coupled to the QD ($|\lambda_2| = 0$), the ET and LAR linear conductances are equal to $e^2/4h$ independently of the magnetic flux phase $\phi$.
When $\phi \neq (2n + 1)\pi$, the linear conductances are zero for finite values of $|\lambda_j|$. 
Note also that the zero-temperature linear conductances do not depend on $\varepsilon_d$. 
This behavior of $\mathcal{G}_{ET}$ and $\mathcal{G}_{LAR}$, defined by Eq.~\eqref{eq:A11}, is analytically verified by using Eqs.~\eqref{eq:30a}-\eqref{eq:30c}. 
If $\varepsilon_M = 0$ with $|\lambda_j|\neq 0$, in the limit $\varepsilon \to 0$, the retarded Green's functions become
\begin{equation}
\label{eq:33}
\begin{aligned}
G_{d11}^{r}(\varepsilon \to 0)&\approx \bigg[
2(\varepsilon + i \Gamma) - \frac{4}{\varepsilon}\frac{|\lambda_1 \lambda_2|^2 \cos^2\frac{\phi}{2}}{|\lambda_1|^2 + |\lambda_2|^2}
\bigg]^{-1},
\\
G_{d12}^r(\varepsilon \to 0)& \approx \frac{\frac{|\lambda_1|^2 - |\lambda_2|^2}{|\lambda_1|^2 + |\lambda_2|^2}\cos\frac{\phi}{2} -i \sin\frac{\phi}{2}}{2(\varepsilon + i \Gamma) - \frac{4}{\varepsilon}\frac{|\lambda_1 \lambda_2|^2 \cos^2\frac{\phi}{2}}{|\lambda_1|^2 + |\lambda_2|^2}}.
\end{aligned}
\end{equation}
We see from Eq.~\eqref{eq:33} that ${G}_{d11}^{r}(\varepsilon \to 0) = {G}_{d12}^{r}(\varepsilon \to 0) = 0$ when $\phi \neq (2n + 1)\pi$, while at $\phi = (2n + 1)\pi$, ${G}_{d11}^{r}(\varepsilon \to 0) \approx 1/2(\varepsilon + i\Gamma)$ and ${G}_{d12}^{r}(\varepsilon \to 0) \approx i (-1)^{n+1}/2(\varepsilon + i\Gamma)$, independent of $\varepsilon_d$ and finite values of $|\lambda_j|$. 
If $|\lambda_1| \neq 0$ and $|\lambda_2| = 0$, the retarded Green's functions are approximated ${G}_{d11}^{r}(\varepsilon \to 0) \approx 1/2(\varepsilon + i\Gamma)$ and
\begin{figure*}[ht]
	\includegraphics[width=1\textwidth]{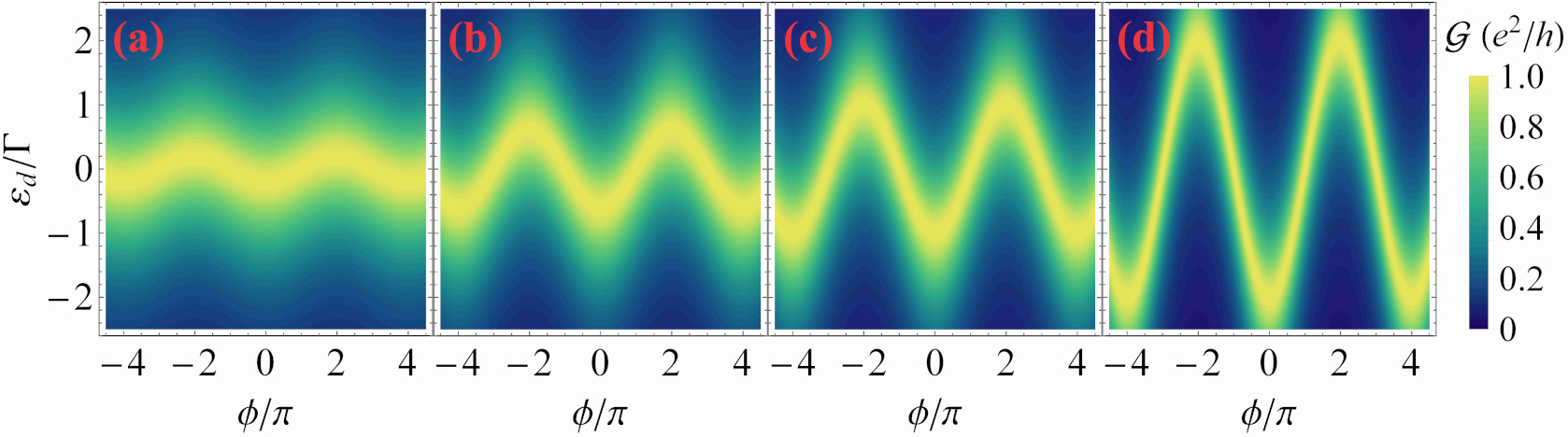}
	\centering
	\caption{The zero-temperature linear conductance $\mathcal{G}$ as a function of magnetic flux phase $\phi$ and QD energy level $\varepsilon_d$ for $\varepsilon_M/\Gamma = 0.3$ when the QD-MBS coupling $|\lambda_1|/\Gamma = 0.3$ is fixed for different strengths of the QD-MBS coupling $|\lambda_2|$. (a) $|\lambda_2|/\Gamma = 0.1$, (b) $|\lambda_2|/\Gamma = 0.3$, (c) $|\lambda_2|/\Gamma = 0.5$, and (d) $|\lambda_2|/\Gamma = 1$.}
	\label{fig:8}
\end{figure*}
${G}_{d12}^{r}(\varepsilon \to 0) \approx e^{-i\phi/2}/2(\varepsilon + i\Gamma)$, independent of $\varepsilon_d$, which results in $\mathcal{G}_{ET} = \mathcal{G}_{LAR} = e^2/4h$.
The independence of $\mathcal{G}$ with respect to $\varepsilon_d$ for $\varepsilon_M = 0$ and at zero temperature was also found in different setups involving QDs connected to TSNWs~\cite{Liu2011,Lopez2014,Lee2013}.
In the case $\varepsilon_M \neq 0$ and $\varepsilon_d=0$ at zero temperature [Fig~\ref{fig:7}(b)], the conductance peaks emerge at $\phi = (2n + 1)\pi$.
For $\varepsilon_M \neq 0$, the retarded Green's function $G_{d12}^r(\varepsilon \to 0) \approx 0$, independent of $\varepsilon_d$, which leads to a vanishing LAR conductance $\mathcal{G}_{LAR} \approx 0$. 
Thus the total linear conductance $\mathcal{G}$ is due to the ET conductance $\mathcal{G}_{ET}$, i.e., $\mathcal{G} \approx \mathcal{G}_{ET}$, and it shows a $2\pi$ periodicity in $\phi$ for $\varepsilon_d = 0$. 
We observe that with the increase of $|\lambda_2|$, the minima of $\mathcal{G}$ located at $\phi = 2n\pi$ gradually decrease and the peaks begin to narrow.
Most remarkably, contrary to the case of $\varepsilon_M = 0$, the maxima of linear conductance for $\varepsilon_M \neq 0$ approach the value $e^2/h$. 
Consequently, in the case of unhybridized MBSs, the linear conductance is attributed equally to the nonlocal (ET) and local (LAR) processes, independently of the $\varepsilon_d$ value. 
However, when the MBSs hybridize, the ET processes contribute dominantly to the linear conductance, with vanishing contribution from the LAR processes.
\begin{figure*}[ht]
	\includegraphics[width =1\linewidth]{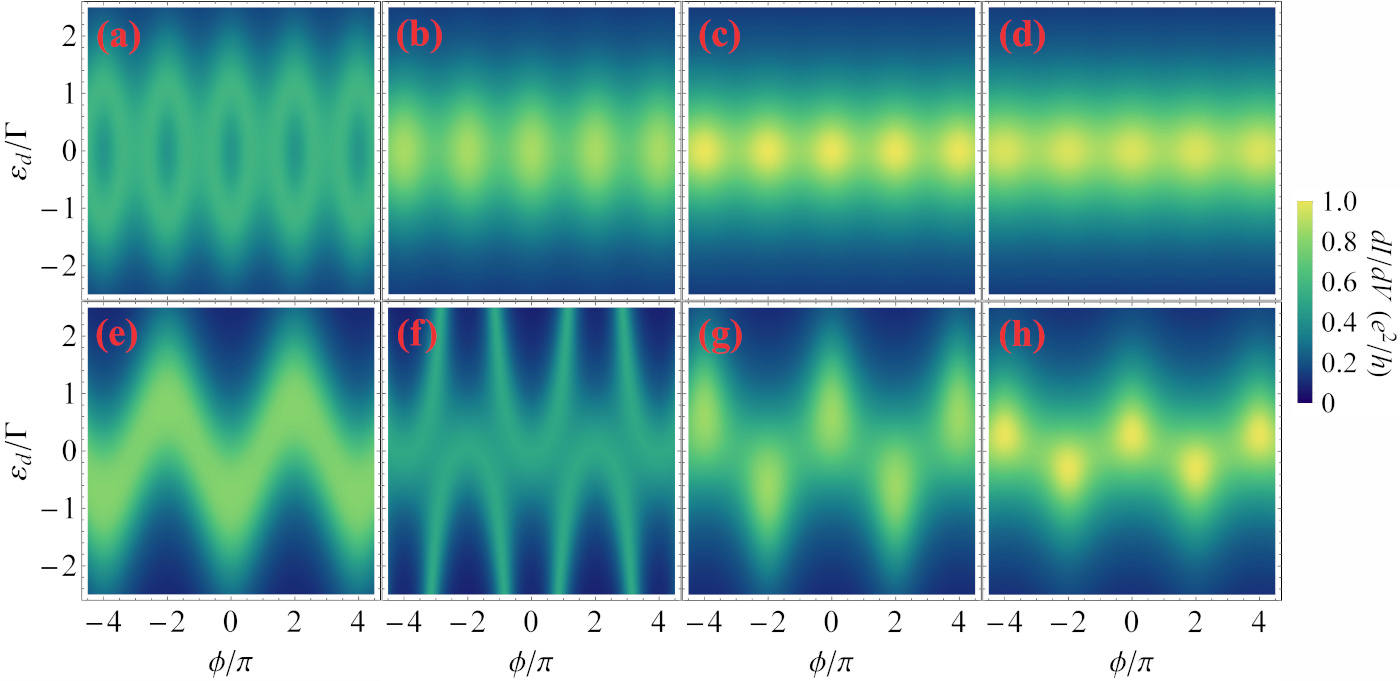}
	\centering
	\caption{The zero-temperature differential conductance $dI/dV$ as a function of magnetic flux phase $\phi$ and dot energy $\varepsilon_d$ for (a)-(d) unhybridized MBSs $\varepsilon_M/\Gamma = 0$ and (e)-(h) $\varepsilon_M/\Gamma = 0.3$ when the symmetrical QD-MBS coupling is $|\lambda|/\Gamma = 0.3$. The rest of the parameters are (a), (e) $eV/\Gamma = 0.28$; (b), (f) $eV/\Gamma = 0.55$; (c), (g) $eV/\Gamma = 0.85$; and (d), (h) $eV/\Gamma = 1.05$.}
	\label{fig:9}
\end{figure*}

To further investigate the regime $\varepsilon_d\neq 0$ and $\varepsilon_M\neq 0$, we plot in Fig.~\ref{fig:8} the zero-temperature total linear conductance $\mathcal{G}\approx \mathcal G_{ET}$ as a function of magnetic flux phase $\phi$ and QD energy level $\varepsilon_d$ for different values of the coupling strength $|\lambda_2|$ for fixed $|\lambda_1|$.
First, we find that the linear conductance depends substantially on the magnetic flux phase and the tuning of QD. It has been established that the linear conductance becomes strongly $\varepsilon_d$-dependent when the overlap energy between MBSs exceeds the system temperature, i.e., $\varepsilon_M \gtrsim T$~\cite{Lopez2014}. According to the results in Fig.~\ref{fig:7}, one sees that when the dot level is at the Fermi level, $\varepsilon_d = 0$, the conductance presents a $2\pi$ periodicity. 
But when the QD energy level moves away from the Fermi level $\varepsilon_d \neq 0$, the linear conductance shows a $4\pi$ periodicity as a function of $\phi$, regardless of the finite values of $|\lambda_2|$. 
The conductance map details further changes with the increase of the coupling strength $|\lambda_2|$.

\begin{figure*}[ht]
	\includegraphics[width =1\linewidth]{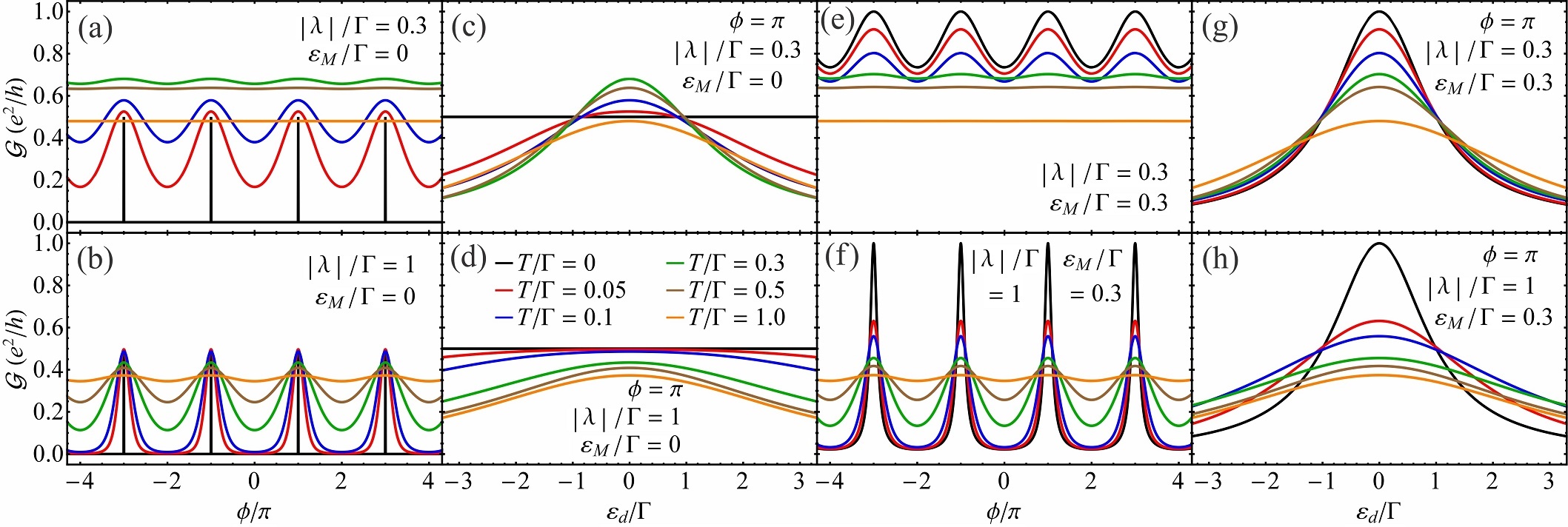}
	\centering
	\caption{(a), (b), (e), (f) The linear conductance $\mathcal{G}$ as a function of magnetic flux phase $\phi$ at $\varepsilon_d/\Gamma = 0$ and (c), (d), (g), (h) as a function of QD energy level $\varepsilon_d$ at $\phi = \pi$ for different temperatures $T$ and  couplings $|\lambda_j| = |\lambda|$. 
	The overlap energy $\varepsilon_M$ is (a)-(d) $\varepsilon_M/\Gamma = 0$ and (e)-(h) $\varepsilon_M/\Gamma = 0.3$. 
	The symmetrical coupling constant $|\lambda|$: (a), (c), (e), (g) $|\lambda|/\Gamma = 0.3$ and (b), (d), (f), (h) $|\lambda|/\Gamma = 1$.}
	\label{fig:10}
\end{figure*}

\begin{figure*}[ht]
	\includegraphics[width =1\linewidth]{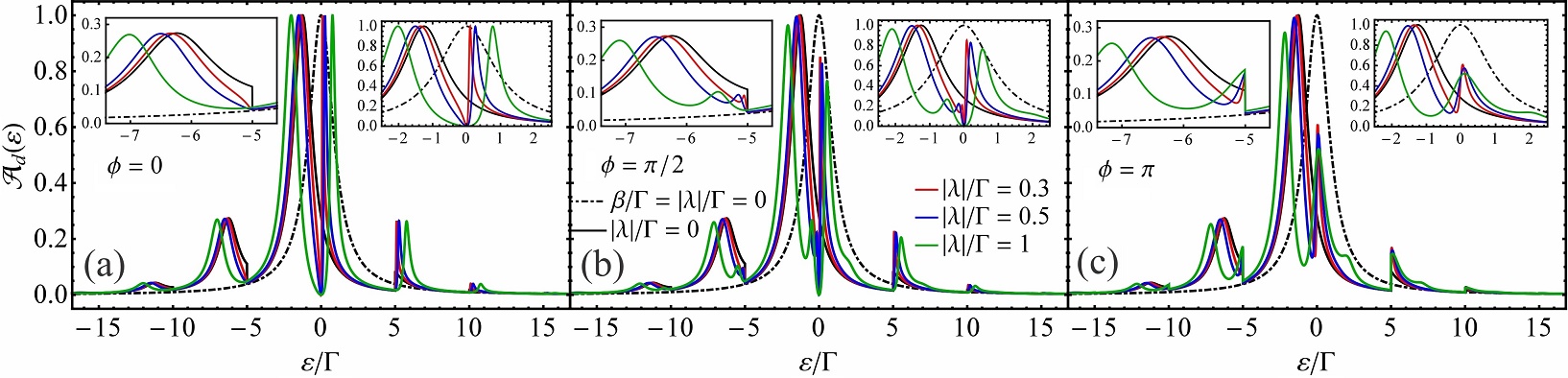}
	\centering
	\caption{The zero-temperature equilibrium spectral function of the QD $\mathcal{A}_d(\varepsilon)$ in the presence of EPI with fixed electron-phonon coupling strength $\beta/\Gamma = 2.5$ at different values of the symmetrical QD-MBS couplings $|\lambda_j| = |\lambda|$ with unhybridized MBSs, $\varepsilon_M/\Gamma = 0$, at the QD energy level $\varepsilon_d/\Gamma = 0$ when the magnetic flux phase is  (a) $\phi = 0$, (b) $\phi = \pi/2$, and (c) $\phi = \pi$.
	The right and left insets in each panel zoom in on the zero-energy features in the spectral function and on the first phonon-induced satellite peak at negative energy, respectively.}
	\label{fig:11}
\end{figure*}

The numerical results for the zero-temperature differential conductance $dI/dV$ as a function of magnetic flux phase $\phi$ and dot energy $\varepsilon_d$ are shown in Fig.~\ref{fig:9} with different finite bias voltages $eV$ and fixed symmetrical QD-MBS couplings $|\lambda|$ for either unhybridized or hybridized MBSs.
We observe that $dI/dV$ has a $2\pi$ periodicity as a function of $\phi$ when the MBSs do not overlap $\varepsilon_M=0$, independent of the dot energy $\varepsilon_d$ [Fig.~\ref{fig:9}(a)-\ref{fig:9}(d)].
In contrast, for $\varepsilon_M\neq 0$, the $2\pi$ periodicity transforms into a $4\pi$ one when $\varepsilon_d$ is not at the Fermi level $\varepsilon_d \neq 0$ [Fig.~\ref{fig:9}(e)-\ref{fig:9}(h)].
Note that the differential conductance maps in the $\varepsilon_M \neq 0$ case are very sensitive to small changes in biasing and dot energy level.
Also note here that the magnitude of ET and LAR differential conductances oscillates with a $2\pi$ period as a function of magnetic flux phase $\phi$ for unhybridized Majoranas regardless of the value of the dot level. Observe that the ET component is dominant compared to the LAR one.

We now investigate the effect of finite temperature on the linear and differential conductance's spectra. 
We plot in Fig.~\ref{fig:10} the linear conductance $\mathcal{G}$ as a function of magnetic flux phase $\phi$ and dot energy level $\varepsilon_d$ at different temperatures $T$ for unhybridized and hybridized MBSs, with two values of the symmetrical QD-MBS couplings $|\lambda|$.
We see that for unhybridized MBSs ($\varepsilon_M = 0$), the conductance is $e^2/2h$ at magnetic flux phase $\phi = (2n+1)\pi$ and zero temperature, independent of the finite strength of $|\lambda|$, in agreement with the results of Fig.~\ref{fig:5}. 
The temperature broadens the conductance peaks, the conductance minima increase significantly with temperature for weak QD-MBS coupling strength $|\lambda|$, and the maxima exceed the value $e^2/2h$ [see Figs.~\ref{fig:10}(a) and~\ref{fig:10}(c)]. 
Moreover, $\mathcal{G}$ grows with $T$ and is suppressed again at higher temperatures in agreement with the literature~\cite{Stefanski2015,Chi2020a}.
For larger values of $|\lambda|$ the conductance peaks are broadened at high $T$ but are more narrow in contrast to the case of weak coupling $|\lambda|$. 
Also, the conductance minima are less sensitive to the change in temperature for strong QD-MBS coupling strengths. 
The enhancement in magnitude of conductance $\mathcal{G}$ peaks with the increase of temperature $T$ is no longer observed [Figs.~\ref{fig:10}(b) and~\ref{fig:10}(d)].
Note that in the case of strong QD-MBS coupling, the conductance does not exceed $e^2/2h$ for any temperature. 
In summary, the conductance has a non-monotonic behavior with temperature at small coupling $|\lambda|$ by surpassing $e^2/2h$, while at large coupling $|\lambda|$ the conductance is limited from above by $e^2/2h$ and decays with temperature.
The finite interaction between MBSs dramatically changes the linear conductance spectrum as shown in Figs.~\ref{fig:10}(e)-\ref{fig:10}(h). 
For $\varepsilon_M \neq 0$, the maxima of $\mathcal{G}$ reach $e^2/h$ at zero temperature regardless of the finite values of $|\lambda|$. 
The linear conductance does not exceed the value $e^2/h$ at any temperature or finite symmetrical QD-MBS couplings $|\lambda|$. 
Moreover, $\mathcal{G}$ decreases rapidly with increasing $T$ when the QD couples strongly to the MBSs. 
In contrast to the $\varepsilon_M=0$ case, the maxima of linear conductance do not increase with temperature for finite values of $\varepsilon_M$.
Note that the differential conductance responds in the same way to the changes in temperature as the linear conductance discussed in Fig.~\ref{fig:10} (not shown here). Namely, for unhybridized Majoranas, in the case of weak QD-MBS coupling strength, the local minima located at $\phi = (2n + 1)\pi$ [see Fig.~\ref{fig:5}(d)] increase with temperature and become plateaus. 
A further increase in temperature leads to the transformation of the differential conductance plateaus into maxima whose amplitude increases with $T$ and after a critical temperature it decreases with temperature.
When the QD-MBS coupling strength grows [see Figs.~\ref{fig:5}(e) and~\ref{fig:5}(f)], the local minima of $dI/dV$ transform into narrow peaks whose amplitude decreases with increasing temperature.
Therefore, in the case of overlapping MBSs, with weak and strong QD-MBS coupling $|\lambda|$, the enhancement of $dI/dV$ peaks with $T$ is not observed as in the case of $\mathcal{G}$.

\subsection{Numerical results in presence of EPI}
\label{sec:IIIB}

In this subsection, we evaluate the influence of EPI on the QD-MBS system transport characteristics.
This problem is relevant since the quantum system is in general coupled to an environment that can induce decoherence effects on quantum computation with MBSs~\cite{Rainis2012,Aseev2018,Aseev2018,Pedrocchi2015,Song2018,Lai2018,Knapp2018}.
In the present case, we analyze the phonon-assisted Majorana-induced transport properties and discuss the resulting equilibrium spectral functions and the linear and differential conductance at zero and finite temperatures. 

\begin{figure*}[ht]
	\includegraphics[width =0.75\linewidth]{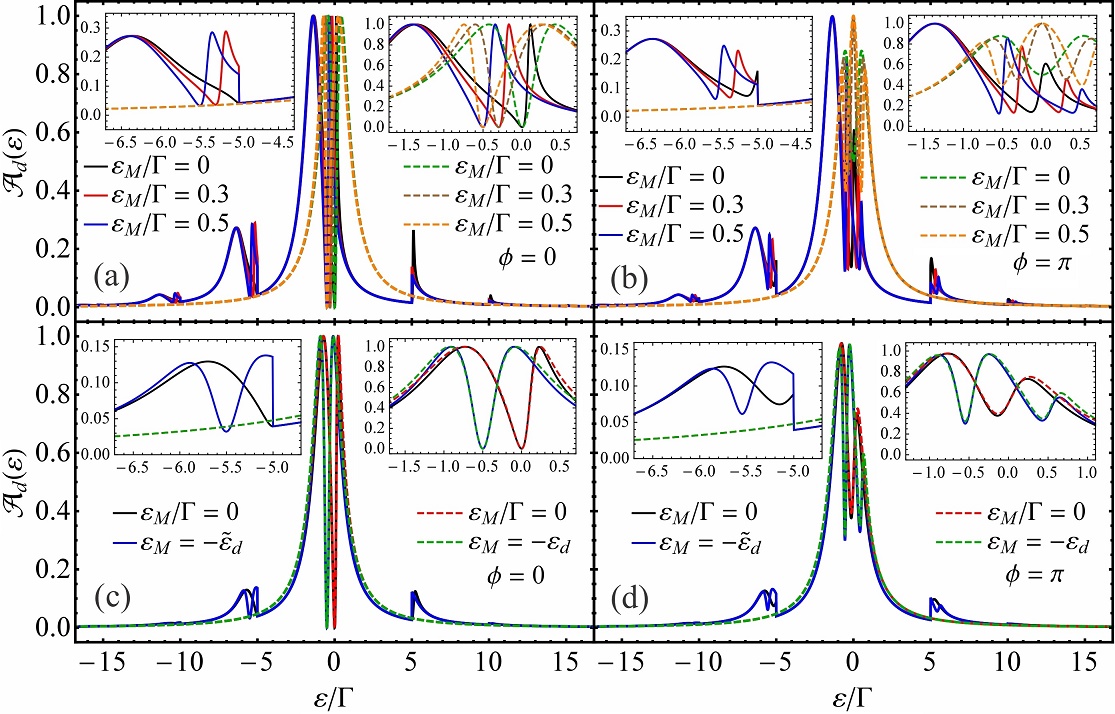}
	\centering
	\caption{(a), (b) The QD zero-temperature equilibrium spectral function $\mathcal{A}_d(\varepsilon)$ at different values of the overlap energy $\varepsilon_M$ with symmetrical QD-MBS couplings $|\lambda|/\Gamma = 0.3$ and electron-phonon coupling strength $\beta/\Gamma = 2.5$ at the QD energy level $\varepsilon_d/\Gamma = 0$ when the magnetic flux phase is (a) $\phi = 0$ and (b) $\phi = \pi$. The solid (dashed) lines denote results in the presence (absence) of EPI.
	(c), (d) The zero-temperature equilibrium spectral function of the QD $\mathcal{A}_d(\varepsilon)$ for nonoverlapping ($\varepsilon_M/\Gamma = 0$) and overlapping MBSs ($\varepsilon_M/\Gamma = 0.5$) with symmetrical QD-MBS couplings $|\lambda|/\Gamma = 0.3$. In presence of EPI (solid lines), at $\varepsilon_d/\Gamma = 0$, the electron-phonon coupling strength $\beta = \sqrt{\varepsilon_M \omega_0}$ with $\varepsilon_M/\Gamma = 0.5$ results in $\beta/\Gamma \approx 1.58$. The dashed lines correspond to EPI absence with $\varepsilon_d/\Gamma = -0.5$. The magnetic flux phase is set to (c) $\phi = 0$ and (d) $\phi = \pi$.
	The right and left insets in each panel zoom in on the zero-energy features in the spectral function and on the first phonon-induced satellite peak at negative energy, respectively.}
	\label{fig:12}
\end{figure*}

\begin{figure*}[ht]
	\includegraphics[width =0.75\linewidth]{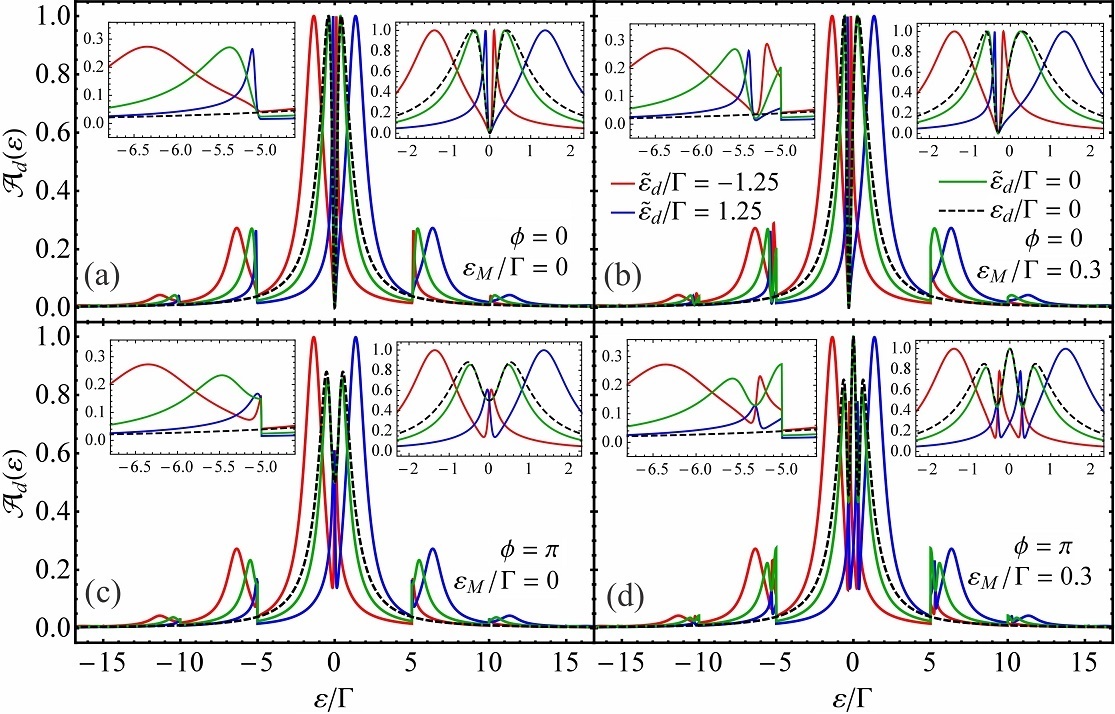}
	\centering
	\caption{The zero-temperature equilibrium spectral function of the QD $\mathcal{A}_d(\varepsilon)$ in the presence of EPI for different $\tilde \varepsilon_d$ with unhybridized and hybridized MBSs. The symmetrical QD-MBS couplings are $|\lambda|/\Gamma = 0.3$ with electron-phonon coupling strength $\beta/\Gamma = 2.5$ for different magnetic flux phases $\phi$. The dashed line corresponds to the case of $\beta/\Gamma = 0$. The rest of parameters are (a) $\phi = 0$, $\varepsilon_M/\Gamma = 0$; (b) $\phi = 0$, $\varepsilon_M/\Gamma = 0.3$; (c) $\phi = \pi$, $\varepsilon_M/\Gamma = 0$; and (d) $\phi = \pi$, $\varepsilon_M/\Gamma = 0.3$. The right and left insets in each panel zoom in on the zero-energy features in the spectral function and on the first phonon-induced satellite peak at negative energy, respectively.}
	\label{fig:13}
\end{figure*}

Note that in contrast to the case when there is no EPI (see Sec.~\ref{sec:IIIA}), the spectral function $\mathcal{A}_d(\varepsilon)$ given by Eq.~\eqref{eq:19} now depends on temperature and bias voltage through the Fermi-Dirac functions appearing in the Keldysh equations~\eqref{eq:29}.
In the following, we investigate the equilibrium ($eV = 0$) spectral function of QD $\mathcal{A}_d(\varepsilon)$. 
We first illustrate the equilibrium spectral function dependence on energy for different strengths of the symmetrical QD-MBS couplings $|\lambda_j| = |\lambda|$, at three values of the magnetic flux phase $\phi$, for $\varepsilon_M = 0$, with fixed electron-phonon coupling strength $\beta/\Gamma = 2.5$, at dot level $\varepsilon_d = 0$, and for zero temperature in Fig.~\ref{fig:11}. 
The spectral function in the absence of EPI and MBSs (denoted with the dot-dashed black line) exhibits a resonant peak with a Lorentzian lineshape located at $\varepsilon =\varepsilon_d$~\cite{Chen2005}. 
When EPI is present in the system, the QD-related parameters are renormalized and polaronic effects are induced. 
In the absence of MBSs ($|\lambda| = 0$, shown by a solid black line), the resonant peak is redshifted at the renormalized QD energy level $\varepsilon = \tilde \varepsilon_d = \varepsilon_d – g \omega_0$ with $g = (\beta/\omega_0)^2$. 
Beside the main peak, a series of phonon-assisted additional channels develop in the spectrum of $\mathcal{A}_d(\varepsilon)$ \cite{Chen2005,Cao2017,Zhang2012,Chi2020}. 
The phonon-assisted channels manifest as satellite peaks emerging in the spectrum at energies $\varepsilon = \tilde \varepsilon_d \pm p \omega_0$ with $p$ an integer. 
The side-bands in the spectrum of $\mathcal{A}_d(\varepsilon)$ correspond to tunneling of electrons through the QD by absorption and emission of phonons with energy $\omega_0$~\cite{Zhu2003}. 
At zero temperature, the coefficients $\mathcal{L}_p$ disappear for $p<0$, the side peaks located in the negative energy region come from $\tilde G_d^<(\varepsilon + p\omega_0)$ in Eq.~\eqref{eq:27}, while those from the positive energy region result from $\tilde G_d^>(\varepsilon - p\omega_0)$~\cite{Chen2005}.
The discontinuity in the side-peak shape is attributed to the Heaviside function $\theta (x)$ which approximates the Fermi-Dirac function at zero temperature. 
Moreover, the height of the side peaks decreases at large $p$ and disappears entirely away from the main peak. 
The phonon side-bands located around the main resonant peak are not symmetric with respect to the renormalized QD energy level $\tilde \varepsilon_d$~\cite{Chen2005}. 
According to the results in Fig.~\ref{fig:2}, when the system is not threaded by a magnetic flux ($\phi = 0$) and the QD couples to the MBSs, two maxima in the spectrum develop at $\varepsilon \approx \pm \sqrt{2} \lambda$ and the zero-energy spectral function reduces to zero regardless of the $|\lambda|$ finite value, in the absence of EPI. 
In the presence of EPI, the spectrum of $\mathcal{A}_d(\varepsilon)$ is modified [see Fig.~\ref{fig:11}(a)]. 
The main peak and the satellite ones located in the negative energy region are redshifted, while those that occur in the positive energy region are blueshifted with an increase of $|\lambda|$.
The zero-energy spectral function remains always unchanged, $\mathcal{A}_d(0) = 0$.
When the magnetic flux phase is switched on [see Figs.~\ref{fig:11}(b) and~\ref{fig:11}(c)], the spectrum of $\mathcal{A}_d(\varepsilon)$ becomes more complicated in the presence of EPI. The value of zero-energy spectral function remains unaffected by EPI, i.e., it is $0$ for the magnetic flux phase $\phi = \pi/2$ and $1/2$ for $\phi = \pi$, respectively. 
For $\phi = \pi/2$, the four-peak structure, which shows up for strong QD-MBS couplings $|\lambda|$ and $\beta = 0$ (see Fig.~\ref{fig:2}), is highly deformed under EPI. 
For $\phi = \pi$, the three-peak structure located around $\varepsilon \approx 0$, due to the presence of MBSs in the system, emerges only for very strong QD-MBS couplings $|\lambda|$.

\begin{figure*}[ht]
	\includegraphics[width =0.75\linewidth]{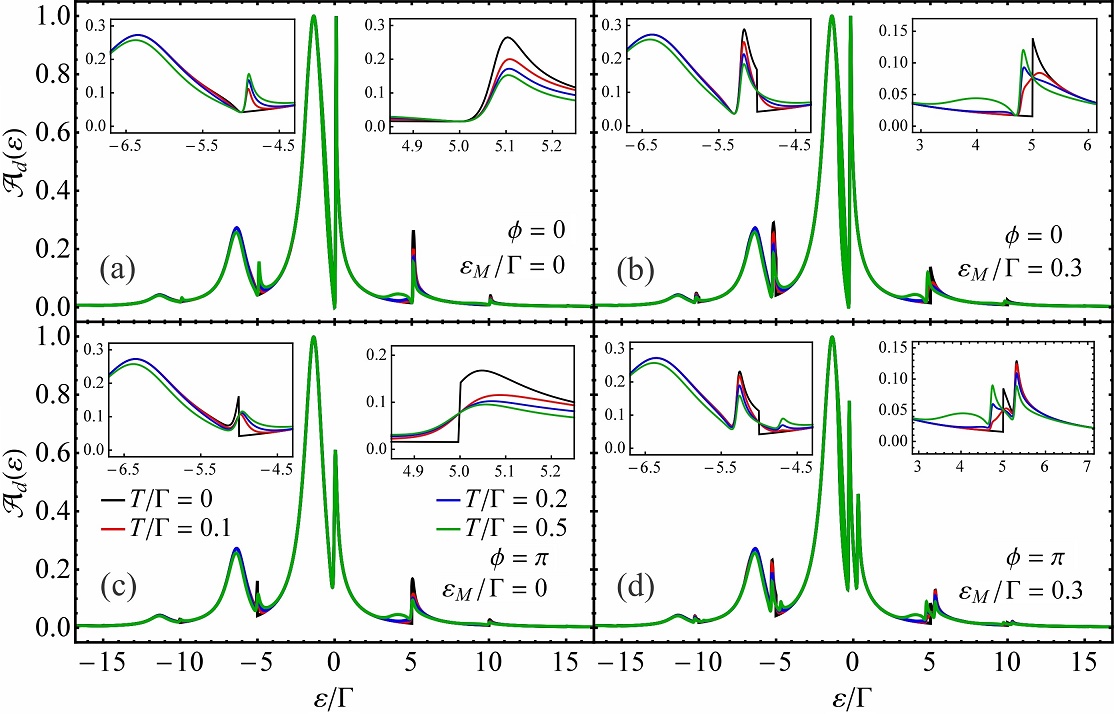}
	\centering
	\caption{The equilibrium spectral function of the QD $\mathcal{A}_d(\varepsilon)$ in the presence of EPI at different temperatures $T$ for $\varepsilon_M/\Gamma = 0$ and $0.3$ with symmetrical QD-MBS couplings $|\lambda|/\Gamma = 0.3$, electron-phonon coupling strength $\beta/\Gamma = 2.5$, QD level $\varepsilon_d/\Gamma=0$, for different magnetic flux phases $\phi$. The rest of parameters are (a) $\phi = 0$, $\varepsilon_M/\Gamma = 0$; (b) $\phi = 0$, $\varepsilon_M/\Gamma = 0.3$; (c) $\phi = \pi$, $\varepsilon_M/\Gamma = 0$; and (d) $\phi = \pi$, $\varepsilon_M/\Gamma = 0.3$.
	The right and left insets in each panel zoom in on the first phonon-induced satellite peak at positive and negative energy, respectively.}
	\label{fig:14}
\end{figure*}

Figures~\ref{fig:12}(a) and~\ref{fig:12}(b) show the numerical results for the equilibrium spectral function $\mathcal{A}_d(\varepsilon)$ in the presence of EPI with the influence of overlap energy $\varepsilon_M$ at two different magnetic flux phases $\phi$, with weak symmetrical QD-MBS couplings $|\lambda_j| = |\lambda|$, for dot level $\varepsilon_d = 0$, and at zero temperature. 
As shown in Fig.~\ref{fig:3}, in the absence of EPI and magnetic flux, the spectral function has an antiresonance point at the energy $\varepsilon \approx -\varepsilon_M$. In the presence of EPI [$\phi = 0$, Fig.~\ref{fig:12}(a)], we observe that the location of the antiresonance point is immune to the EPI. The two main peaks located around the antiresonance point, corresponding to the MBSs, are redshifted. The broadened peak position is mainly determined by the electron-phonon coupling strength $\beta$ via the renormalized dot level $\tilde \varepsilon_d$ and is weakly affected by the overlap energy $\varepsilon_M$. The location of the narrow peak depends strongly on $\varepsilon_M$. 
The magnitude of the main peaks reaches the limit $1$. 
The satellite peaks are also modified by the finite overlap energy $\varepsilon_M$.
In the positive energy region, the height of the side peaks decreases with increasing $\varepsilon_M$. 
In the negative energy region, the satellite peaks split into two peaks, a broadened and a narrow one.
In the presence of magnetic flux phase $\phi = \pi$ [Fig.~\ref{fig:12}(b)], the two main peaks located around the zero-energy in the spectral function evolve into a three-peak structure for TSNWs with finite overlap energy, $\varepsilon_M \neq 0$, providing the fingerprints of MBSs in the system. 
Most remarkably, in the case of $\varepsilon_M = 0$, the zero-energy spectral function is immune to the EPI, $\mathcal{A}_d(\varepsilon = 0) = 1/2$. 
When $\varepsilon_M \neq 0$, the zero-energy spectral function differs from the unitary limit, i.e., $\mathcal{A}_d(\varepsilon = 0) \neq 1$. Besides, the main sharp peak due to the MBSs is shifted to the lower energies and its height is reduced relative to the $\beta = 0$ case. 
The broadened peak again is weakly $\varepsilon_M$-dependent and takes the unitary limit.
The position of antiresonance points, which originally occurred in the absence of EPI at $\varepsilon \approx  \pm \varepsilon_M$, changes at finite electron-phonon coupling. 
The three-peak structures at the side peaks are truncated at $\varepsilon \approx \pm |p|\omega_0$ and local minima emerge at $\varepsilon \approx \pm (|p|\omega_0 + \varepsilon_M)$.
Note that for $\tilde \varepsilon_d = - \varepsilon_M$ and $\varepsilon_d = 0$, the following relation holds for electron-phonon coupling strength $\beta = \sqrt{\omega_0\varepsilon_M}$. 
For stronger MBS-MBS coupling, i.e., $\varepsilon_M/\Gamma = 0.5$, the electron-phonon coupling strength becomes $\beta/\Gamma \approx 1.58$, which is sufficiently strong ($\beta > \Gamma$). 
The numerical results are shown in Figs.~\ref{fig:12}(c) and~\ref{fig:12}(d) for the special case when $\varepsilon_M = -\tilde \varepsilon_d$ with $\varepsilon_d = 0$ in the presence of EPI and with $\varepsilon_d \neq 0$ in the absence of it. 
In the absence of magnetic flux [$\phi = 0$, Fig.~\ref{fig:12}(c)], when $\varepsilon_M = -\tilde \varepsilon_d$, the two main peaks in the vicinity of $\varepsilon \approx 0$, due to the MBSs, become approximately identical, as was found in the case without EPI (see also Fig.~\ref{fig:3}). In the presence of a magnetic flux [$\phi = \pi$, Fig.~\ref{fig:12}(d)], when $\varepsilon_M = -\tilde \varepsilon_d$, the main sharp peak of the three-peak structure shows an adequate match relative to the $\beta = 0$ case, as for $\phi = 0$. Therefore, the symmetry of the two peaks located around the main peak is also broken even for $\tilde \varepsilon_d = - \varepsilon_M$ in the presence of EPI, as was discussed in Fig.~\ref{fig:3} in the absence of EPI.
Moreover, the truncation of side peaks is also present in this special case.
Consequently, the signatures of MBSs depend on the variation of electron-phonon coupling strength $\beta$, MBS-MBS coupling $\varepsilon_M$, or the dot energy $\varepsilon_d$. 

Figure~\ref{fig:13} shows the zero-temperature equilibrium spectral function $\mathcal{A}_d(\varepsilon)$ in the presence of EPI when tuning the renormalized dot level $\tilde\varepsilon_d=\varepsilon_d-\beta^2/\omega_0$ for $\varepsilon_M = 0$ and $\varepsilon_M \neq 0$, with symmetrical QD-MBS couplings $|\lambda_j| = |\lambda|$, at two different magnetic flux phases $\phi$.
The features of the main peaks ($p=0$) in the $\mathcal A_d(\varepsilon)$ reproduce the results seen in the absence of EPI (Fig.~\ref{fig:3}) under the mapping $\varepsilon_d\to \tilde\varepsilon_d$.
Independent of the other parameters, absence or presence of EPI, the antiresonance point remains for $\phi = 0$ at $\varepsilon \approx -\varepsilon_M$ [see Figs.~\ref{fig:13}(a) and~\ref{fig:13}(b)].
When $\phi = \pi$ and $\varepsilon_M = 0$ [Fig.~\ref{fig:13}(c)], the zero-energy spectral function is $\mathcal{A}_d(\varepsilon = 0) = 1/2$, and the Fano line shape spectral asymmetry in the vicinity of $\varepsilon \approx 0$ vanishes again when the dot level $\varepsilon_d$ matches the energy shift $g\omega_0$, i.e., $\tilde\varepsilon_d = 0$. 
Therefore, for finite $\varepsilon_M$, the characteristic three-peak structure becomes manifest [see Fig.~\ref{fig:13}(d)]. Additionally, the shapes of the side peaks ($p\neq 0$) are identical but mirrored when the renormalized QD level matches the Fermi energy $\tilde \varepsilon_d \approx 0$ when $\varepsilon_M = 0$, at any magnetic flux phase $\phi$, or when $\varepsilon_M \neq 0$ at $\phi = (2n+1)\pi$.
As before, in the absence of EPI, the broadened peaks are caused by the renormalized QD energy level $\tilde \varepsilon_d$ while the sharp peaks are attributed to the regular fermionic states originating from the MBSs.
Moreover, if both $\tilde \varepsilon_d$ and $\varepsilon_M$ vanish, the main peaks become approximately symmetric at any $\phi$ [see Figs.~\ref{fig:13}(a) and~\ref{fig:13}(c)], as discussed above in the presence of EPI in Fig.~\ref{fig:12}, or without EPI in Fig.~\ref{fig:3}.

\begin{figure}[ht]
	\includegraphics[width =1\linewidth]{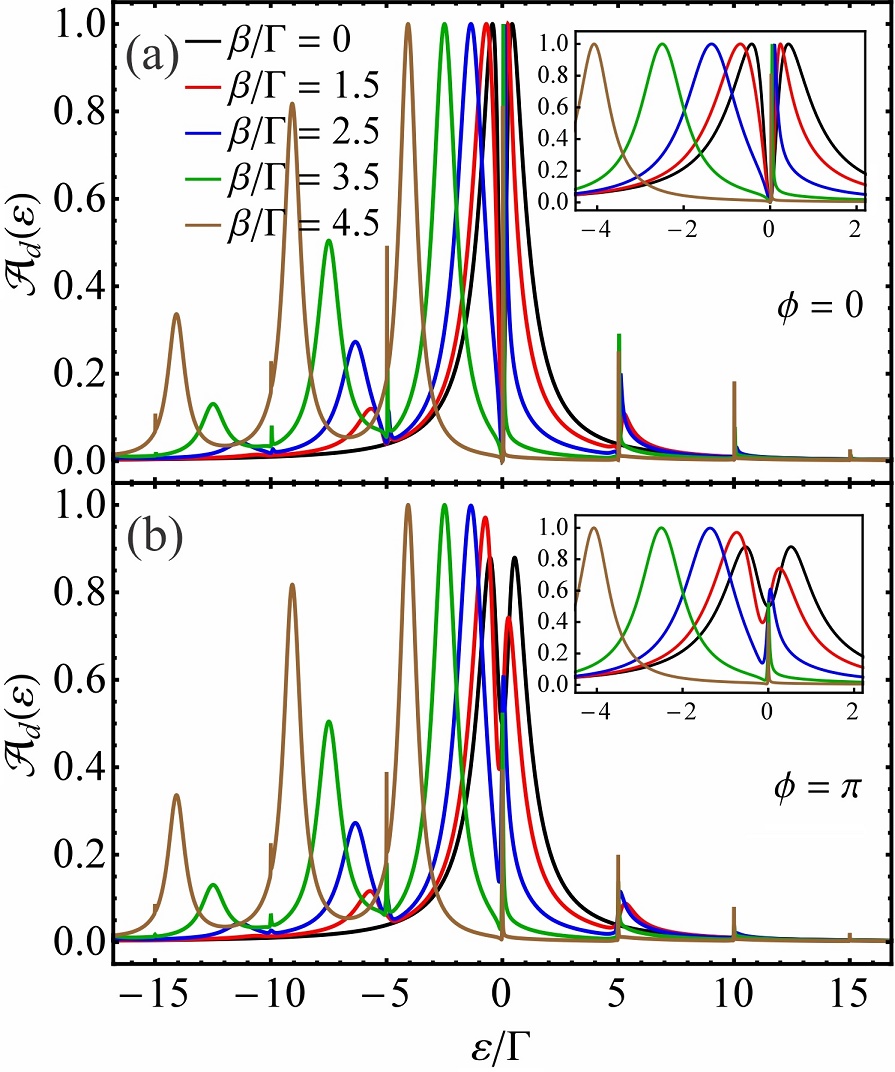}
	\centering
	\caption{The equilibrium spectral function of the QD $\mathcal{A}_d(\varepsilon)$ at different values of the electron-phonon coupling strength $\beta$ for unhybridized MBSs $\varepsilon_M/\Gamma = 0$, with symmetrical QD-MBS couplings $|\lambda|/\Gamma = 0.3$, at the QD energy level $\varepsilon_d/\Gamma = 0$, and temperature $T/\Gamma = 0.1$, when the magnetic flux phase is (a) $\phi = 0$ and (b) $\phi = \pi$.
	The inset in each panel zooms in on the zero-energy features in the spectral function.}
	\label{fig:15}
\end{figure}

As we have seen in Sec.~\ref{sec:IIIA}, the spectral function is temperature-independent when there is no EPI in the system.  
We now investigate the influence of temperature on the spectral function in the presence of EPI. The equilibrium spectral function $\mathcal{A}_d(\varepsilon)$ at different temperatures $T$ for unhybridized and hybridized MBSs, with weak symmetrical QD-MBS couplings $|\lambda_j| = |\lambda|$, at the dot level $\varepsilon_d = 0$, for two different flux phases $\phi$, is plotted in Fig.~\ref{fig:14}. 
We see that the main peaks are less sensitive to the change in temperature than the satellite peaks.
The spectral function in Eq.~\eqref{eq:C1} is proportional to the coefficients $\mathcal{L}_p$, such that the peak heights are directly influenced by $\mathcal{L}_p$.
For low temperatures, the $\mathcal L_p$ coefficients are weakly dependent on $T$, which leads only to extremely small changes in the peak height in the temperature regime considered here. 
However at large temperatures, all $\mathcal L_p$ start to decay as $1/\sqrt{T}$, which eventually leads to suppressing all the peaks.
The discontinuity in the shape of the side peaks is smoothed at finite temperatures because of the Fermi-Dirac distribution function. 
The peaks width is determined by the renormalized QD-lead coupling $\tilde \Gamma = \Gamma e^{-g(2N_{ph}+1)}$ with $\tilde \Gamma$ decreasing as $T$ increases. This leads to a narrowing of peaks with the increase of temperature~\cite{Dai2019}.
Moreover, the location of peaks is also weakly influenced by the temperature via the renormalized QD-MBS couplings $|\tilde \lambda| = |\lambda|e^{-g(N_{ph}+1/2)}$. 

\begin{figure}[ht]
	\includegraphics[width =1\linewidth]{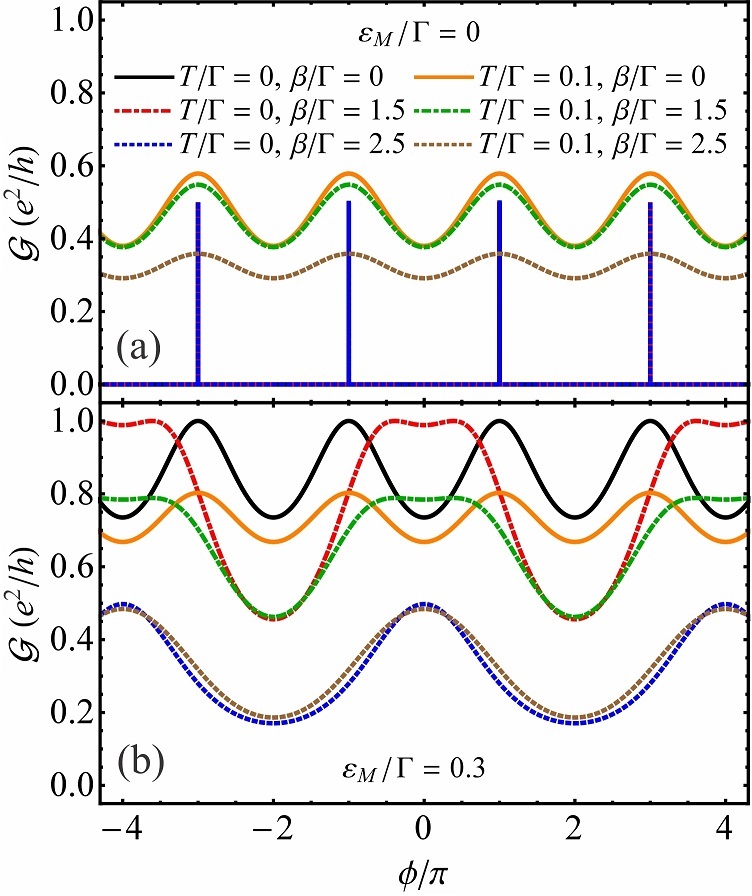}
	\centering
	\caption{The linear conductance $\mathcal{G}$ as a function of magnetic flux phase $\phi$ for different electron-phonon coupling strengths $\beta$, for (a) unhybridized $\varepsilon_M/\Gamma = 0$ and (b) hybridized MBSs $\varepsilon_M/\Gamma = 0.3$. We use symmetrical QD-MBS couplings $|\lambda|/\Gamma = 0.3$, QD energy level $\varepsilon_d/\Gamma = 0$, and zero or finite $T/\Gamma=0.1$ temperature.}
	\label{fig:16}
\end{figure}

In Fig.~\ref{fig:15}, we plot the equilibrium spectral function $\mathcal{A}_d(\varepsilon)$ as a function of energy for different values of the electron-phonon coupling strength $\beta$, for symmetrical QD-MBS couplings $|\lambda_j| = |\lambda|$, at two values of the magnetic flux phase $\phi$, for $\varepsilon_M = 0$, at dot level $\varepsilon_d = 0$, and at finite temperature. 
According to the results in Fig.~\ref{fig:2}, the zero-energy equilibrium spectral function is $\mathcal{A}_d(0) = 0$ and $\mathcal{A}_d(0) = 1/2$ for magnetic flux phase $\phi = 0$ and $\phi = \pi$, respectively. 
In the presence of EPI and at $\phi = 0$, the position of the antiresonance point at $\varepsilon=0$ is not affected by the electron-phonon coupling strength $\beta$. 
The main peak developing in the negative energy region of the spectrum shifts towards lower energies with increasing electron-phonon coupling strength $\beta$ due to the renormalized QD energy level $\tilde \varepsilon_d = \varepsilon_d - \beta^2/\omega_0$.
The main peak at the positive energy region, caused by MBSs, shifts towards zero-energy with the increase of $\beta$.
The position of satellite peaks in the spectral function is also modified by $\beta$. 
Due to the fact that the QD-lead couplings $\Gamma$ are renormalized in the presence of EPI, i.e., $\tilde \Gamma = \Gamma e^{-g(2 N_{ph} + 1)}$ at finite temperature, the height and width of the resonant peaks are directly affected by the electron-phonon coupling strength $\beta$ and temperature $T$. 
The increase of $\beta$ suppresses the QD-lead couplings, which leads to narrow peaks of increased height. 
This results in a redistribution of side-peaks weights~\cite{Dai2019}. 
Moreover, the enhancement of side-peak amplitudes with the increase of $\beta$ is more significant in the negative energy region than in the positive energy one.
Note that for each $p$, the coefficient $\mathcal{L}_p$ increases with $\beta$ up to a maximum, after which it starts to decrease. 
The QD-MBS couplings are also renormalized in presence of EPI and become smaller when the electron-phonon coupling strength gets stronger, which in turn affects the position of resonant peaks.  

\begin{figure}[ht]
	\includegraphics[width =1\linewidth]{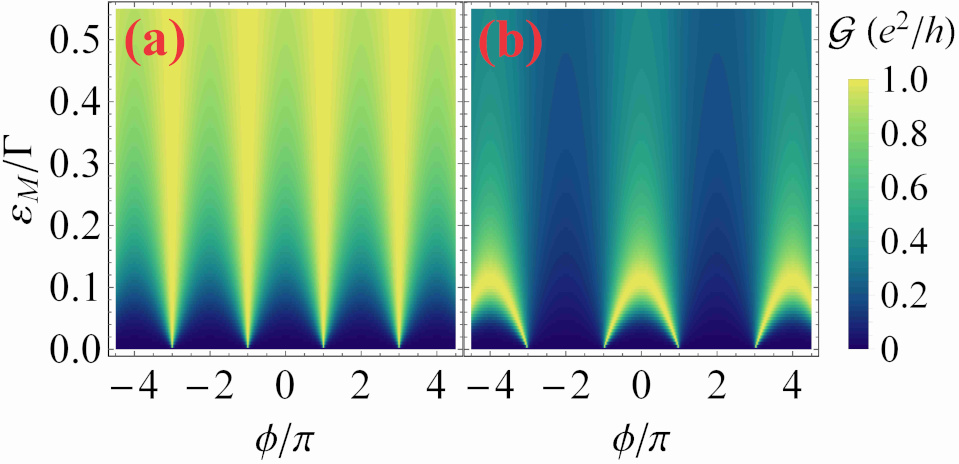}
	\centering
	\caption{The zero-temperature linear conductance $\mathcal{G}$ as a function of magnetic flux phase $\phi$ and MBSs overlap energy $\varepsilon_M$, for symmetrical QD-MBS couplings $|\lambda|/\Gamma = 0.3$, at the dot level $\varepsilon_d/\Gamma = 0$, in the (a) absence $\beta/\Gamma = 0$ and (b) presence of EPI $\beta/\Gamma = 2.5$.}
	\label{fig:17}
\end{figure}

In the following, we investigate the linear conductance $\mathcal{G}$ in presence of EPI at zero and finite temperature. We first plot in Fig.~\ref{fig:16} the linear conductance $\mathcal{G}$ as a function of magnetic flux phase $\phi$, for dot level $\varepsilon_d = 0$, at zero and finite temperature, with and without EPI, with either $\varepsilon_{M}=0$, or $\varepsilon_{M}\neq 0$.
As discussed in Sec.~\ref{sec:IIIA} (see Figs.~\ref{fig:7} and~\ref{fig:8}), in absence of EPI and for $\varepsilon_M=0$, $\mathcal{G}$ becomes independent of $\varepsilon_d$ and is a $2\pi$-periodic function in magnetic flux phase $\phi$ at zero temperature. 
In the presence of EPI, when $\varepsilon_M = 0$, in the limit $\varepsilon \to 0$, the retarded Green's functions $\tilde G_{d11}^r(\varepsilon\to 0)$ and $\tilde G_{d12}^r(\varepsilon\to 0)$ have the approximate expression from Eqs.~\eqref{eq:33} with $|\lambda_j|$ replaced with the renormalized $|\tilde \lambda_j|$.
Therefore, parallel to the discussion in the case without EPI, they do not depend on the renormalized dot level $\tilde \varepsilon_d$ or finite values of $|\tilde \lambda_j|$.
Therefore, at zero temperature when $\varepsilon_M = 0$, $\mathcal{G}$ becomes independent of the electron-phonon coupling strength $\beta$.
Such behavior was also observed in a similar setup where a single MBS is connected to a normal lead through a QD~\cite{Wang2021}.
At finite temperatures, the conductance peaks are broadened and suppressed with $\beta$. 
The periodicity of $\mathcal{G}$ remains $2\pi$ in presence of EPI. 
As is discussed above, in the absence of EPI, $\mathcal{G}$ has $4\pi$ periodicity as a function of $\phi$ when $\varepsilon_M \neq 0$ and $\varepsilon_d \neq 0$. 
In the presence of EPI, at $\varepsilon_M \neq 0$, the linear conductance depends on $\varepsilon_d$ 
and $\beta$ both at zero and finite temperatures. 
The electron-phonon coupling strength $\beta$ suppresses the magnitude of $\mathcal{G}$ at fixed $\varepsilon_d$. 
Most remarkably, when $\varepsilon_M \neq 0$, the $2\pi$ periodicity of $\mathcal{G}$ in $\phi$ transforms into $4\pi$ in the presence of EPI due to the renormalized dot level $\tilde \varepsilon_d$.

We now plot in Fig.~\ref{fig:17} the linear conductance $\mathcal{G}$ as a function of magnetic flux phase $\phi$ and MBSs overlap energy  $\varepsilon_M$ with symmetrical QD-MBS couplings $|\lambda_j| = |\lambda|$ at the QD energy level $\varepsilon_d = 0$, zero temperature, with and without EPI. 
The effect of electron-phonon coupling strength $\beta$ on the linear conductance $\mathcal{G}$ is clearly visible. One observes from Fig.~\ref{fig:17}(a) that the linear conductance $\mathcal{G}$ shows a $2\pi$ periodicity as a function of magnetic flux phase $\phi$ in absence of EPI regardless of the value of $\varepsilon_M$, in agreement with the results of Fig.~\ref{fig:5}. 
Moreover, for strong overlap energy $\varepsilon_M$, the linear conductance is significantly enhanced while preserving its $2\pi$ periodicity. 
When the EPI is switched on [see Fig.~\ref{fig:17}(b)], the periodicity of $\mathcal{G}$ as a function of $\phi$ changes from $2\pi$ to $4\pi$, provided that $\varepsilon_M$ is finite. 
At $\varepsilon_M=0$, the linear conductance shows $2\pi$ periodicity with minima at $\phi \neq (2n + 1)\pi$, with $n$ an integer. 
Moreover, the linear conductance takes the value $e^2/2h$ at magnetic flux phases $\phi = (2n+1)\pi$, independent of the electron-phonon coupling strength $\beta$.
For $\varepsilon_M>0$, $\mathcal{G}$ has $4\pi$ periodicity as a function of $\phi$. 
At $\phi = 4n\pi$, $\mathcal{G}$ has local minima for small values of $\varepsilon_M$ while for larger values of $\varepsilon_M$, $\mathcal{G}$ presents maxima. Moreover, the further increase of $\varepsilon_M$, leads to a decrease in the peaks of $\mathcal{G}$. 
In conclusion, in the presence of EPI, there is a transition from a $2\pi$-periodic linear conductance at $\varepsilon_{M}=0$ to a $4\pi$ one at $\varepsilon_M>0$.

\begin{figure}[ht]
\includegraphics[width =1\linewidth]{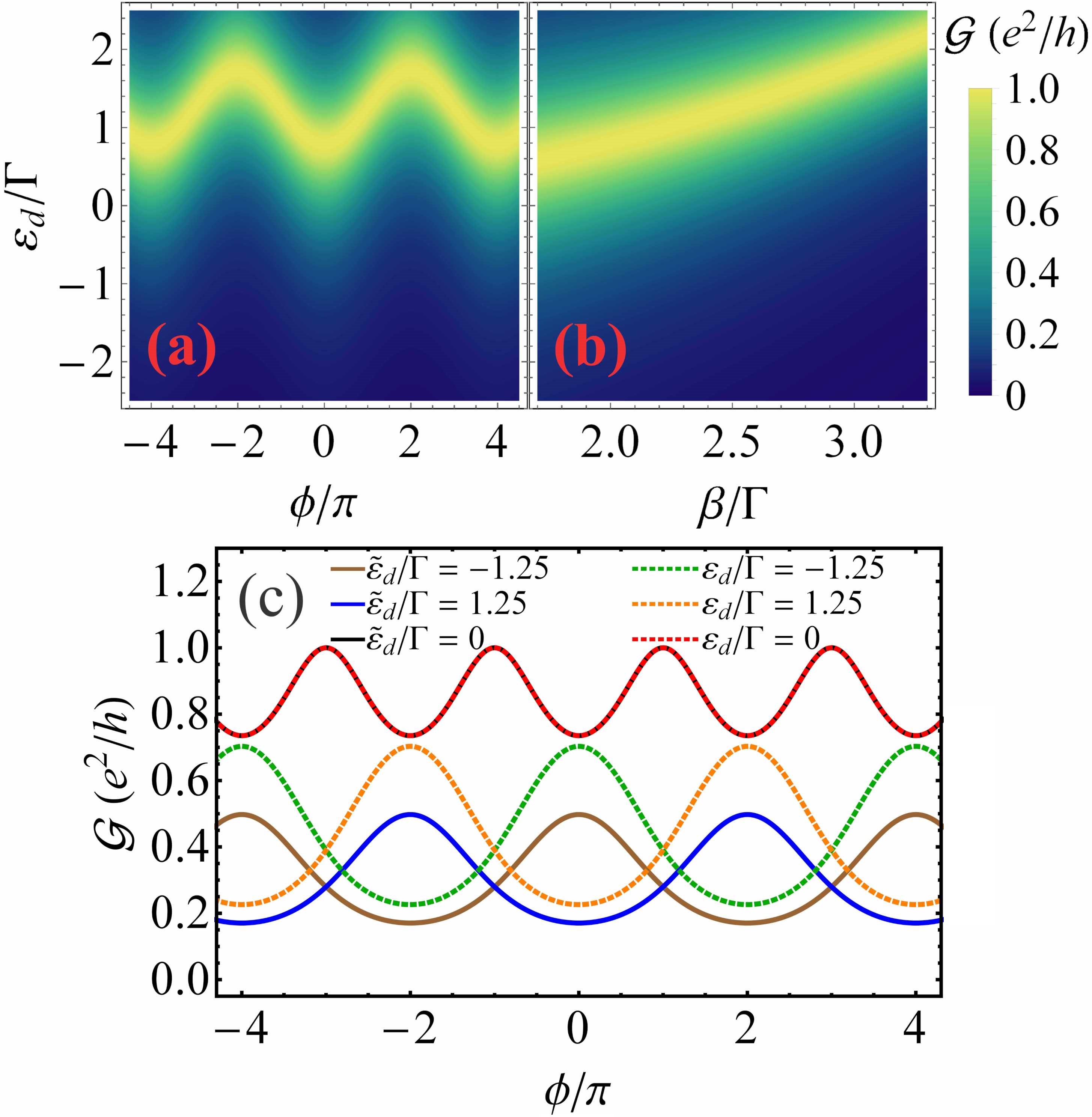}
\centering
\caption{(a) The zero-temperature linear conductance $\mathcal{G}$ as a function of magnetic flux phase $\phi$ and QD energy level $\varepsilon_d$ with electron-phonon coupling strength $\beta/\Gamma = 2.5$. (b) The zero-temperature linear conductance $\mathcal{G}$ as a function of electron-phonon coupling strength $\beta$ and QD energy level $\varepsilon_d$ at magnetic flux phase $\phi = \pi$. (c) The zero-temperature linear conductance $\mathcal{G}$ as a function of magnetic flux phase $\phi$ for different values of $\varepsilon_d$, with and without EPI ($\tilde \varepsilon_d=\varepsilon_d-\beta^2/\omega_0$, $\beta/\Gamma=2.5$), where the solid lines correspond to the case $\beta/\Gamma = 2.5$ and the dashed lines to $\beta/\Gamma = 0$, respectively.
The rest of parameters are overlap energy $\varepsilon_M/\Gamma = 0.3$ and symmetrical QD-MBS couplings $|\lambda|/\Gamma = 0.3$.}
\label{fig:18}
\end{figure}

\begin{figure*}[ht]
	\includegraphics[width =1\linewidth]{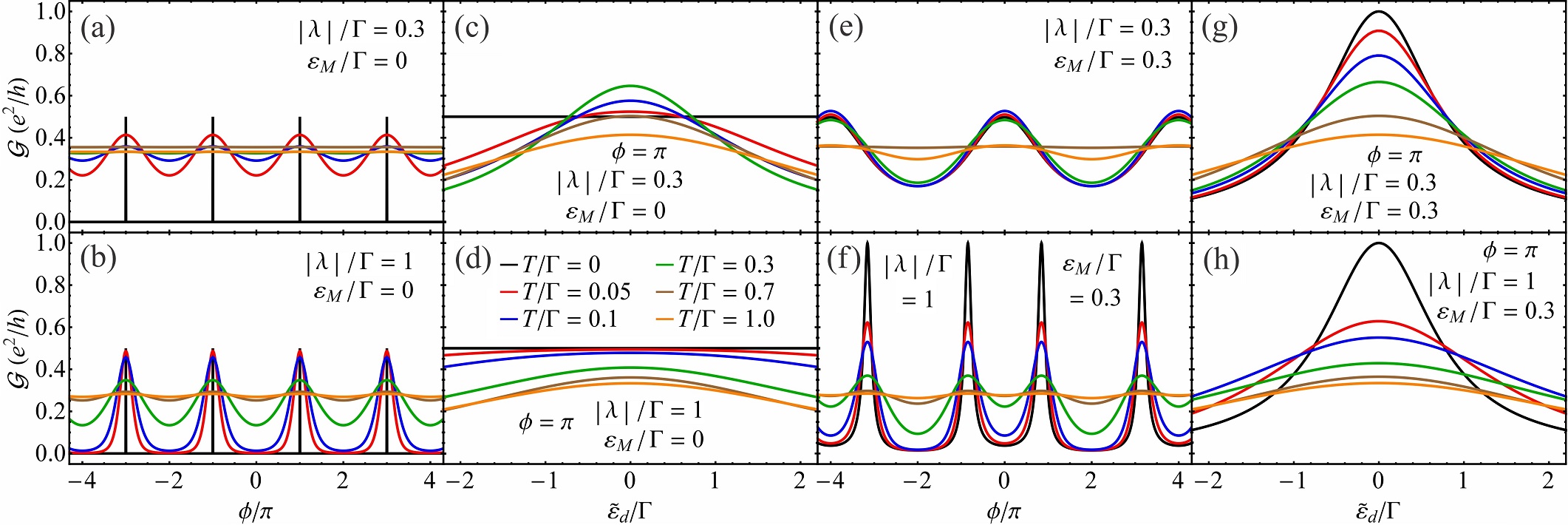}
	\centering
	\caption{(a), (b), (e), (f) The linear conductance $\mathcal{G}$ as a function of magnetic flux phase $\phi$ at $\varepsilon_d/\Gamma = 0$, and (c), (d), (g), (h), as a function of renormalized QD energy level $\tilde \varepsilon_d$, at $\phi = \pi$, for different temperatures $T$ and couplings $|\lambda_j| = |\lambda|$, with fixed electron-phonon coupling strength $\beta/\Gamma = 2.5$. 
	The overlap energy $\varepsilon_M$ is (a)-(d) $\varepsilon_M/\Gamma = 0$ and (e)-(h) $\varepsilon_M/\Gamma = 0.3$ and the symmetrical coupling constant $|\lambda|$, (a), (c), (e), (g) $|\lambda|/\Gamma = 0.3$ and (b), (d), (f), (h) $|\lambda|/\Gamma = 1$.}
	\label{fig:19}
\end{figure*}

Now we investigate the  QD energy level effect on the linear conductance by plotting the zero-temperature linear conductance $\mathcal{G}$ as a function of magnetic flux phase $\phi$ and dot energy $\varepsilon_d$ for $\varepsilon_M \neq 0$ with symmetrical QD-MBS couplings $|\lambda_j| = |\lambda|$ in the presence of EPI [see Fig.~\ref{fig:18}(a)].
The features of $\mathcal G$ are readily understood by recalling that in the absence of EPI (see Fig.~\ref{fig:8}), the linear conductance has $2\pi$ periodicity as a function of $\phi$ when $\varepsilon_d = 0$, and it transforms to $4\pi$ for $\varepsilon_d \neq 0$ in the case of $\varepsilon_M \neq 0$.
In the case when the EPI is switched on [Fig.~\ref{fig:18}(a)], the dot energy becomes renormalized $\varepsilon_d \to \tilde\varepsilon_d = \varepsilon_d - \beta^2/\omega_0$, which therefore leads to a shift in the conductance map in the positive direction on the $\varepsilon_d$ axis. 
Thus, at $\varepsilon_d = \beta^2/\omega_0$, $\mathcal{G}$ has $2\pi$ periodicity as a function of $\phi$. 
Otherwise, for $\varepsilon_d \neq \beta^2/\omega_0$, the linear conductance is a $4\pi$-periodic function of $\phi$.
Figure~\ref{fig:18}(b) shows the numerical results for the zero-temperature linear conductance $\mathcal{G}$ as a function of electron-phonon coupling strength $\beta$ and QD energy level $\varepsilon_d$, at magnetic flux phase $\phi = \pi$, with symmetrical QD-MBS couplings $|\lambda_j| = |\lambda|$, and $\varepsilon_M \neq 0$. 
Note that the $2\pi$-periodic $\mathcal G$ is obtained only if the effective dot energy $\tilde\varepsilon_d$ vanishes, which requires $\varepsilon_d > 0$, since $\beta^2/\omega_0>0$.
We plot in Fig.~\ref{fig:18}(c) the linear conductance $\mathcal{G}$ against the magnetic flux phase $\phi$ for different values of the dot energy $\varepsilon_d$, without EPI and in the presence of EPI with fixed electron-phonon coupling strength $\beta$ and $\varepsilon_M \neq 0$ at zero temperature. 
It is readily observed again that the effect of EPI on $\mathcal{G}$ is eliminated at $\tilde\varepsilon_d=0$ by tuning $\varepsilon_d = \beta^2/\omega_0$. 
Also note that the oscillations for two opposite values of $\tilde\varepsilon_d$ are in antiphase as seen also in the absence of EPI (see also Fig.~\ref{fig:8}).
In conclusion, the main message of Fig.~\ref{fig:18} is that in the presence of EPI, for $\varepsilon_M\neq0$, the linear conductance periodicity switches between $2\pi$ and $4\pi$ when varying $\tilde \varepsilon_d$ through $\varepsilon_d$ or electron-phonon coupling $\beta$.
\begin{figure}[ht]
	\includegraphics[width =0.97\linewidth]{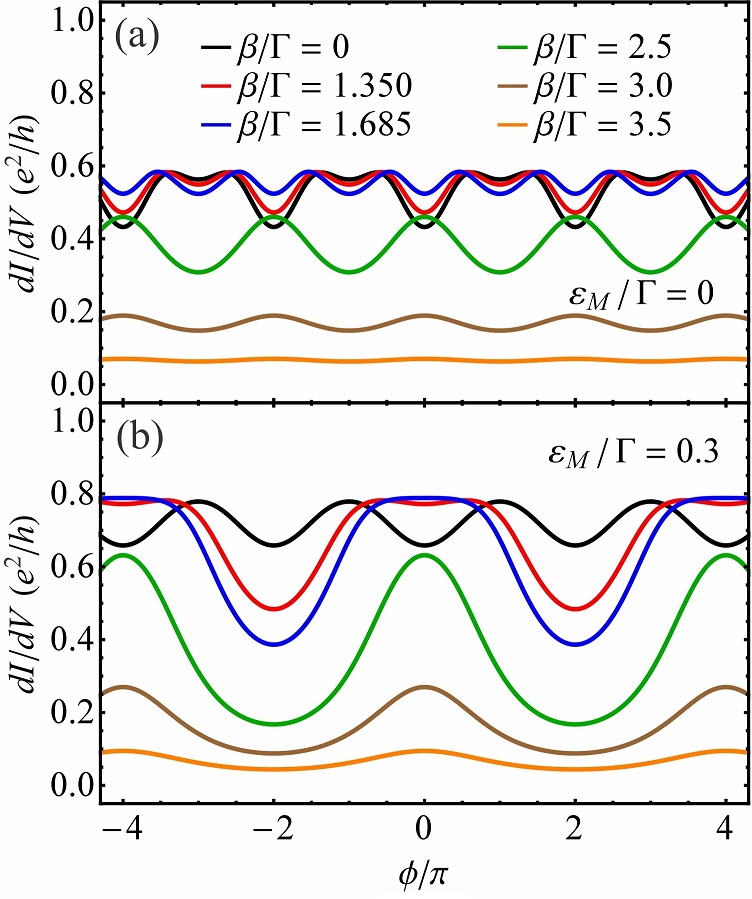}
	\centering
	\caption{The zero-temperature differential conductance $dI/dV$ as a function of magnetic flux phase $\phi$ for different values of the electron-phonon coupling strength $\beta$, with bias voltage $eV/\Gamma = 0.28$, QD energy level $\varepsilon_d/\Gamma = 0$, symmetrical QD-MBS couplings $|\lambda|/\Gamma = 0.3$, and overlap energy (a) $\varepsilon_M/\Gamma = 0$ and (b) $\varepsilon_M/\Gamma = 0.3$.}
	\label{fig:20}
\end{figure}

In the following, we analyze the temperature effect on linear conductance in the presence of EPI.
In general, the temperature has a strong effect on the linear conductance since $\mathcal{G} \propto T^{-1} \int d\varepsilon \, \mathcal{A}_d(\varepsilon) f(\varepsilon)[1-f(\varepsilon)]$, which decays with an overall prefactor $T^{-1}$.
The results from Fig.~\ref{fig:19} expand on the study in Fig.~\ref{fig:16} where a single finite temperature was considered.
Figure~\ref{fig:19} shows the linear conductance $\mathcal{G}$ as a function of magnetic flux phase $\phi$ and renormalized dot energy level $\tilde \varepsilon_d$, in the presence of EPI, at different temperatures, for unhybridized and hybridized MBSs, with symmetrical QD-MBS couplings $|\lambda_j| = |\lambda|$. 
According to the results in Fig.~\ref{fig:10}, we see that the increasing temperature broadens the Dirac-comb structure of the resonant peaks at zero temperature for $\varepsilon_{M}=0$. 
In the presence of EPI, beside the increase of minima with the temperature for weak QD-MBS coupling strength $|\lambda|$, an enhancement of maxima in the spectrum of $\mathcal{G}$ is observed at $\tilde \varepsilon_d = 0$ [Figs.~\ref{fig:19}(a) and~\ref{fig:19}(c)]. 
Thus the increase of $T$ causes $\mathcal G$ to exceed $e^2/2h$ in the case $\varepsilon_M = 0$, after which $\mathcal G$ drops below $e^2/2h$ at higher temperatures. 
Moreover, at higher temperatures the peaks of $\mathcal{G}$ are suppressed, in agreement with the results in the absence of EPI from Fig.~\ref{fig:10}. 
In contrast, for strong $|\lambda|$, the peaks of $\mathcal{G}$ decrease monotonically with $T$ from $e^2/2h$ at $T=0$ and $\tilde \varepsilon_d = 0$ [see Figs.~\ref{fig:19}(b) and~\ref{fig:19}(d)].
Therefore, the widths of the peaks are more narrow compared to the small $|\lambda|$ case. At high temperatures, the peaks are flattened to the point of masking the periodicity of $\mathcal{G}$.
In the case when $\varepsilon_M \neq 0$, the conductance has a $4\pi$ periodicity due to the finite renormalized dot level $\tilde \varepsilon_d$, regardless of the value of $|\lambda|$ [see Figs.~\ref{fig:19}(e) and~\ref{fig:19}(f)].
Also, the peaks are sharper when $|\lambda|$ is stronger.
The maxima of $\mathcal{G}$ reach the value $e^2/h$ at $T = 0$ and $\tilde \varepsilon_d = 0$, after which decrease with increasing temperature [see Figs.~\ref{fig:19}(g) and~\ref{fig:19}(h)]. Note that the maxima of $\mathcal{G}$ do not exceed the value $e^2/h$ with the increase of $T$ as was found in the case of the absence of EPI (Fig.~\ref{fig:10}).

We now turn our attention to the differential conductance $dI/dV$ in the presence of EPI. We plot in Fig.~\ref{fig:20} the zero-temperature differential conductance $dI/dV$ as a function of magnetic flux phase $\phi$, at different values of the electron-phonon coupling strength $\beta$, for unhybridized ($\varepsilon_M = 0$) and hybridized ($\varepsilon_M \neq 0$) MBSs, with weak symmetrical QD-MBS couplings $|\lambda_j| = |\lambda|$, at dot level $\varepsilon_d = 0$, with a fixed bias voltage $eV$. 
The differential conductance has a $2\pi$ periodicity as a function of magnetic flux phase $\phi$ when $\varepsilon_M = 0$, in the absence of EPI [see also Figs.~\ref{fig:5}(d)-\ref{fig:5}(f)].
When the EPI is taken into account, the minima of $dI/dV$, located without EPI at $\phi = 2n\pi$, with $n$ an integer, begin to increase with the electron-phonon coupling strength $\beta$.
When $\beta$ reaches the value $\beta/\Gamma \approx 1.685$ [see Fig.~\ref{fig:20}(a), blue line], the local minima emerging at $\phi = (2n+1)\pi$ are at the same value as the minima at $\phi = 2n\pi$ such that $dI/dV$ is approximately $\pi$-periodic. 
A further increase of $\beta$ results in recovering the $2\pi$ periodicity, but the maxima are located now at $\phi=2n\pi$.
Finally, the differential conductance is completely suppressed at large $\beta$.
In the case of finite overlap energy $\varepsilon_M \neq 0$ [see Fig.~\ref{fig:20}(b)], the differential conductance without EPI starts with a $2\pi$ periodicity as a function of $\phi$, with minima at $\phi = 2n\pi$, which in the presence of EPI changes to $4\pi$. 
Half of the $\beta=0$ minima, namely those occurring at $\phi=4n\pi$, transform into local plateaus when the
\begin{figure}[ht]
	\includegraphics[width =0.98\linewidth]{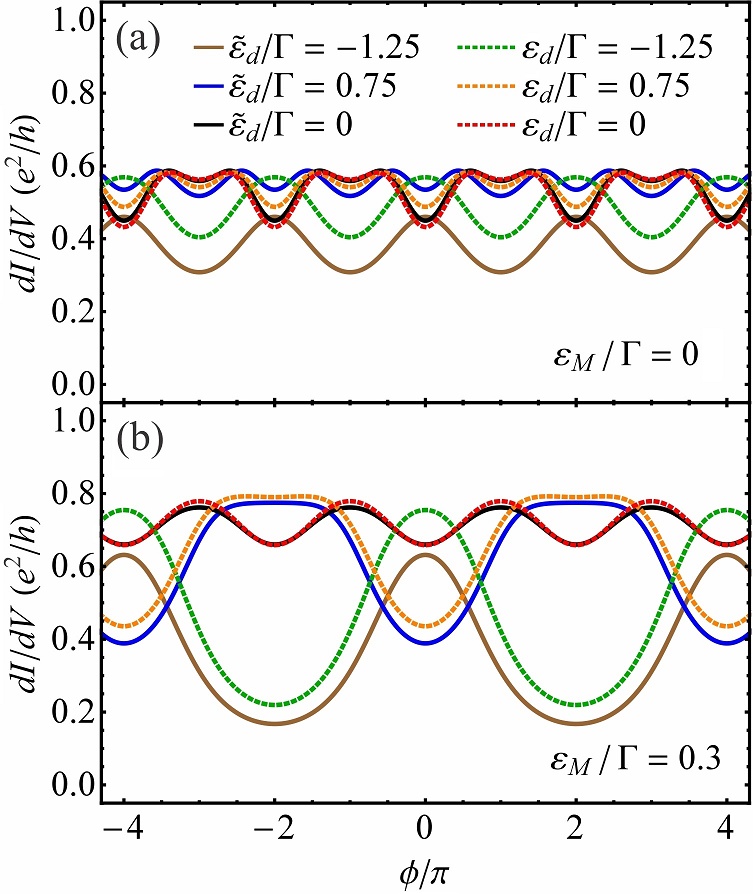}
	\centering
	\caption{The zero-temperature differential conductance $dI/dV$ as a function of magnetic flux phase $\phi$ for different values of the QD energy level $\varepsilon_d$, in the presence and absence of EPI ($\tilde \varepsilon_d=\varepsilon_d-\beta^2/\omega_0$, $\beta/\Gamma=2.5$), with bias voltage $eV/\Gamma = 0.28$, symmetrical QD-MBS couplings $|\lambda|/\Gamma = 0.3$, and overlap energy (a) $\varepsilon_M/\Gamma = 0$ and (b) $\varepsilon_M/\Gamma = 0.3$. The solid (dashed) lines denote results in the presence (absence) of EPI.}
	\label{fig:21}
\end{figure}
electron-phonon coupling strength $\beta$ reaches the critical value $\beta/\Gamma \approx 1.5$.
A further enhancement of $\beta$ leads to the plateaus changing into maxima of $dI/dV$ and finally to the suppression of the differential conductance. 
In short, under the variation of electron-phonon coupling strength $\beta$, we observe a $\pi$-phase shift in $dI/dV$ for $\varepsilon_{M}=0$ and a transition in the periodicity of $dI/dV$ from $2\pi$ to $4\pi$ for $\varepsilon_{M}\neq0$.

\begin{figure*}[ht]
	\includegraphics[width =0.75\linewidth]{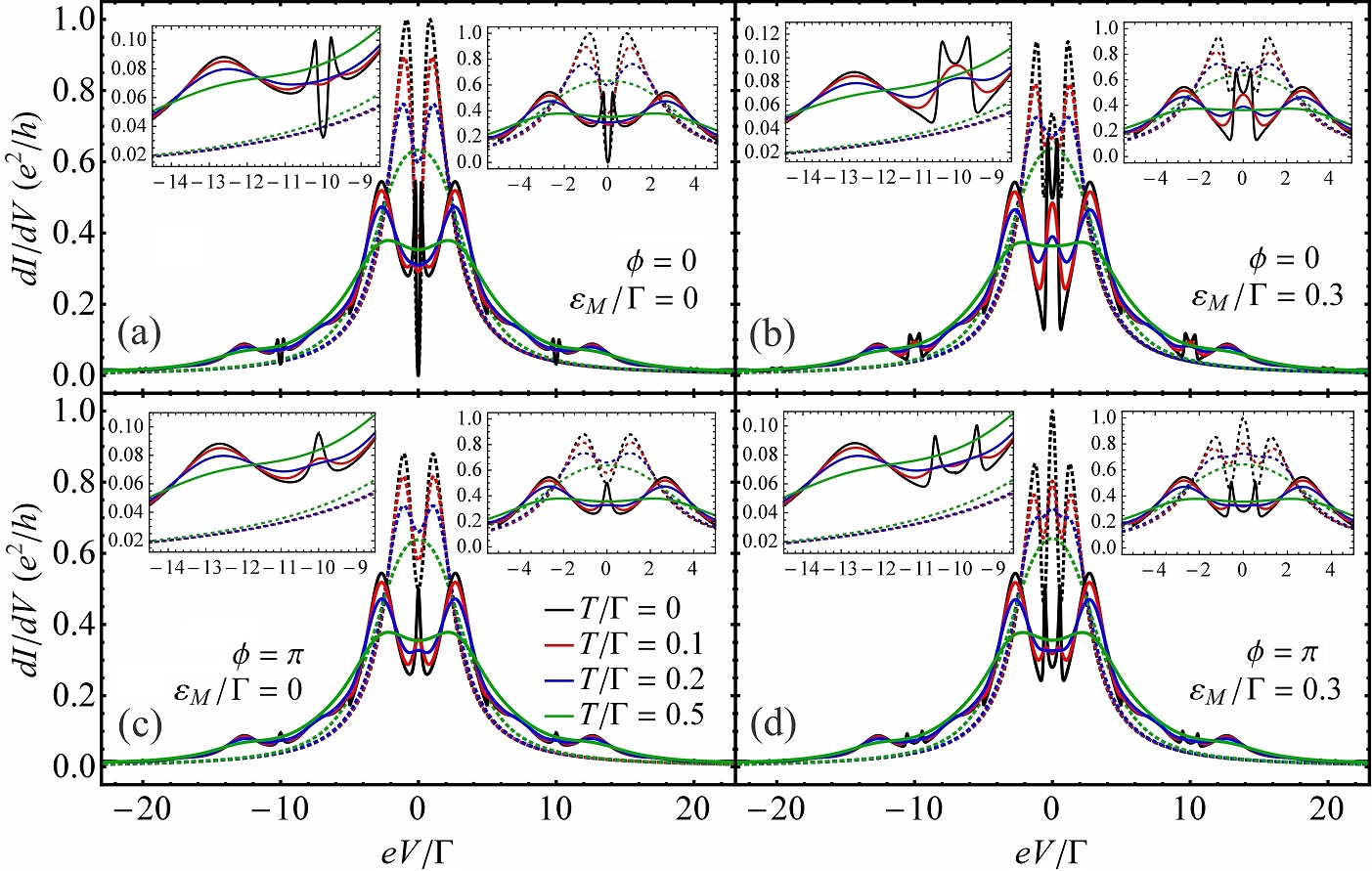}
	\centering
	\caption{The differential conductance $dI/dV$ as a function of bias voltage $eV$ in the presence of EPI for different values of temperature $T$ and magnetic flux phase $\phi$. The QD energy level is $\varepsilon_d/\Gamma = 0$ and the symmetrical QD-MBS couplings are $|\lambda|/\Gamma = 0.3$. The dashed lines correspond to the case of $\beta/\Gamma = 0$ and the solid lines correspond to $\beta/\Gamma = 2.5$. The rest of parameters are indicated in the figure.
	The right and left insets in each panel zoom in on the central and first phonon-induced satellite peak at negative energy, respectively.}
	\label{fig:22}
\end{figure*}
 
We now illustrate in Fig.~\ref{fig:21} the zero-temperature differential conductance $dI/dV$ as a function of magnetic flux phase $\phi$, with and without EPI, with the variation of the QD energy level $\varepsilon_d$, for unhybridized ($\varepsilon_M = 0$) and hybridized ($\varepsilon_M \neq 0$) MBSs, with symmetrical QD-MBS couplings $|\lambda_j| = |\lambda|$, at a fixed electron-phonon coupling strength $\beta$, and bias voltage $eV$. 
We observe that when there is a sufficiently strong electron-phonon coupling strength $\beta/\Gamma = 2.5$, the differential conductance is a $2\pi$-periodic function of $\phi$ when $\tilde \varepsilon_d = 0$, regardless of the value of $\varepsilon_M$, with minima at the points of $\phi = 2n \pi$ [see Figs.~\ref{fig:21}(a) and~\ref{fig:21}(b)].
For finite values of $\varepsilon_M$ and $\tilde \varepsilon_d$ [see Fig.~\ref{fig:21}(b)], the $dI/dV$ periodicity becomes $4\pi$. 
When the QD energy level matches $\varepsilon_d = \beta^2/\omega_0$, the renormalized dot level vanishes $\tilde \varepsilon_d = 0$, leading to the suppression of EPI effects on $dI/dV$ periodicity. 
Thus the $dI/dV$ periodicity as a function of $\phi$ approaches the case of $\beta = 0$ with the QD level $\varepsilon_d=0$. 
Consequently, the periodicity of $dI/dV$ changes when tuning $\varepsilon_d$.

We finally examine the differential conductance $dI/dV$ as a function of bias voltage $eV$, in the presence of EPI, at zero and finite temperatures, for different values of overlap energy between MBSs $\varepsilon_M$, magnetic flux phase $\phi$, and electron-phonon coupling strength $\beta$, when the QD weakly couples to the MBSs with symmetrical couplings $|\lambda_j| = |\lambda|$ and at $\varepsilon_d = 0$. The numerical results are presented in Fig.~\ref{fig:22}.
In the absence of EPI, we find at weak $|\lambda|$ that three resonant peaks appear in the $dI/dV$ spectra for $\varepsilon_M \neq 0$ [see Figs.~\ref{fig:22}(b) and~\ref{fig:22}(d), dashed lines], and only two peaks appear for $\varepsilon_M = 0$ [see Figs.~\ref{fig:22}(a) and~\ref{fig:22}(c), dashed lines]. Namely, at $eV = 0$ a peak develops for $\varepsilon_M \neq 0$ in $dI/dV$, while there is a minimum at $eV=0$ for $\varepsilon_M = 0$.
With increasing temperature, the resonant peaks decrease in amplitude. The central peak vanishes first and eventually the remaining peaks merge into a single peak at temperatures higher than the overlap energy between MBSs ($T > \varepsilon_M$), as was found in~\cite{Schuray2017} for weak $|\lambda|$.
Recall that the behavior of $dI/dV$ at $eV = 0$ and zero temperature is determined from the zero-energy spectral function $\mathcal{A}_d(0)$.
In the presence of EPI, we observe in Fig.~\ref{fig:22}(a), at $\varepsilon_{M}/\Gamma=0$ and $\phi=0$, that the two main peaks associated with the MBSs become narrow compared to the $\beta = 0$ case. 
Beside the sharp peaks, two broadened resonances appear in the spectrum at $eV \approx \pm 2 |\tilde \varepsilon_d|$ due to the EPI.
Therefore, a series of additional phonon-assisted channels emerge in the spectrum of $dI/dV$ at $eV \approx  \pm 2(|\tilde \varepsilon_d| + |p|\omega_0)$, with $p$ an integer, corresponding to tunneling of electrons through the QD by absorption and emission of phonons with frequency $\omega_0$. Near the side-band resonances, characteristic MBS-induced peaks develop at voltages $eV \approx \pm 2 |p| \omega_0$.
Note that at zero temperature, the zero-bias differential conductance is not affected by the EPI, as discussed above for $\mathcal{G}$. 
When the MBSs overlap [see Fig.~\ref{fig:22}(b)], the main three-peak structure (at $\beta = 0$) evolves into a spectrum of two narrow MBS-induced peaks with broadened side peaks at $eV \approx \pm 2 |\tilde\varepsilon_d|$ due to the presence of EPI. With increasing $T$, the two narrow peaks merge into one.
The peaks are totally smeared when the temperature exceeds the overlap energy, i.e., $T > \varepsilon_M$. 
The phonon-assisted peaks are similar to the central peak shape. 
In the case of $\phi = \pi$ with $\varepsilon_M = 0$ [see Fig.~\ref{fig:22}(c)], the main two-peak structure, which develops in the absence of EPI in the spectrum of $dI/dV$, transforms into a peak-structure that consists of a narrow peak at $eV = 0$ and two side-peaks at $eV \approx \pm 2 |\tilde\varepsilon_d|$. The zero-bias peak is immune to the EPI at zero temperature. In the same way, the phonon side-bands in the spectrum are characterized by the main MBS-induced peaks. Therefore, when $\varepsilon_M \neq 0$ [see Fig.~\ref{fig:22}(d)], the main three-peak structure evolves into a four-peak one, formed by two sharp MBS peaks and two broadened resonances. 
The satellite peaks follow again the central peak shape. At larger temperatures, the peaks are completely smeared.

\section{Summary}
\label{sec:IV}
\setcounter{equation}{0}

In this work, we have investigated the transport properties in a quantum dot system formed by two Majorana bound states located at the edges of a topological superconducting nanowire coupled laterally to a dot and in the presence of electron-phonon interaction. 
The dot-topological superconducting nanowire forms a ring system that is threaded by a tunable magnetic flux, allowing the control of transport properties. 
The electrons in the quantum dot interact with a single long-wave optical phonon mode, which leads to phonon-assisted processes in electron transport. 
The electrodes are oppositely biased, which results in a tunneling current through the dot with two components, one from the electron tunneling and the other from the local Andreev reflection. 
Crossed Andreev reflection is present only when the bias is asymmetrically applied. 
The electron-phonon interaction was incorporated in the model using a canonical transformation within the nonequilibrium Green's function formalism. 
The relevant retarded Green's functions have been determined by using the equation of motion method. 
The influence of electron-phonon interaction on transport characteristics in the presence of Majorana bound states has been analyzed in detail.
For comparison, we have studied a system both with and without electron-phonon interaction.

In the absence of electron-phonon interaction, the quantum dot spectral function shows a typical lineshape characterizing the presence of Majorana bound states.
The location of antiresonance points that occur in the spectrum is determined by the overlap energy between Majorana bound states. 
We have found that when the Majorana bound states do not overlap, the linear conductance is in equal manner attributed to electron tunneling and local Andreev reflection processes at zero temperature. 
Therefore, the linear conductance and its components are $2\pi$-periodic functions of magnetic flux phase and do not depend on dot energy or finite values of dot-Majorana bound states couplings at zero temperature.
We have also established that the electron tunneling and local Andreev reflection components of linear conductance respond differently to the change of temperature. 
Namely, the electron tunneling conductance exceeds the value $e^2/4h$, while the local Andreev reflection does not, for weak dot-Majorana couplings, at finite temperatures.
At finite temperatures, the linear conductance peaks are broadened with the increase of temperature and depend on dot energy and finite values of dot-Majorana couplings even for unhybridized Majorana bound states. 
The conductance $2\pi$ periodicity becomes $4\pi$ at finite dot energy $\varepsilon_d$ for hybridized Majorana bound states.
We have also established that the differential conductance has $2\pi$ periodicity for unhybridized Majoranas regardless of the value of dot energy, and it has $4\pi$ periodicity when the dot level is tuned for hybridized Majorana bound states. 
The differential conductance spectrum exhibits an asymmetric lineshape as a function of the dot energy. 
The differential conductance peaks attributed to the hybridized Majorana bound states are smeared when the temperature exceeds the value of Majorana overlap energy, in agreement with the literature results~\cite{Schuray2017}.
The periodicity of the linear and differential conductances holds also for an asymmetrically coupled quantum dot system, which would facilitate the experimental verification.

In the presence of electron-phonon interaction, the spectral function of the dot is modified. 
Alongside the peaks attributed to the Majorana bound states, a series of peaks, due to new phonon-assisted transport channels, develop in the spectrum, with shapes influenced by the electron-phonon coupling strength, temperature, overlap energy, and quantum dot energy. 
Most remarkably, when the Majorana bound states do not overlap, the linear conductance shows a $2\pi$ periodicity as a function of magnetic flux phase and is immune to the electron-phonon interaction, dot energy, and finite values of dot-Majorana bound states couplings at zero temperature.
Moreover, at finite temperatures, the linear conductance depends on dot energy and electron-phonon coupling strength for unhybridized Majoranas. 
In the case of hybridized Majorana bound states, the linear conductance has a $4\pi$ periodicity even for dot level $\varepsilon_d= 0$, due to dot energy being renormalized by electron-phonon coupling at zero and finite temperatures.
Consequently, the linear conductance periodicity may change to $4\pi$ for hybridized Majorana wave functions when varying the gate voltage or the electron-phonon coupling strength.
Therefore, a transition in the differential conductance periodicity between $2\pi$ and $\pi$ for perfectly degenerate Majorana bound states, as well as between $2\pi$ and $4\pi$ for hybridized Majorana bound states, can be achieved by changing the electron-phonon coupling strength.
The differential conductance exhibits phonon-assisted satellite peaks due to the electron-phonon interaction, whose height is sensitive to the change of electron-phonon coupling strength and temperature. 
The phonon-assisted peaks of differential conductance decrease with the increase of temperature and are eventually suppressed at high temperatures. 

The devices studied in the present paper are relevant for the quest of realizing topological quantum computation, and they stand within experimental reach. 
The topological superconducting nanowire loop structures can be engineered using molecular beam epitaxy techniques~\cite{Gazibegovic2017}. 
The integration of quantum dots in such semiconducting-superconducting heterostructures is also feasible within current technology~\cite{Deng2016,Lutchyn2018,Deng2018,Razmadze2020}.
The effects of an environmental phonon bath, which limits the coherence times in such devices, could thus be accounted first within theoretical toy models such as those proposed in Fig.~\ref{fig:1}.
We hope our work sheds light on the expected behavior for Majorana bound state signatures in more complex transport experiments where electron-phonon interaction plays a central role.\\
\begin{acknowledgements}
This work was supported by a grant of the Ministry of Research, Innovation and Digitization, CNCS/CCCDI – UEFISCDI, under Projects No. PN-III-P1-1.1-TE-2019-0423, and PN-III-P4-ID-PCCF-2016-0047 within PNCDI III.
\end{acknowledgements} 
\appendix
\onecolumngrid
\section{Current and conductances in the absence of EPI}
\label{sec:A}
In this Appendix, we derive formulas for current, and differential and linear conductances in the absence of EPI. We consider that the SC is grounded, $\mu_S = 0$, and an opposite bias voltage is applied to the leads, $\mu_L = -\mu_R = \frac{eV}{2}$. Following Eq.~\eqref{eq:14}, with the introduced conditions $\Gamma_{\alpha}^{e(h)} = \Gamma$, the current is symmetrized as
\begin{equation}
I=\frac{I_L - I_R}{2} = \frac{I_L^{ET}-I_R^{ET}}{2} + \frac{I_L^{LAR} - I_R^{LAR}}{2}
\equiv I_{ET} + I_{LAR},
\label{eq:A1}
\end{equation}
where the currents from the ET and LAR processes read
\begin{eqnarray}
I_{ET}&=& \frac{e}{h} \int d\varepsilon \, \mathcal{T}_{LR}^{ee}(\varepsilon) [f_{L}^{e} (\varepsilon)-f_{R}^{e} (\varepsilon)],
\label{eq:A2}\\
I_{LAR}&=& \frac{e}{h} \int d\varepsilon \, \mathcal{T}_{LL}^{eh}(\varepsilon) [f_{L}^{e} (\varepsilon)-f_{R}^{e} (\varepsilon)],
\label{eq:A3}
\end{eqnarray}
where $f_{\alpha}^e(\varepsilon) = f_{\alpha'}^h(\varepsilon)$. The corresponding transmission probabilities are
\begin{eqnarray}
\mathcal{T}_{LR}^{ee}(\varepsilon)&=&\mathcal{T}_{RL}^{ee}(\varepsilon)= \Gamma^2 |G_{d11}^r(\varepsilon)|^2\\
\mathcal{T}_{LL}^{eh}(\varepsilon) &=& \mathcal{T}_{RR}^{eh}(\varepsilon)=\Gamma^2 |G_{d12}^r(\varepsilon)|^2.
\label{eq:A4}
\end{eqnarray}
Moreover, the differential conductance corresponding to the ET and LAR currents is expressed as $dI/dV = dI_{ET}/dV + dI_{LAR}/dV$ with:

\begin{eqnarray}
\frac{dI_{ET}}{dV}&=& \frac{e^2}{2h} \frac{1}{T} \int d\varepsilon \, \mathcal{T}_{LR}^{ee}(\varepsilon)
\big\{ f_{L}^{e}(\varepsilon)[1-f_{L}^{e}(\varepsilon)]+f_{R}^{e}(\varepsilon)[1-f_{R}^{e}(\varepsilon)] \big\},
\label{eq:A5}\\
\frac{dI_{LAR}}{dV}&=& \frac{e^2}{2h} \frac{1}{T} \int d\varepsilon \, \mathcal{T}_{LL}^{eh}(\varepsilon)
\big\{ f_{L}^{e}(\varepsilon)[1-f_{L}^{e}(\varepsilon)]+f_{R}^{e}(\varepsilon)[1-f_{R}^{e}(\varepsilon)] \big\},
\label{eq:A6}
\end{eqnarray}
where we applied the identity $\frac{\partial f_{L(R)}^{e}(\varepsilon)}{\partial V}=\pm \frac{e}{2T} f_{L(R)}^{e}(\varepsilon)[1-f_{L(R)}^{e}(\varepsilon)]$. 
Therefore, the linear conductances read
\begin{eqnarray}
\mathcal{G}_{ET}&=&\frac{dI_{ET}}{dV}\bigg|_{V \to 0}
= \frac{e^2}{h} \frac{1}{T} \int d\varepsilon \, \mathcal{T}_{LR}^{ee}(\varepsilon) f(\varepsilon)[1-f(\varepsilon)],
\label{eq:A7}\\
\label{eq:A8}
\mathcal{G}_{LAR}&=&\frac{dI_{LAR}}{dV}\bigg|_{V \to 0}
=\frac{e^2}{h} \frac{1}{T} \int d\varepsilon \, \mathcal{T}_{LL}^{eh}(\varepsilon) f(\varepsilon)[1-f(\varepsilon)],
\end{eqnarray}
with the equilibrium Fermi-Dirac function $f(\varepsilon) = f_{\alpha}^{e}(\varepsilon)$. The zero-temperature differential conductances become
\begin{eqnarray}\label{eq:A9}
\frac{dI_{ET}}{dV} &=& \frac{e^2}{2h} \big[ \mathcal{T}_{LR}^{ee}(eV/2) + \mathcal{T}_{LR}^{ee}(-eV/2) \big],\\
\label{eq:A10}
\frac{dI_{LAR}}{dV} &=& \frac{e^2}{2h} \big[ \mathcal{T}_{LL}^{eh}(eV/2) + \mathcal{T}_{LL}^{eh}(-eV/2) \big],
\end{eqnarray}
where we introduced the notations $\mathcal{T}_{LR}^{ee}(\pm eV/2) = \mathcal{T}_{LR}^{ee}(\varepsilon = \pm eV/2)$ and $\mathcal{T}_{LL}^{eh}(\pm eV/2) = \mathcal{T}_{LL}^{eh}(\varepsilon = \pm eV/2)$ and used the property $f_{\alpha}^{e} (\varepsilon) = \theta(\mu_\alpha - \varepsilon)$, with $\theta(\mu_\alpha - \varepsilon)$ the Heaviside function. The corresponding zero-temperature linear conductances are
\begin{equation}
\mathcal{G}_{ET} = \frac{e^2}{h}\mathcal{T}_{LR}^{ee}(0), \,\,\, \, \mathcal{G}_{LAR} = \frac{e^2}{h}\mathcal{T}_{LL}^{eh}(0).
\label{eq:A11}
\end{equation}

In the following, we consider that the system is biased as follows: $\mu_L=q eV$ and $\mu_R = (q-1)eV$ such that $\mu_L - \mu_R = eV$ with the choice of $0 \le q \le 1$. This leads to the current
\begin{equation}
I_L = \frac{e}{h}\int d\varepsilon \mathcal{T}_{LR}^{ee}(\varepsilon)[f_{L}^{e}(\varepsilon)-f_{R}^{e}(\varepsilon)]
 + \frac{e}{h}\int d\varepsilon \mathcal{T}_{LL}^{eh}(\varepsilon)[f_{L}^{e} (\varepsilon)-f_{L}^{h} (\varepsilon)]
 + \frac{e}{h}\int d\varepsilon \mathcal{T}_{LR}^{eh}(\varepsilon)[f_{L}^{e} (\varepsilon)-f_{R}^{h} (\varepsilon)].
\label{eq:A12}
\end{equation}
Thus, the zero-temperature differential conductance is expressed as
\begin{equation}
\begin{split}
\frac{dI_L}{dV}&= \frac{e^2}{h}\big[q \mathcal{T}_{LR}^{ee}(qeV) + (1-q) \mathcal{T}_{LR}^{ee}((q-1)eV) \big]
+ \frac{e^2}{h}q\big[\mathcal{T}_{LL}^{eh}(qeV) + \mathcal{T}_{LL}^{eh}(-qeV) \big]\\
&\quad+ \frac{e^2}{h}\big[q \mathcal{T}_{LR}^{eh}(qeV) + (q-1) \mathcal{T}_{LR}^{eh}((1-q)eV) \big].
\end{split}
\label{eq:A13}
\end{equation}

\section{Green's functions of the dot}
\label{sec:B}
In this Appendix, we calculate the dot retarded Green's functions for arbitrary values of $\varepsilon_M$, $\lambda_1$, and $\lambda_2$ by using the EOM technique. To do this, we define the retarded Green's function for fermionic operators $A$ and $B$ as $\langle \langle A(t)|B(0)\rangle \rangle _t ^r =-i\theta(t)\langle \{A(t),B(0)\} \rangle$, where $\theta(t)$ is the Heaviside function~\cite{Kashcheyevs2006,VanRoermund2010,Mathe2020}. 
The Fourier transform for $\langle \langle A(t)|B(0)\rangle \rangle _t ^r$ is given by $\langle \langle A|B \rangle \rangle _\varepsilon^r$. 
We write down the EOM for the retarded Green's function in energy space as $\varepsilon^+ \langle \langle A|B \rangle \rangle_\varepsilon^r + \langle \langle [\mathcal{\bar{H}_{\text{el}}},A]|B \rangle \rangle _ \varepsilon^r=\langle \{A,B\} \rangle$, where $\varepsilon^+ = \varepsilon+i\delta$, with $\delta$ a positive infinitesimal~\cite{Kashcheyevs2006,VanRoermund2010,Mathe2020}. 
We introduce the electron-electron and electron-hole retarded Green's functions for the dot as $\tilde{G}_{d11}^r(\varepsilon)=\langle \langle d|d^{\dagger}\rangle \rangle_\varepsilon^r$ and $\tilde{G}_{d12}^r(\varepsilon)=\langle \langle d|d\rangle \rangle_\varepsilon^r$ by replacing $A(B)$ with $d (d^{\dagger})$ and $d (d)$, respectively.

The EOM for $\tilde{G}_{d11}^r(\varepsilon)$ is expressed as
\begin{equation}
(\varepsilon^+-\tilde \varepsilon_{d})\langle \langle d|d^{\dagger} \rangle \rangle_\varepsilon ^r + \frac{1}{\sqrt 2}(\tilde\lambda_1^* + \tilde\lambda_2^*)\langle\langle f^{\dagger}|d^{\dagger} \rangle\rangle_\varepsilon ^r + \frac{1}{\sqrt 2}(\tilde\lambda_1^* - \tilde\lambda_2^*)\langle\langle f|d^{\dagger} \rangle\rangle_\varepsilon ^r - \sum_{\alpha,k}\tilde V_{\alpha k}^* \langle \langle c_{\alpha k}|d^{\dagger}\rangle \rangle _\varepsilon ^r=1.
\label{eq:B1}
\end{equation}
To determine $\langle \langle d|d^{\dagger} \rangle \rangle_\varepsilon ^r$, we need to calculate the new higher-order correlation functions that appear in Eq.~\eqref{eq:B1}. The equation for the term $\langle \langle c_{\alpha k}|d^{\dagger}\rangle \rangle_\varepsilon ^r$ reads
\begin{equation}
\langle \langle c_{\alpha k}|d^{\dagger}\rangle \rangle_\varepsilon ^r = {\frac{\tilde V_{\alpha k}}{\varepsilon^+-\varepsilon_{\alpha k}}} \langle \langle d|d^{\dagger}\rangle \rangle_\varepsilon ^r.
\label{eq:B2}
\end{equation}
Using Eq.~\eqref{eq:B2}, in the wide-band limit, we write~\cite{Jauho1994}
\begin{equation}
\sum_{\alpha,k} \tilde V_{\alpha k}^*\langle \langle c_{\alpha k}|d^{\dagger}\rangle \rangle_\varepsilon ^r \approx -i\tilde\Gamma \langle \langle d|d^{\dagger}\rangle \rangle_\varepsilon ^r,
\label{eq:B3}
\end{equation}
where $\tilde\Gamma=\frac{1}{2}\sum_{\alpha}\tilde\Gamma_{\alpha}$ with $\tilde\Gamma_{\alpha}= 2\pi\sum_k |\tilde V_{\alpha k}|^2\delta(\varepsilon - \varepsilon_{\alpha k})$ being the coupling between the QD and the lead $\alpha$, where we assumed electron-hole symmetry in the system ($\Gamma_{\alpha}=\Gamma_{\alpha}^{e}=\Gamma_{\alpha}^{h}$), as mentioned in Sec.~\ref{sec:II}. Substituting Eq.~\eqref{eq:B3} into Eq.~\eqref{eq:B1}, we have 
\begin{equation}
(\varepsilon^+-\tilde \varepsilon_{d}+i\tilde\Gamma)\langle \langle d|d^{\dagger} \rangle \rangle_\varepsilon ^r + \frac{1}{\sqrt 2}(\tilde\lambda_1^* + \tilde\lambda_2^*)\langle\langle f^{\dagger}|d^{\dagger} \rangle\rangle_\varepsilon ^r + \frac{1}{\sqrt 2}(\tilde\lambda_1^* - \tilde\lambda_2^*)\langle\langle f|d^{\dagger} \rangle\rangle_\varepsilon ^r=1.
\label{eq:B4}
\end{equation}
The equations of motion for terms $\langle\langle f^{\dagger}|d^{\dagger} \rangle\rangle_\varepsilon ^r$ and $\langle\langle f|d^{\dagger} \rangle\rangle_\varepsilon ^r$ are
\begin{equation}
(\varepsilon^+ + \varepsilon_M)\langle\langle f^{\dagger}|d^{\dagger} \rangle\rangle_\varepsilon ^r - \frac{1}{\sqrt{2}}(\tilde\lambda_1^* - \tilde\lambda_2^*)\langle \langle d^{\dagger}|d^{\dagger}\rangle \rangle_\varepsilon^r + \frac{1}{\sqrt{2}}(\tilde\lambda_1 + \tilde\lambda_2)\langle \langle d|d^{\dagger}\rangle \rangle_\varepsilon^r =0,
\label{eq:B5}
\end{equation}
\begin{equation}
(\varepsilon^+ - \varepsilon_M)\langle\langle f|d^{\dagger} \rangle\rangle_\varepsilon ^r - \frac{1}{\sqrt{2}}(\tilde\lambda_1^* + \tilde\lambda_2^*)\langle \langle d^{\dagger}|d^{\dagger}\rangle \rangle_\varepsilon^r + \frac{1}{\sqrt{2}}(\tilde\lambda_1 - \tilde\lambda_2)\langle \langle d|d^{\dagger}\rangle \rangle_\varepsilon^r =0.
\label{eq:B6}
\end{equation}
We now write an equation for the correlation function $\langle \langle d^{\dagger}|d^{\dagger}\rangle \rangle_\varepsilon^r$ appearing in Eqs. \eqref{eq:B5} and \eqref{eq:B6},
\begin{equation}
(\varepsilon^+ + \tilde\varepsilon_d)\langle \langle d^{\dagger}|d^{\dagger}\rangle \rangle_\varepsilon^r -\frac{1}{\sqrt{2}}(\tilde\lambda_1 - \tilde\lambda_2)\langle \langle f^{\dagger}|d^{\dagger}\rangle \rangle_\varepsilon^r -\frac{1}{\sqrt{2}}(\tilde\lambda_1 + \tilde\lambda_2)\langle \langle f|d^{\dagger}\rangle \rangle_\varepsilon^r + \sum_{\alpha, k} \tilde{V}_{\alpha k} \langle\langle c^{\dagger}_{\alpha k}|d^{\dagger}\rangle\rangle_\varepsilon^r=0,
\label{eq:B7}
\end{equation}
where the equation for $\langle \langle c^{\dagger}_{\alpha k}|d^{\dagger}\rangle\rangle_\varepsilon^r$ is expressed as
\begin{equation}
\langle \langle c^{\dagger}_{\alpha k}|d^{\dagger}\rangle\rangle_\varepsilon^r =-\frac{\tilde{V}_{\alpha k}^{*}}{\varepsilon^+ + \varepsilon_{\alpha k}}\langle \langle d^{\dagger}|d^{\dagger} \rangle \rangle_\varepsilon^r.
\label{eq:B8}
\end{equation}
Here, assuming again electron-hole symmetry in the system, and substituting Eq.~\eqref{eq:B8} into Eq.~\eqref{eq:B7}, one obtains
\begin{equation}
(\varepsilon^+ + \tilde\varepsilon_d+i\tilde\Gamma)\langle \langle d^{\dagger}|d^{\dagger}\rangle \rangle_\varepsilon^r - \frac{1}{\sqrt{2}}(\tilde\lambda_1 - \tilde\lambda_2)\langle \langle f^{\dagger}|d^{\dagger}\rangle \rangle_\varepsilon^r - \frac{1}{\sqrt{2}}(\tilde\lambda_1 + \tilde \lambda_2)\langle \langle f|d^{\dagger}\rangle \rangle_\varepsilon^r =0.
\label{eq:B9}
\end{equation}
Now, we have a complete set of equations based on the higher-order correlation functions, which leads to expressing an analytical formula for the retarded Green's function $\langle \langle d|d^{\dagger} \rangle \rangle_\varepsilon ^r$. To do this, we substitute the terms $\langle \langle f^{\dagger}|d^{\dagger} \rangle \rangle _{\varepsilon}^r$ from Eq.~\eqref{eq:B5} and $\langle \langle f|d^{\dagger}\rangle \rangle_\varepsilon^r$ from Eq.~\eqref{eq:B6} into Eqs.~\eqref{eq:B4} and~\eqref{eq:B9}, and we obtain
\begin{equation}
\Big\{ (\varepsilon^+ - \tilde \varepsilon_d + i \tilde \Gamma) - \Big[ (|\tilde\lambda_1|^2 + |\tilde\lambda_2|^2) - \frac{\varepsilon_M}{\varepsilon} (\tilde\lambda_1 \tilde\lambda_2^* + \tilde\lambda_1^* \tilde\lambda_2) \Big] \mathcal{K}(\varepsilon)\Big\} \langle \langle d|d^{\dagger} \rangle \rangle_\varepsilon ^r + (\tilde\lambda_1^*\tilde\lambda_1^* - \tilde\lambda_2^*\tilde\lambda_2^*)\mathcal{K}(\varepsilon)\langle\langle d^{\dagger}|d^{\dagger} \rangle\rangle_\varepsilon ^r=1,
\label{eq:B10}
\end{equation}
\begin{equation}
\Big\{ (\varepsilon^+ + \tilde \varepsilon_d + i \tilde \Gamma) - \Big[ (|\tilde\lambda_1|^2 + |\tilde\lambda_2|^2) + \frac{\varepsilon_M}{\varepsilon} (\tilde\lambda_1 \tilde\lambda_2^* + \tilde\lambda_1^* \tilde\lambda_2) \Big] \mathcal{K}(\varepsilon)\Big\} \langle \langle d^{\dagger}|d^{\dagger} \rangle \rangle_\varepsilon ^r + (\tilde\lambda_1 \tilde\lambda_1 - \tilde\lambda_2 \tilde\lambda_2)\mathcal{K}(\varepsilon)\langle\langle d|d^{\dagger} \rangle\rangle_\varepsilon ^r=0,
\label{eq:B11}
\end{equation}
where we introduced the following notation:
\begin{equation}
\mathcal{K}(\varepsilon) = \frac{\varepsilon}{\varepsilon ^2 - \varepsilon_M^2}.
\label{eq:B12}
\end{equation}
Expressing $\langle \langle d^{\dagger}|d^{\dagger} \rangle \rangle_{\varepsilon}^r$ from Eq.~\eqref{eq:B11} and replacing it into Eq.~\eqref{eq:B10}, we find for the electron-electron contribution of the dot retarded Green's function
\begin{equation}
\tilde{G}_{d11}^r(\varepsilon)=\Big\{ (\varepsilon + \tilde \varepsilon_d + i \tilde \Gamma) - \Big[ (|\tilde\lambda_1|^2 + |\tilde\lambda_2|^2) + \frac{\varepsilon_M}{\varepsilon}(\tilde\lambda_1 \tilde\lambda_2^* + \tilde\lambda_1^* \tilde\lambda_2) \Big]\mathcal{K}(\varepsilon) \Big\} \mathcal{\tilde F}(\varepsilon)^{-1},
\label{eq:B13}
\end{equation}
where
\begin{equation}
\begin{split}
\mathcal{\tilde F}(\varepsilon)&=(\varepsilon - \tilde \varepsilon_d + i \tilde \Gamma)(\varepsilon + \tilde \varepsilon_d + i \tilde \Gamma) - 2(\varepsilon + i \tilde \Gamma)(|\tilde\lambda_1|^2 + |\tilde\lambda_2|^2)\mathcal{K}(\varepsilon) \\
&+ \frac{2\varepsilon_M}{\varepsilon}\tilde\varepsilon_d (\tilde\lambda_1 \tilde\lambda_2^* + \tilde\lambda_1^* \tilde\lambda_2)\mathcal{K}(\varepsilon) + \frac{1}{\varepsilon}(2|\tilde \lambda_1|^2|\tilde\lambda_2|^2 + \tilde\lambda_1 \tilde\lambda_1 \tilde\lambda_2^* \tilde\lambda_2^* + \tilde\lambda_1^* \tilde\lambda_1^* \tilde\lambda_2 \tilde\lambda_2)\mathcal{K}(\varepsilon).
\label{eq:B14}
\end{split}
\end{equation}
The EOM for $\tilde{G}_{d12}^r(\varepsilon)$ reads
\begin{equation}
(\varepsilon^+ - \tilde \varepsilon_{d})\langle \langle d|d \rangle \rangle_\varepsilon ^r + \frac{1}{\sqrt 2}(\tilde\lambda_1^* + \tilde\lambda_2^*)\langle\langle f^{\dagger}|d \rangle\rangle_\varepsilon ^r + \frac{1}{\sqrt 2}(\tilde\lambda_1^* - \tilde\lambda_2^*)\langle\langle f|d\rangle\rangle_\varepsilon ^r - \sum_{\alpha,k}\tilde V_{\alpha k}^* \langle \langle c_{\alpha k}|d\rangle \rangle _\varepsilon ^r = 0,
\label{eq:B15}
\end{equation}
where the equation for $\langle \langle c_{\alpha k}|d\rangle \rangle_\varepsilon ^r$ becomes
\begin{equation}
\langle \langle c_{\alpha k}|d\rangle \rangle_\varepsilon ^r = {\frac{\tilde V_{\alpha k}}{\varepsilon^+-\varepsilon_{\alpha k}}} \langle \langle d|d\rangle \rangle_\varepsilon ^r.
\label{eq:B16}
\end{equation}
Substituting Eq.~\eqref{eq:B16} into Eq.~\eqref{eq:B15}, we have
\begin{equation}
(\varepsilon^+ - \tilde \varepsilon_{d}+i\tilde\Gamma)\langle \langle d|d \rangle \rangle_\varepsilon ^r + \frac{1}{\sqrt 2}(\tilde\lambda_1^* + \tilde\lambda_2^*)\langle\langle f^{\dagger}|d \rangle\rangle_\varepsilon ^r + \frac{1}{\sqrt 2}(\tilde\lambda_1^* - \tilde\lambda_2^*)\langle\langle f|d \rangle\rangle_\varepsilon ^r=0.
\label{eq:B17}
\end{equation}
The equations for terms $\langle\langle f^{\dagger}|d \rangle\rangle_\varepsilon ^r$ and $\langle\langle f|d \rangle\rangle_\varepsilon ^r$ read
\begin{equation}
(\varepsilon^+ + \varepsilon_M)\langle\langle f^{\dagger}|d \rangle\rangle_\varepsilon ^r - \frac{1}{\sqrt{2}}(\tilde\lambda_1^* - \tilde\lambda_2^*)\langle \langle d^{\dagger}|d \rangle \rangle_\varepsilon^r + \frac{1}{\sqrt{2}}(\tilde\lambda_1 + \tilde\lambda_2)\langle \langle d|d \rangle \rangle_\varepsilon^r =0,
\label{eq:B18}
\end{equation}
\begin{equation}
(\varepsilon^+ - \varepsilon_M)\langle\langle f|d \rangle\rangle_\varepsilon ^r - \frac{1}{\sqrt{2}}(\tilde\lambda_1^* + \tilde\lambda_2^*)\langle \langle d^{\dagger}|d \rangle \rangle_\varepsilon^r + \frac{1}{\sqrt{2}}(\tilde\lambda_1 - \tilde\lambda_2)\langle \langle d|d \rangle \rangle_\varepsilon^r =0.
\label{eq:B19}
\end{equation}
Here, the equation for the term $\langle \langle d^{\dagger}|d \rangle \rangle_\varepsilon^r$ appearing in Eqs.~\eqref{eq:B18} and~\eqref{eq:B19} is expressed as
\begin{equation}
(\varepsilon^+ + \tilde\varepsilon_d)\langle \langle d^{\dagger}|d \rangle \rangle_\varepsilon^r - \frac{1}{\sqrt{2}}(\tilde\lambda_1 - \tilde\lambda_2)\langle \langle f^{\dagger}|d \rangle \rangle_\varepsilon^r - \frac{1}{\sqrt{2}}(\tilde\lambda_1 + \tilde \lambda_2)\langle \langle f|d \rangle \rangle_\varepsilon^r + \sum_{\alpha,k}\tilde V_{\alpha k} \langle \langle c^{\dagger}_{\alpha k}|d\rangle \rangle _\varepsilon ^r = 1,
\label{eq:B20}
\end{equation}
where the EOM for $\langle \langle c^{\dagger}_{\alpha k}|d\rangle \rangle_\varepsilon ^r$ reads
\begin{equation}
\langle \langle c^{\dagger}_{\alpha k}|d\rangle\rangle_\varepsilon^r =-\frac{\tilde{V}_{\alpha k}^{*}}{\varepsilon^+ + \varepsilon_{\alpha k}}\langle \langle d^{\dagger}|d\rangle \rangle_\varepsilon^r.
\label{eq:B21}
\end{equation}
Substituting Eq.~\eqref{eq:B21} into Eq.~\eqref{eq:B20}, one obtains
\begin{equation}
(\varepsilon^+ + \tilde\varepsilon_d+i\tilde\Gamma)\langle \langle d^{\dagger}|d \rangle \rangle_\varepsilon^r - \frac{1}{\sqrt{2}}(\tilde\lambda_1 - \tilde\lambda_2)\langle \langle f^{\dagger}|d \rangle \rangle_\varepsilon^r - \frac{1}{\sqrt{2}}(\tilde\lambda_1 + \tilde \lambda_2)\langle \langle f|d \rangle \rangle_\varepsilon^r =1.
\label{eq:B22}
\end{equation}
Now, achieving a complete set of equations, we express an analytical formula for the retarded Green's function $\langle \langle d|d \rangle \rangle_\varepsilon^r $. First, we substitute $\langle \langle f^{\dagger}|d \rangle \rangle _{\varepsilon}^r$ from Eq.~\eqref{eq:B18} and $\langle \langle f|d \rangle \rangle_\varepsilon^r$ from Eq.~\eqref{eq:B19} into Eqs.~\eqref{eq:B17} and~\eqref{eq:B22}, and we find
\begin{equation}
\Big\{ (\varepsilon^+ - \tilde \varepsilon_d + i \tilde \Gamma) - \Big[ (|\tilde\lambda_1|^2 + |\tilde\lambda_2|^2) -\frac{\varepsilon_M}{\varepsilon} (\tilde\lambda_1 \tilde\lambda_2^* + \tilde\lambda_1^* \tilde\lambda_2) \Big] \mathcal{K}(\varepsilon)\Big\} \langle \langle d|d \rangle \rangle_\varepsilon ^r + (\tilde\lambda_1^* \tilde\lambda_1^* - \tilde\lambda_2^* \tilde\lambda_2^*)\mathcal{K}(\varepsilon)\langle\langle d^{\dagger}|d \rangle\rangle_\varepsilon ^r=0,
\label{eq:B23}
\end{equation}
\begin{equation}
\Big\{ (\varepsilon^+ + \tilde \varepsilon_d + i \tilde \Gamma) - \Big[ (|\tilde\lambda_1|^2 + |\tilde\lambda_2|^2) +\frac{\varepsilon_M}{\varepsilon} (\tilde\lambda_1 \tilde\lambda_2^* + \tilde\lambda_1^* \tilde\lambda_2) \Big] \mathcal{K}(\varepsilon)\Big\} \langle \langle d^{\dagger}|d \rangle \rangle_\varepsilon ^r + (\tilde\lambda_1 \tilde\lambda_1 - \tilde\lambda_2 \tilde\lambda_2)\mathcal{K}(\varepsilon)\langle\langle d|d \rangle\rangle_\varepsilon ^r=1.
\label{eq:B24}
\end{equation}
Substituting $\langle \langle d^{\dagger}|d \rangle \rangle_\varepsilon^r $ from Eq.~\eqref{eq:B24} into Eq.~\eqref{eq:B23} yields the electron-hole contribution of the dot retarded Green's function,
\begin{equation}
\tilde{G}_{d12}^r(\varepsilon)=-(\tilde\lambda_1^* \tilde\lambda_1^* - \tilde\lambda_2^* \tilde\lambda_2^*)\mathcal{K}(\varepsilon) \mathcal{\tilde F}(\varepsilon)^{-1}.
\label{eq:B25}
\end{equation}

Finally, maxima of the spectral function in the absence of EPI are determined from the poles of the retarded Green's function $G_{d11}^r(\varepsilon)$ in the limit $\Gamma \to 0$~\cite{Gorski2018}. 
The expression for the poles of ${G}_{d11}^r(\varepsilon)$ when $ \lambda_1=| \lambda_1|e^{i\phi/4}$ and $ \lambda_2=| \lambda_2|e^{-i\phi/4}$ reads
\begin{equation}
	\begin{split}
		2\varepsilon^2&\approx 2(|\lambda_1|^2 + |\lambda_2|^2) + \varepsilon_d^2 + \varepsilon_M^2 \\
		&\pm \sqrt{[2(|\lambda_1|^2 + |\lambda_2|^2) + \varepsilon_d^2 + \varepsilon_M^2]^2 - 4\{\varepsilon_d^2 \varepsilon_M^2 + 4|\lambda_1||\lambda_2| \cos(\phi/2)[|\lambda_1||\lambda_2| \cos(\phi/2) + \varepsilon_d\varepsilon_M]\}}.
	\end{split}
	\label{eq:B26}
\end{equation}

\section{Current and conductances in the presence of EPI}
\label{sec:C}
The relations~\eqref{eq:27} and \eqref{eq:29} with the constraint $\tilde \Gamma_{\alpha}^{e} = \tilde \Gamma_{\alpha}^{h}=\tilde \Gamma_{\alpha}=\tilde \Gamma$ determine the spectral function
\begin{equation}
\begin{split}
\mathcal{A}_d(\varepsilon) =\frac{\Gamma \tilde \Gamma}{2} \sum_{p=-\infty}^{\infty} \mathcal{L}_p &\big\{ \tilde{\mathcal{P}}_{d}(\varepsilon + p\omega_0) \big[f_{L}^{e}(\varepsilon + p\omega_0) + f_{R}^{e}(\varepsilon + p\omega_0)\big]\\
&+ \tilde{\mathcal{P}}_{d}(\varepsilon - p\omega_0) \big[2 - f_{L}^{e}(\varepsilon - p\omega_0) - f_{R}^{e}(\varepsilon - p\omega_0)\big] \big\},
\end{split}
\label{eq:C1}
\end{equation}
where the identity $f_{\alpha}^{e}(x)=f_{\alpha '}^{h}(x)$ and the following notation are used:
\begin{equation}
\tilde{\mathcal{P}}_{d}(x) = |\tilde{G}_{d11}^{r}(x)|^{2} + |\tilde{G}_{d12}^{r}(x)|^{2}.
\label{eq:C2}
\end{equation}
At zero temperature, the spectral function given by~\eqref{eq:C1} reduces to
\begin{equation}
\begin{split}
\mathcal{A}_d(\varepsilon) =\frac{\Gamma \tilde \Gamma}{2} e^{-g} \sum_{p=0}^{\infty}\frac{g^p}{p!}&\big\{ \tilde{\mathcal{P}}_{d}(\varepsilon + p\omega_0) \big[\theta (eV/2-(\varepsilon + p\omega_0)) + \theta (-eV/2-(\varepsilon + p\omega_0))\big]\\
&+ \tilde{\mathcal{P}}_{d}(\varepsilon - p\omega_0) \big[2 - \theta (eV/2-(\varepsilon - p\omega_0)) - \theta (-eV/2-(\varepsilon - p\omega_0))\big] \big\}.
\end{split}
\label{eq:C3}
\end{equation}
Moreover, the current formula given by Eq.~\eqref{eq:18} in the presence of EPI becomes
\begin{equation}
\begin{split}
I=\frac{e}{2h} \Gamma \tilde \Gamma \sum_{p=-\infty}^{\infty} \mathcal{L}_p \int &d\varepsilon \big[f_{L}^{e}(\varepsilon) - f_{R}^{e}(\varepsilon)\big]\big\{ \tilde{\mathcal{P}}_{d}(\varepsilon + p\omega_0) \big[f_{L}^{e}(\varepsilon + p\omega_0) + f_{R}^{e}(\varepsilon + p\omega_0)\big]\\
&+ \tilde{\mathcal{P}}_{d}(\varepsilon - p\omega_0) \big[2 - f_{L}^{e}(\varepsilon - p\omega_0) - f_{R}^{e}(\varepsilon - p\omega_0)\big] \big\}.
\end{split}
\label{eq:C4}
\end{equation}
At finite temperature, the differential conductance is expressed as
\begin{equation}
\begin{split}
\frac{dI}{dV}=\frac{e^2}{4h}\frac{\Gamma \tilde \Gamma}{T} \sum_{p=-\infty}^{\infty}& \mathcal{L}_p \int d\varepsilon \big[ \tilde{\mathcal{P}}_{d}(\varepsilon + p\omega_0)\big\{ \varrho_{+}^{e}(\varepsilon)\big[f_{L}^{e}(\varepsilon + p\omega_0)+ f_{R}^{e}(\varepsilon + p\omega_0)\big]+\varrho_{-}^{e}(\varepsilon + p\omega_0)\big[f_{L}^{e}(\varepsilon) - f_{R}^{e}(\varepsilon)\big]\big\}\\
&+\tilde{\mathcal{P}}_{d}(\varepsilon - p\omega_0)\big\{ \varrho_{+}^{e}(\varepsilon) \big[2 - f_{L}^{e}(\varepsilon - p\omega_0) - f_{R}^{e}(\varepsilon - p\omega_0)\big] -\varrho_{-}^{e}(\varepsilon - p\omega_0)\big[f_{L}^{e}(\varepsilon) - f_{R}^{e}(\varepsilon)\big]\big\}\big],
\end{split}
\label{eq:C5}
\end{equation}
where we introduced the notation
\begin{equation}
\varrho_{\pm}^{e}(x) = f_{L}^{e}(x) \big[ 1 - f_{L}^{e}(x) \big] \pm f_{R}^{e}(x) \big[ 1 - f_{R}^{e}(x) \big].
\label{eq:C6}
\end{equation}
In the absence of EPI, the relation~\eqref{eq:C5} reduces to Eqs.~\eqref{eq:A5}-\eqref{eq:A6}. The linear conductance reads
\begin{equation}
\mathcal{G}=\frac{dI}{dV}\bigg|_{V \to 0}=\frac{e^2}{h}\frac{\Gamma \tilde \Gamma}{T} \sum_{p=-\infty}^{\infty} \mathcal{L}_p \int d\varepsilon f(\varepsilon)[1 - f(\varepsilon)] \big\{ \tilde{\mathcal{P}}_{d}(\varepsilon + p\omega_0) f(\varepsilon + p\omega_0)+\tilde{\mathcal{P}}_{d}(\varepsilon - p\omega_0) [1 - f(\varepsilon - p\omega_0)]\big\},
\label{eq:C7}
\end{equation}
with the equilibrium Fermi-Dirac function $f(x)=f_{\alpha}^{e}(x)$. For $\beta = 0$, Eq.~\eqref{eq:C7} changes to the relations~\eqref{eq:A7}-\eqref{eq:A8}. The zero-temperature differential conductance reads
\begin{eqnarray}
\label{eq:C8}
&&\frac{dI}{dV}=\frac{e^2}{4h}\Gamma \tilde \Gamma e^{-g} \bigg\{ 2\bigg[ \tilde{\mathcal{P}}_{d}\Big(\frac{eV}{2}\Big) + \tilde{\mathcal{P}}_{d}\Big(-\frac{eV}{2}\Big) \bigg] + \sum_{p=1}^{\infty} \frac{g^p}{p!} \bigg[ \theta(-eV-p\omega_0) \tilde{\mathcal{P}}_{d}\Big(\frac{eV}{2} + p\omega_0\Big)\notag\\
&&+\theta(eV-p\omega_0) \tilde{\mathcal{P}}_{d}\Big(-\frac{eV}{2} + p\omega_0\Big) + [1- \theta(-eV+p\omega_0)] \tilde{\mathcal{P}}_{d}\Big(\frac{eV}{2} - p\omega_0\Big) + [1- \theta(eV+p\omega_0)] \tilde{\mathcal{P}}_{d}\Big(-\frac{eV}{2} - p\omega_0\Big)\\
&& + [1- \theta(-eV+p\omega_0) + \theta(-eV-p\omega_0)]\tilde{\mathcal{P}}_{d}\Big(\frac{eV}{2}\Big) + [1- \theta(eV+p\omega_0) + \theta(eV-p\omega_0)]\tilde{\mathcal{P}}_{d}\Big(-\frac{eV}{2}\Big)\bigg]\bigg\},\notag
\end{eqnarray}
where $f_{\alpha}^{e} (x) = \theta(\mu_\alpha - x)$, with $\theta(\mu_\alpha - x)$ the Heaviside function.
In the absence of EPI, the relation \eqref{eq:C8} reduces to Eqs.~\eqref{eq:A9} and~\eqref{eq:A10}. From Eq.~\eqref{eq:C8}, the zero-temperature linear conductance becomes
\begin{equation}
\mathcal{G}=\frac{dI}{dV}\bigg|_{V \to 0}
=\frac{e^2}{h}\Gamma \tilde \Gamma e^{-g} \tilde{\mathcal{P}}_{d}(0),
\label{eq:C9}
\end{equation}
which then reduces to Eq.~\eqref{eq:A11} for $\beta = 0$.
\twocolumngrid
\bibliographystyle{apsrev4-2}
\bibliography{References}
\end{document}